\documentclass[pdflatex,sn-mathphys-num]{my-sn-jnl}% Math and Physical Sciences Numbered Reference Style
\usepackage{graphicx}%
\usepackage{multirow}%
\usepackage{amsmath,amssymb,amsfonts}%
\usepackage{amsthm}%
\usepackage{mathrsfs}%
\usepackage[title]{appendix}%
\usepackage{xcolor}%
\usepackage{textcomp}%
\usepackage{manyfoot}%
\usepackage{booktabs}%
\usepackage{algorithm}%
\usepackage{algorithmicx}%
\usepackage{algpseudocode}%
\usepackage{listings}%
\usepackage{url}%
\usepackage{hyperref}%
\usepackage[acronym]{glossaries}%
\usepackage{pifont}%
\usepackage{tikz}%
\usepackage{subfigure}%
\usetikzlibrary{shapes.geometric, arrows, calc, positioning,}%
\usepackage{float}%
\usepackage[acronym]{glossaries}
\makeglossaries
\newacronym{RS}{RS}{Recommender System}
\newacronym{RPRS}{RP-RS}{Research-Paper Recommender System}
\newacronym{POI}{POI}{Paper of Interest}
\newacronym{FoS}{FoS}{Field of Study}
\newacronym{HIN}{HIN}{Heterogeneous Information Network}
\newacronym{CN}{CN}{Citation Network}
\newacronym{VSM}{VSM}{Vector Space Model}
\newacronym{LLM}{LLM}{Large Languages Model}
\newacronym{RNN}{RNN}{Recurrent Neural Networks}
\newacronym{CNN}{CNN}{Convolution Neural Network}
\newacronym{GNN}{GNN}{Graph Neural Network}
\newacronym{LFV}{LFV}{Latent Feature Vector}
\newacronym{MF}{MF}{Matrix Factorization}
\newacronym{FV}{FV}{Feature Vector}
\newacronym{TD}{TD}{Topic Distribution}
\newacronym{LDA}{LDA}{Latent Dirichlet Allocation}
\newacronym{MLP}{MLP}{Multi Layer Perceptron}
\newacronym{GCN}{GCN}{Graph Convolution Network}
\newacronym{RW}{RW}{Random Walk}
\newacronym{TM}{TM}{Topic Modeling}
\newacronym{ML}{ML}{Machine Learning}
\newacronym{ER}{ER}{Evidential Reasoning}
\newacronym{HAN}{HAN}{Hierarchical Attention Network}
\newacronym{SVM}{SVM}{Support Vector Machine}
\newacronym{NN}{NN}{Neural Network}
\newacronym{MSE}{MSE}{Mean Square Error}
\newacronym{JSD}{JSD}{Jensen-Shannon Divergence}
\newacronym{SSN}{SSN}{Scientific Social Network}
\newacronym{MAP}{MAP}{Mean Average Precision}
\newacronym{NDCG}{NDCG}{Normalized Discounted Cumulative Gain }
\newacronym{MRR}{MRR}{Mean Reciprocal Rank}
\newacronym{TNR}{TNR}{True Negative Rate}
\newacronym{CAGR}{CAGR}{Compound Annual Growth Rate}
\newacronym{HR}{HR}{Hit Rate}
\newacronym{AUC}{AUC}{Area Under the Curve}
\newacronym{ACL}{ACL}{Association for Computational Linguistics}
\newacronym{ARC}{ARC}{Anthology Reference Corpus}
\newacronym{AAN}{AAN}{ACL Anthology Network}
\newacronym{NLP}{NLP}{Natural Language Processing}
\newacronym{IR}{IR}{Information Retrieval}
\newacronym{WoS}{WoS}{Web of Science}
\newacronym{KDM}{KDM}{Knowledge Discovery Model}
\newacronym{AI}{AI}{Artificial Intelligence}
\newacronym{WMS}{WMS}{Weighted Mean Squared}
\newacronym{WARP}{WARP}{Weighted Approximate-Rank Pairwise}
\newacronym{PMRA}{PMRA}{PubMed Related Article}
\newacronym{DL}{DL}{Deep Learning}
\newacronym{ALS}{ALS}{Alternating Least Squares}
\newacronym{BPR}{BPR}{Bayesian Personalized Ranking}
\newacronym{PMF}{PMF}{Probabilistic Matrix Factorizatio}

\def\RSS{\glspl*{RS}}
\def\RPRS{\gls*{RPRS}}
\def\RPRSS{\glspl*{RPRS}}
\def\POI{\gls*{POI}}

\def\FoS{\gls*{FoS}}

\def\HIN{\gls*{HIN}}
\def\HINS{\glspl*{HIN}}
\def\CNS{\glspl*{CN}}
\def\VSM{\gls*{VSM}}

\def\LLMS{\glspl*{LLM}}
\def\RNN{\gls*{RNN}}
\def\RNNS{\glspl*{RNN}}
\def\CNN{\gls*{CNN}}
\def\CNNS{\glspl*{CNN}}
\def\GNN{\gls*{GNN}}
\def\GNNS{\glspl*{GNN}}
\def\LFV{\gls*{LFV}}
\def\LFVS{\glspl*{LFV}}
\def\MF{\gls*{MF}}

\def\FV{\gls*{FV}}

\def\TD{\gls*{TD}}

\def\LDA{\gls*{LDA}}

\def\MLP{\gls*{MLP}}
\def\MLPS{\glspl*{MLP}}
\def\GCN{\gls*{GCN}}
\def\GCNS{\glspl*{GCN}}
\def\RW{\gls*{RW}}

\def\TM{\gls*{TM}}

\def\ML{\gls*{ML}}

\def\ER{\gls*{ER}}

\def\HAN{\gls*{HAN}}

\def\WARP{\gls*{WARP}}

\def\SVM{\gls*{SVM}}

\def\NN{\gls*{NN}}

\def\MSE{\gls*{MSE}}

\def\JSD{\gls*{JSD}}

\def\SSN{\gls*{SSN}}
\def\SSNS{\glspl*{SSN}}
\def\MAP{\gls*{MAP}}

\def\NDCG{\gls*{NDCG}}

\def\MRR{\gls*{MRR}}

\def\TNR{\gls*{TNR}}

\def\CAGR{\gls*{CAGR}}

\def\HR{\gls*{HR}}

\def\AUC{\gls*{AUC}}

\def\ACL{\gls*{ACL}}

\def\ARC{\gls*{ARC}}
\def\AAN{\gls*{AAN}}
\def\LLMS{\glspl*{LLM}}
\def\NLP{\gls*{NLP}}
\def\IR{\gls*{IR}}
\def\WoS{\gls*{WoS}}
\def\KDM{\gls*{KDM}}
\def\AI{\gls*{AI}}
\def\DL{\gls*{DL}}
\def\PMF{\gls*{PMF}}
\def\WMS{\gls*{WMS}}
\def\WARP{\gls*{WARP}}
\def\PMRA{\gls*{PMRA}}

\def\ALS{\gls*{ALS}}
\def\BPR{\gls*{BPR}}
\theoremstyle{thmstyleone}%
\theoremstyle{thmstyletwo}%

\theoremstyle{thmstylethree}%

\begin{document}

\title[Article Title]{Recent Advances and Trends in Research Paper Recommender Systems: A Comprehensive Survey}


\author[1]{\fnm{Iratxe} \sur{Pinedo}}\email{iratxe.pinedo@ehu.eus}

\author[1]{\fnm{Mikel} \sur{Larrañaga}}\email{mikel.larranaga@ehu.eus}

\author*[1]{\fnm{Ana} \sur{Arruarte}}\email{a.arruarte@ehu.eus}

\affil[1]{\orgdiv{Languages and Information Systems}, \orgname{University of the Basque Country UPV/EHU}, \orgaddress{\street{Manuel Lardizabal Ibilbidea, 1}, \city{Donostia}, \postcode{20018}, \state{Basque Country}, \country{Spain}}}


\abstract{As the volume of scientific publications grows exponentially, researchers increasingly face difficulties in locating relevant literature. Research Paper Recommender Systems have become vital tools to mitigate this information overload by delivering personalized suggestions. This survey provides a comprehensive analysis of Research Paper Recommender Systems developed between November 2021 and December 2024, building upon prior reviews in the field. It presents an extensive overview of the techniques and approaches employed, the datasets utilized, the evaluation metrics and procedures applied, and the status of both enduring and emerging challenges observed during the research. Unlike prior surveys, this survey goes beyond merely cataloguing techniques and models, providing a thorough examination of how these methods are implemented across different stages of the recommendation process. By furnishing a detailed and structured reference, this work aims to function as a consultative resource for the research community, supporting informed decision-making and guiding future investigations in the advances of effective Research Paper Recommender Systems.}

\keywords{Research Paper Recommender Systems, Academic Information Retrieval, Literature Review, Research Trends}



\maketitle

\section{Introduction}\label{sec1}

Global scientific publications are projected to continue growing significantly in the coming years. The overall market for scientific publishing is expected to grow at a \CAGR\ of 3.3\% from 2024 to 2030 \cite{OrbisResearch2023}. For example, computer science publications are projected to reach 580,000 annually by 2025, reflecting a surge driven by advances in fields such as artificial intelligence and quantum computing \cite{PlutoInsights2025}. In the case of the medical sector, it is expected to see a 5-7\% annual increase in publications through 2025, driven by \AI\ and biotechnology. Meanwhile, the percentage of publications available through open access is forecasted to reach 45\% by 2025, up from 38\% in 2023 \cite{STM2024}. These trends reflect a broader, robust expansion in scientific research output across disciplines, particularly in computer science, fueled by technological advances and growing research investments.

 While this expansion makes research more accessible, it also presents challenges such as information overload. Furthermore, this overwhelming growth makes it increasingly difficult for researchers and scholars to find relevant papers aligned with their specific interests. However, tools such as \RPRSS\ help address these issues by offering tailored research suggestions, facilitating the discovery of high-quality works, and allowing researchers to stay focused and up-to-date.

\RPRSS\ have evolved by adapting to new technologies, and although some of the underlying ideas  and procedures remain, current systems are quite different from the first systems developed in the early 1990s.
\DL \cite{IJ:DSM:Sivasankari:2023, IJ:ESWA:XIAO:2023, IJ:AS:Wang:2024, IC:DASFAA:Wang:2022, arXiv:KARIMI:2022}, and \LLMS \cite{IJ:Scientometrics:ALI:2022, IC:WIST:Roßner:2023,IJ:AI:GAO:2023, IC:ASSP:Shen:2024,IJ:Scientometrics:Thierry:2023} are more and more present in new proposals.
In addition, \SSNS \cite{IJ:ASC:Wang:2024,IJ:SMC:WANG:2022,IJ:mathematics:GUO:2024} also offer new opportunities that are undoubtedly being exploited. Despite the successful incorporation of all these advances, there is still significant room for improvement in both the evaluation approaches 
and the adoption of these systems in real world scenarios \cite{IJ:KIS:STERGIOPOULOS:2024}.

In this survey, an in-depth analysis of the current state of research on \RPRSS\ is provided, using as a starting point one of the most recent, complete and comprehensive surveys on the subject \cite{IJ:DL:KREUTZ:2022}.  Since the work by \cite{IJ:DL:KREUTZ:2022} encompasses data up to October 2021, this new review extends the analysis to cover the period from November 2021 to December 2024,  maintaining consistency by employing the same bibliographic review methodology. The main goal of this work is to answer the following research question:
\begin{itemize}
    \item \textbf{RQ1}: How do the \RPRSS\ generate the recommendations?
    \item \textbf{RQ2}: What datasets are used to train and evaluate the \RPRSS?
    \item \textbf{RQ3}: How are the \RPRSS\ evaluated?
    \item \textbf{RQ4}: What are the main challenges addressed by the \RPRSS?
    \item \textbf{RQ5}: What new challenges have arisen?
\end{itemize}

The survey is structured as follows. Section~\ref{sec:Methodology} introduces the methodology and approach followed in designing the survey, including the data collection techniques and the overall research framework. Section~\ref{sec:Results} provides an in-depth analysis of the works found in the literature review. Section~\ref{sec:Discussion} presents a discussion on challenges identified in previous surveys, examines their current status, and highlights new challenges emerging from the present survey. Finally, Section~\ref{sec:Conclusion} concludes the survey, summarizing the key findings and offering an overall conclusion, including suggestions for future research.

\section{Methodology}\label{sec:Methodology}

The main goal of this work is to provide insight into the field of \RPRSS, analyzing the trends of the last few years and comparing its evolution with previous systematic literature reviews. To this end, this work aims to answer the following research questions:

\begin{itemize}
\item \textbf{RQ1}: How do the \RPRSS\ generate the recommendations?

The aim of this research question is to classify the works presented and to gain an understanding of how the systems generate the recommendations. In the literature, \RPRSS\ have traditionally been classified according to the taxonomy proposed by \cite{burkeHybridRecommenderSystems2002}, even in literature reviews such as \cite{IJ:DL:BEEL:2015}. However, as stated by \cite{IJ:DL:KREUTZ:2022}, this taxonomy has limitations to categorize \RPRSS\ that have been developed considering newer approaches and techniques. Therefore, this work classifies the \RPRSS\ considering the technology used and the purpose of the technique. To this end, \textit{RQ1} has been divided into the following questions:
\subitem $\circ$ \textbf{RQ1.1}: What type of input do the \RPRSS\ use to generate the recommendations?
\subitem $\circ$ \textbf{RQ1.2}: What kinds of information do the \RPRSS\ use to generate the recommendations?
\subitem $\circ$ \textbf{RQ1.3}: How do the \RPRSS\ represent the papers?
\subitem $\circ$ \textbf{RQ1.4}: How do the \RPRSS\ represent the users?
\subitem $\circ$ \textbf{RQ1.5}: How do the \RPRSS\ generate the recommendations?
\item \textbf{RQ2}: What datasets are used to train and evaluate \RPRSS?

Understanding which datasets are used to train and evaluate \RPRSS\ is fundamental for the advance of research in the field. This includes examining the origins, characteristics, modifications and availability of these datasets to assess their suitability and impact on reproducibility and real-world applicability.

\item \textbf{RQ3}: How are the \RPRSS\ evaluated?

The main goal of this research question is to let the readers know how the \RPRSS\ are evaluated. To this end, the procedure followed, including the baselines and the metrics the authors used in the evaluation of their proposals, are analyzed. This question is, therefore, divided into the following questions:
\subitem $\circ$ \textbf{RQ3.1}: What are the relevance criteria used to evaluate the recommendations?
\subitem $\circ$ \textbf{RQ3.2}: What metrics are used in the evaluation process?
\subitem $\circ$ \textbf{RQ3.3}: How are the evaluations conducted?  
\item \textbf{RQ4}:  What are the main challenges addressed by the \RPRSS?

Many  of the works aim to solve, or at least mitigate, challenges such as scalability or cold start. In this question, the challenges addressed, if any, are analyzed.
\item \textbf{RQ5}: What new challenges have arisen?

The adoption of new technology opens new opportunities but may also lead to new challenges. This question aims at identifying those new challenges.
\end{itemize}

To answer these questions and to compare the current state of the research with previous surveys, the methodology implemented in  \cite{IJ:DL:KREUTZ:2022}, has been followed. The literature search has been conducted using the same digital libraries: \textit{ACM DL}\footnote{\url{https://dl.acm.org/}}, \textit{DBLP}\footnote{\url{https://dblp.uni-trier.de/}}, \textit{Google Scholar}\footnote{\url{https://scholar.google.com/}}, and \textit{Springer}\footnote{\url{https://link.springer.com/}}. The search query requested papers that contain the terms \textit{paper}, \textit{article} or \textit{publication} in the title along with some form of \textit{recommendation}, as in  \cite{IJ:DL:KREUTZ:2022}, and were published  between 2021 and 2024. As the period considered in this work spans from  November 2021 to December 2024, all the works published beyond that period were discarded.

Initially, 228 records were identified from \textit{ACM DL}, \textit{DBLP}, \textit{Google Scholar}, and \textit{Springer Link}. Records that did not meet the query terms, language, time frame requirements, or were duplicates were first removed. Then, a screening process considering the title and abstract was carried out, resulting in 140 papers. After this screening, papers whose full text was inaccessible or those which did not align with the scope of this work were discarded, leaving 97 articles to be analyzed for eligibility assessment. Additionally, references cited by these 97 papers and meeting the criteria were screened, resulting in 2 additional papers included for further analysis. This led to a total of 99 full-text articles assessed for eligibility.

Subsequently, papers that did not provide sufficient information to describe their approach, \textit{i.e.}, to answer \textit{RQ1}, or did not explain how their system was evaluated, \textit{i.e.}, to answer \textit{RQ3}, were excluded from the review. Consequently, 36 articles were discarded during the final analysis step, resulting in a total of 63 papers analyzed.

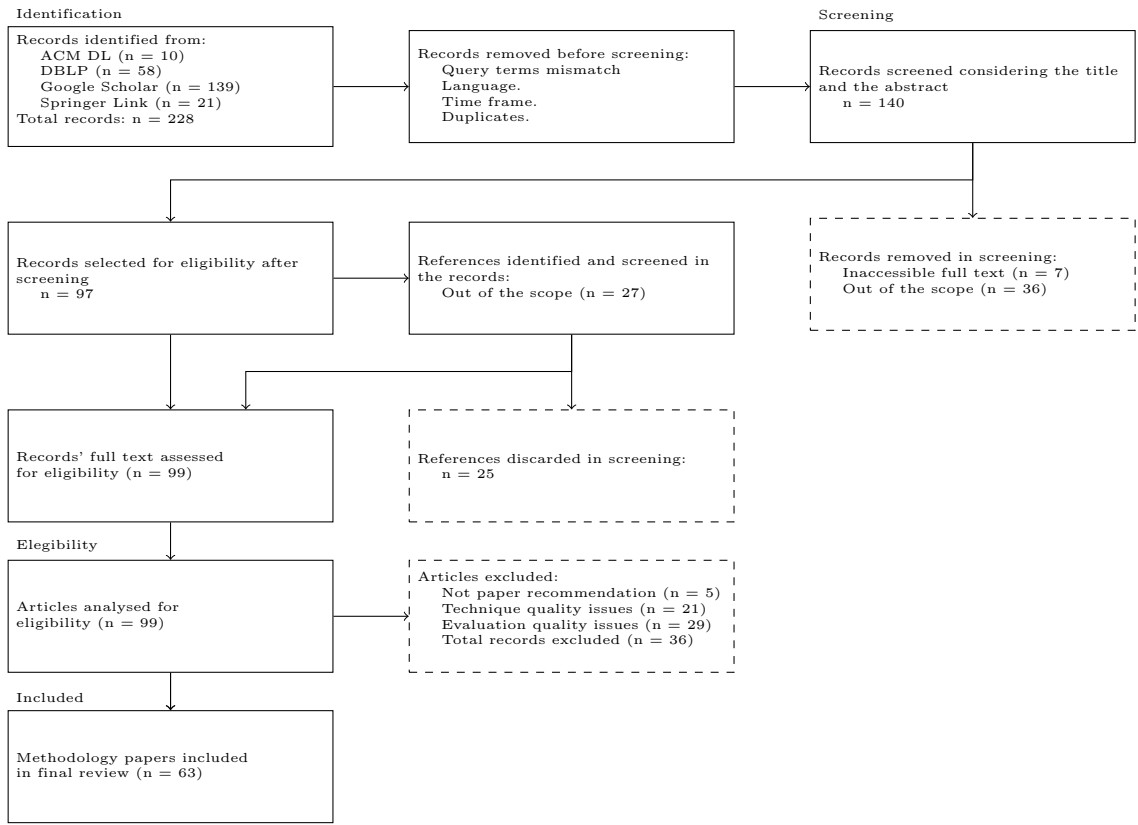
\begin{figure}[H]
		\begin{center}
			%% Scalebox si quieres reducir el tamaño
			\scalebox{.99}{% Here begins the workflow
\tikzstyle{anstep} = [rectangle, draw=black, font=\tiny,   minimum height=1.5cm]
\tikzstyle{discarded} = [rectangle, dashed, draw=black, font=\tiny,  minimum height=1.5cm]
\begin{tikzpicture}
    \tikzset{node distance = 1cm, every state/.style = {semithick}, every edge/.style = {draw, ->, > = stealth', semithick}}
    \tikzset{every label/.style={font=\tiny}}
    \tikzset{every node/.style={text width=4.1cm}}
    % Identification
    % Nodes
    \node[anstep, label=above:Identification] (id_1) {Records identified from: \newline 
        \-\ \-\ \-\ \-\ ACM DL (n = 10) \newline
        \-\ \-\ \-\ \-\ DBLP (n = 58) \newline
        \-\ \-\ \-\ \-\ Google Scholar (n = 139) \newline
        \-\ \-\ \-\ \-\ Springer Link (n = 21) \newline
        Total records: n = 228 \newline
        };
    \node[anstep, right=of id_1] (id_2) {Records removed before screening: \newline 
        \-\ \-\ \-\ \-\ Query terms mismatch \newline
        \-\ \-\ \-\ \-\ Language. \newline
        \-\ \-\ \-\ \-\ Time frame. \newline
        \-\ \-\ \-\ \-\ Duplicates.};
    \node[anstep, right=of id_2, label=above:Screening] (id_3) {Records screened considering the title and the abstract\newline 
        \-\ \-\ \-\ \-\ n = 140 
    };
    % Arrows
    \draw[->] (id_1) -- (id_2);
    \draw[->] (id_2) -- (id_3);
    
    % Screening 1
    % Nodes
    \node[anstep, below= of id_1] (sc_1_1) {Records selected for eligibility after screening\newline 
        \-\ \-\ \-\ \-\ n = 97 
    
    };
    \node[discarded, below=of id_3] (sc_1_2) {Records removed in screening: \newline 
        \-\ \-\ \-\ \-\ Inaccessible full text (n = 7) \newline
        \-\ \-\ \-\ \-\ Out of the scope (n = 36)};

    \node[anstep, right=of sc_1_1] (sc_122) {References identified  and screened in the records: \newline 
        \-\ \-\ \-\ \-\ Out of the scope (n = 27)};
    % Arrows    
    \draw[->] (id_3) -- (sc_1_2);
    \draw[->] (sc_1_1) -- (sc_122);
    \draw[->] (id_3) -- ++(down:1.25cm) -| (sc_1_1.north);

    \node[discarded, below=of sc_122] (rc-disc) {References discarded in screening: \newline 
        \-\ \-\ \-\ \-\ n = 25};
    
    % Screening 2
    % Nodes
    \node[anstep,below= of  sc_1_1] (sc_2_1) {Records’ full text assessed for
     eligibility (n = 99)};
    % Arrows 
    \draw[->] (sc_122) -- ++(down:1.25cm) -| ($(sc_2_1.north) + (right:1cm)$);
    \draw[->] (sc_1_1) -- (sc_2_1);
    \draw[->] (sc_122) --(rc-disc);
    
    % Screening 3
    % Nodes
    \node[anstep, below= 0.5cm of  sc_2_1, label=above:Elegibility] (sc_3_1) {Articles analysed for eligibility (n = 99)};

    \node[discarded, right=of sc_3_1] (sc_2_2) {Articles excluded: \newline 
        \-\ \-\ \-\ \-\ Not paper recommendation (n = 5)\newline
        \-\ \-\ \-\ \-\ Technique quality issues (n = 21) \newline
        \-\ \-\ \-\ \-\ Evaluation quality issues (n = 29) \newline
        \-\ \-\ \-\ \-\ Total records excluded (n = 36) \newline};

    % Included
    % Nodes
    \node[anstep, below= 0.5cm of sc_3_1, label=above:Included] (in_1) {Methodology papers included in final review (n = 63)};
    % Arrows
    \draw[->] (sc_3_1) -- (in_1);
    \draw[->] (sc_2_1) -- (sc_3_1);
    \draw[->] (sc_3_1) -- (sc_2_2);
    \draw[->] (sc_3_1) -- (in_1);
    
\end{tikzpicture}
% Here ends the workflow
}
		\end{center}
		%\vspace{-.5cm}
	\caption{Literature review process}
	\label{fig:BiblioReviewProcess}
\end{figure}

Figure~\ref{fig:BiblioReviewProcess} graphically shows the bibliographic review process carried out.

\section{Results}\label{sec:Results}

In this section, the works found in the literature are categorized considering three key aspects: \textit{approach}, which discusses the main methodologies and techniques used; \textit{datasets}, which covers the different datasets employed and their characteristics; and \textit{evaluation}, which examines the evaluation methods, metrics, and baselines used to assess the effectiveness of the approaches.
In addition, the current status of the \textit{challenges} highlighted in previous surveys has been analyzed. Finally, the new challenges that have emerged due to the adoption of newer technologies have been identified.

\subsection{How do the \RPRSS\ generate the recommendations?}\label{sec:Approach}

In order to answer \textit{RQ1}, different aspects were considered and, therefore, the research question was formulated into five questions. The results for each questions are presented in the corresponding subsection.
%This section is divided into five sub-sections, each focusing on a specific aspect of \RPRSS\ methodologies. 
\textit{Section~\ref{sec:Input}} discusses and analyzes the different input types used in \RPRSS. \textit{Section~\ref{sec:Used information}} explores the information employed in the process to generate the recommendations. \textit{Section~\ref{sec:PaperRepresentation}} reviews how content (titles, abstracts, keywords, etc.) and/or structural relations (interactions, citations, co-authorships, etc.) are processed to generate the representation of the papers. \textit{Section~\ref{sec:User representation}} addresses user representations, examining how user profiles and preferences are modelled. Finally, \textit{Section~\ref{sec:Recommendation procedure}} looks at the recommendation generation processes, detailing how recommendations are generated and optimized based on the aforementioned components. 

\subsubsection{What type of input do the \RPRSS\ use to generate the recommendations?}\label{sec:Input}

Three primary kinds of inputs, namely \textit{paper}, \textit{user} and \textit{text}, were identified in the literature analysis performed. The works which use papers as inputs were labeled as \textit{paper}.
Regarding the works that employ user information as input, in the analysis carried out it was observed that a remarkable percentage of them actually use paper authors as users. Therefore, in this analysis the works that use authors as inputs are classified into the \textit{user(a)} category while the remaining works that use users as inputs are classified as \textit{user}. Works which use any kind of text input were categorized as \textit{text}. Some works also combine different kinds of inputs. 
\tablename~\ref{tab:InputType} summarizes the classification of the works according to the type of inputs used. 

30.16\% of the works use a \textit{\POI} as the input for the recommendation process. The goal of the \RPRSS\ that use this type of input is to generate recommendations based on the content of the \textit{\POI}, using its content and other relevant features to identify similar or related relevant papers.

\RPRSS\ which provide personalized recommendations must be aware of the interests of the users to provide them with appropriate recommendations. 53.97\% of the systems analyzed use this kind of input. 28.57\% of them classified as \textit{user} and 25.4\% as \textit{user(a)}.

The third major kind of input is textual content (\textit{text}), where a text that represents the interests of the users is used as input for the recommendation process. This approach has been followed by three works (4.76\%). \cite{EC:ECIR:Takahashi:2022} uses paper abstracts as input. \cite{IJ:DSM:Sivasankari:2023} takes a search query string as input. Finally, \cite{IJ:ELECTRONICS:Blazevic:2023} uses different sections of the paper.

To a much lesser extent, combinations of the aforementioned types have also been found;  \cite{IJ:SMC:WANG:2022}  uses information of the research groups (\textit{group}) in order to predict \SSN\ group-paper ratings. \cite{IC:MLBDBI:GUAN:2022} uses both the information of the paper and the user
(\textit{user\&paper}) to predict the likelihood that user $u$ will cite paper $p$. 
\cite{IJ:KBS:Xiao:2023} uses the information of authors and papers
(\textit{user(a)\&paper}) to recommend papers by predicting the occurrence of co-authorship given a co-author ternary $(a_1, p_1, a_2)$. Some works use author information and text
(\textit{user(a)\&text}) to construct a query vector \cite{IJ:Scientometrics:Huang:2024} or to search domain literature that reflects the characteristics and hot information of the domain \cite{IJ:AS:Chen:2023}. 
\cite{IJ:AI:Xiao:2023} allows the users to specify their interests by introducing the \textit{authors}, \textit{keywords}, \textit{venues}, \textit{organizations} or \textit{funds} (\textit{author$|$keyword$|$venue$|$organization$|$fund}).
Finally, \cite{IJ:KIS:STERGIOPOULOS:2024} relies on user information along with text or papers and the publication year to represent the interests of the users (\textit{user\&(text$|$papers)\&publication year}).

\begin{footnotesize}
\begin{longtable}{lp{5cm}p{3cm}l}
\caption{Classification according to the input type}\label{tab:InputType}\\
\toprule
\textbf{Personalization} & \textbf{Input type} & \textbf{Works} & \textbf{pct}\\
\midrule  
\endfirsthead

\multicolumn{4}{c}%
{{\bfseries \tablename\ \thetable{} - continued from previous page - }} \\
\toprule
\textbf{Personalization} & \textbf{Input type} & \textbf{Works} & \textbf{pct}\\
\midrule
\endhead

\midrule
\multicolumn{4}{c}{{Continued on next page}} \\ 
\bottomrule
\endfoot

\bottomrule
\endlastfoot

& paper & \cite{ARXIV:Shen:2024, IJ:AI:Thierry:2023, IJ:EIT:Smail:2023, IJ:ESWA:MEI:2022, IJ:Heliyon:LI:2024, IJ:IEEE:HADHIATMA:2023, IJ:IEEE:SARWAR:2021, IJ:IEEE:STERLING:2022, IJ:IPM:Long:2024, IJ:JOI:Yadav:2022, IJ:Scientometrics:ALI:2022, IJ:Scientometrics:GUNDOGAN:2022, IJ:Scientometrics:Kanwal:2024, IJ:Scientometrics:Thierry:2023, IJ:SECS:Hamisu:2024, IC:WIST:Roßner:2023,IJ:SHTI:GUO:2022,IJ:AMEC:KUS:2023, IJ:Scientometrics:TAN:2023} & 30.16\%\\
\ding{51} & user & \cite{IC:IKT:JAFARI:2021,IJ:ASC:Wang:2024, IJ:AUT:Jafari:2023, IJ:DATABASE:Kart:2022, IJ:ESWA:XI:2023, IJ:IS:Chaudhuri:2022, IJ:SC:XI:2024, J:EIJ:Huang:2024, IJ:CDS:AHMEDI:2022,IC:KSEM:YU:2022,arXiv:KARIMI:2022, IJ:JS:Huang:2023,IS:CSET:WU:2022,J:JIS:MUNGEN:2022,IJ:IS:PINEDO:2024,IC:ICITACEE:Switrayana:2022, ARXIV:Mohamed:2022, PP:Mahesh:2023} & 28.57\%\\
\ding{51} & user(a) & \cite{IC:DASFAA:Wang:2022, IJ:AI:GAO:2023, IJ:ESWA:Wang:2023, IJ:ESWA:XIAO:2023,IJ:IEEE:REN:2022,IJ:Scientometrics:Jiang:2023,IJ:TIIS:GUO:2024, CF:CSCW:LI:2022,IJ:AS:Wang:2024, IJ:mathematics:GUO:2024, IC:ASSP:Shen:2024,IC:EITCE:NIU:2023,IJ:JES:LI:2024,AC:ACIIDS:PAN:2024,IC:ICDE:XIE:2022, IC:SEAI:Wang:2024} & 25.4\%\\
& text & \cite{IJ:ELECTRONICS:Blazevic:2023, IJ:DSM:Sivasankari:2023,EC:ECIR:Takahashi:2022} & 4.76\%\\
\ding{51} & user(a)\&text & \cite{IJ:Scientometrics:Huang:2024, IJ:AS:Chen:2023} & 3.17\%\\
\ding{51} & user\&paper & \cite{IC:MLBDBI:GUAN:2022} & 1.59\%\\
\ding{51} & user(a)\&paper & \cite{IJ:KBS:Xiao:2023} & 1.59\%\\
\ding{51} & group & \cite{IJ:SMC:WANG:2022} & 1.59\%\\
\ding{51} & user\&(text$|$papers)\&publication year & \cite{IJ:KIS:STERGIOPOULOS:2024} & 1.59\%\\
& author$|$keyword$|$venue$|$organization$|$fund & \cite{IJ:AI:Xiao:2023} & 1.59\%\\
\end{longtable}
\end{footnotesize}
% Table comments: In IJ:Scientometrics:Huang:2024 the author is provided if available

\subsubsection{What kinds of information do the \RPRSS\ use to generate the recommendations?}\label{sec:Used information}

\RPRSS\ can be classified into two main categories according to the information they use to recommend research papers, in addition to user information: \RPRSS\ that use the information explicitly contained in the article (25\%), and \RPRSS\ that, in addition to the information explicitly included in the paper, use additional relational data (75\%).  
  
Only the information available in the dataset or data sources used by the analyzed works is presented in this analysis. Any information extracted by the approach, which is considered part of the representation or recommendation process, is analyzed in the corresponding section.

\tablename~\ref{tab:PureContentInformation} summarizes the papers that rely on the content explicitly included in the paper, along with information about the users, in the case of personalized recommendations \cite{IJ:IS:Chaudhuri:2022}, indicating the type of information used:  title (\textit{Tit.}), abstract (\textit{Abs.}), keyword (\textit{Key}), text (\textit{Txt.}), author (\textit{Aut.}), user (\textit{Us.}), venue (\textit{Ven.}), others (\textit{Oth.}). 
The \textit{text} category correspond to any textual content in the paper other than the \textit{title}, \textit{abstract}, \textit{authors} and \textit{keywords}.

Most of the works make use of the textual information contained in the papers. 68.75\% of the papers use the \textit{title}, 68.75\% the \textit{abstract}, and  18.75\%  use the \textit{keywords}. 31.25\% of the works use other kinds or unspecified type of \textit{text} inputs. \cite{IJ:EIT:Smail:2023, IJ:AS:Chen:2023} do not specify the kind of text used, while \cite{IJ:IEEE:SARWAR:2021, AC:ACIIDS:PAN:2024} use the \textit{full text} of the paper and  \cite{IJ:ELECTRONICS:Blazevic:2023} experiments with different parts of the text to examine the impact of text length (short, medium, and long texts are represented by \textit{abstract}, \textit{abstract–introduction} combination, and \textit{full text}, respectively).

\textit{Author} information is also widely used (37.5\%). In particular,  \cite{AC:ACIIDS:PAN:2024, IJ:JES:LI:2024, IJ:AS:Chen:2023} use \textit{authors} as input (\textit{user(a)}) for the recommendation process.  \cite{IJ:Scientometrics:Thierry:2023, IJ:SECS:Hamisu:2024} use the information of the \textit{authors} to generate the representations of the papers. \cite{IJ:AI:Xiao:2023} uses  \textit{authors} as nodes in a graph-based approach. The use of \textit{venues} (12.5\%) is residual.

\begin{footnotesize}
\fontsize{6}{8}\selectfont
\begin{longtable}{p{0.5cm}p{0.5cm}p{0.5cm}p{0.5cm}p{0.5cm}p{0.5cm}p{0.5cm}p{0.7cm}l}
\caption{Kinds of information used in \RPRSS\ that rely on the content explicitly included in the paper}\label{tab:PureContentInformation}\\
\toprule
\textbf{Tit.} & \textbf{Abs.} & \textbf{Key.} & \textbf{Txt.} & \textbf{Aut.} & \textbf{Us.} & \textbf{Ven.} & \textbf{Oth.} & \textbf{Works}\\
\midrule
\endfirsthead

\multicolumn{9}{c}%
{{\bfseries \tablename\ \thetable{} -- continued from previous page}} \\
\toprule
\textbf{Tit.} & \textbf{Abs.} & \textbf{Key.} & \textbf{Txt.} & \textbf{Aut.} & \textbf{Us.} & \textbf{Ven.} & \textbf{Oth.} & \textbf{Works}\\
\midrule
\endhead

\midrule
\multicolumn{9}{c}{{Continued on next page}} \\ 
\bottomrule
\endfoot

\endlastfoot

\ding{51} & \ding{51} &  &  &  & & &  & \cite{IJ:Scientometrics:GUNDOGAN:2022, IC:WIST:Roßner:2023,IJ:AMEC:KUS:2023}\\
\ding{51} & \ding{51} &  &  &  & &  & \ding{51} & \cite{IJ:SMC:WANG:2022, IJ:Scientometrics:TAN:2023}\\
&  &  & \ding{51} & & & &  & \cite{IJ:EIT:Smail:2023, IJ:IEEE:SARWAR:2021}\\
&  &  & \ding{51} & \ding{51} & & &  & \cite{IJ:AS:Chen:2023, AC:ACIIDS:PAN:2024}\\
\ding{51} & \ding{51} &  & \ding{51} & & & &  & \cite{IJ:ELECTRONICS:Blazevic:2023}\\
\ding{51} & \ding{51} & \ding{51} &  & \ding{51} & & & & \cite{IJ:SECS:Hamisu:2024}\\
\ding{51} & \ding{51} & \ding{51} &  & & \ding{51} & & & \cite{IJ:IS:Chaudhuri:2022}\\
\ding{51} & \ding{51} &  &  & \ding{51} & & & \ding{51} & \cite{IJ:Scientometrics:Thierry:2023}\\
\ding{51} & \ding{51} &  &  & \ding{51} & & \ding{51} &  & \cite{IJ:JES:LI:2024}\\
\ding{51} &  & \ding{51} &  & \ding{51} & & \ding{51} & \ding{51} & \cite{IJ:AI:Xiao:2023}\\
 & \ding{51} &  & & &  &  &  & \cite{EC:ECIR:Takahashi:2022}\\
\bottomrule
68.75\% & 68.75\% & 18.75\% & 31.25\% & 37.5\% & 6.25\% & 12.5\% & 25\% & \\
\end{longtable}
\end{footnotesize}

\tablename~\ref{tab:RelationalInformation} describes the \RPRSS\ which, besides the content of the paper, consider relational data to achieve their goal. Therefore, in addition to the information explicitly contained in the article, the use of relational data, such as interactions (\textit{Int.}), citations (\textit{Cit.}) and co-authors (\textit{Co-Aut.}) is also analyzed.
These relational data are used to improve the representation of papers and users, and therefore, generate better recommendations. For example, \cite{IC:EITCE:NIU:2023, IJ:ESWA:MEI:2022,IJ:Scientometrics:Jiang:2023,IJ:ASC:Wang:2024} use the collaboration network of the authors to this end. 
\cite{IJ:IEEE:STERLING:2022} uses a hybrid similarity metric that combines text similarity and citation similarity to determine how close are the candidate paper and the \POI. To  achieve this goal, the citation network is used to calculate the similarity between the candidate papers and \POI\ according to the articles that each of them refers to.

As in \tablename~\ref{tab:PureContentInformation}, most of the works that also use relational information, and since they rely on a solution that is totally or partially content-based, make extensive use of fields with more substantial text content, such as the \textit{title} (78.72\%) or the \textit{abstract} (72.34\%). Regarding the rest of the content-based data, it is worth highlighting the greater use of \textit{venues} (25\%) and \textit{others} (48.93\%), largely due to the growing reliance on graph-based solutions, in which these types of data are significantly utilized. 

The most commonly used relational data are \textit{citations} (78.72\%), probably because the information is easily accessible.  38.29\% of the works rely on user information and 34.04\% use interactions to generate the recommendations. In fact, most works offering personalized solutions based on users rely on interactions \cite{IJ:ESWA:XI:2023, IJ:SC:XI:2024,IS:CSET:WU:2022, IJ:IS:PINEDO:2024,
ARXIV:Mohamed:2022}, such as browsing history or tagging, to extract user interests. In contrast, when the users are the authors themselves, \textit{citations} serve as the primary source for extracting these interests \cite{IC:DASFAA:Wang:2022, IJ:AI:GAO:2023, 
IJ:TIIS:GUO:2024, IJ:ESWA:Wang:2023, IJ:Scientometrics:Huang:2024, IC:ICDE:XIE:2022}.  The \textit{co-author} data  is the least used of the three relational types (23.40\%), and in most cases it is used together with \textit{citations} \cite{IJ:AI:Thierry:2023, IJ:Scientometrics:Kanwal:2024, IJ:AS:Wang:2024, IC:EITCE:NIU:2023, IJ:mathematics:GUO:2024, IC:SEAI:Wang:2024}.

\begin{footnotesize}
\fontsize{6}{8}\selectfont
\begin{longtable}{p{0.5cm}p{0.5cm}p{0.5cm}p{0.5cm}p{0.5cm}p{0.5cm}p{0.5cm}p{0.5cm}p{0.5cm}p{0.5cm}l}
\caption{Kinds of information used in \RPRSS\ that rely on  relational information}\label{tab:RelationalInformation}\\

\toprule
\textbf{Tit.} & \textbf{Abs.} & \textbf{Key.} & \textbf{Aut.} & \textbf{Us.} & \textbf{Ven.} & \textbf{Oth.} & \textbf{Int.} & \textbf{Cit.} & \textbf{Co-Aut.} & \textbf{Works}\\
\midrule
\endfirsthead

\multicolumn{11}{c}%
{{\bfseries \tablename\ \thetable{} -- continued from previous page}} \\
\toprule
\textbf{Tit.} & \textbf{Abs.} & \textbf{Key.} & \textbf{Aut.} & \textbf{Us.} & \textbf{Ven.} & \textbf{Oth.} & \textbf{Int.} & \textbf{Cit.} & \textbf{Co-Aut.} & \textbf{Works}\\
\midrule
\endhead

\midrule
\multicolumn{11}{c}{{Continued on next page}} \\ 
\bottomrule
\endfoot

\endlastfoot

\ding{51} & \ding{51} & & & \ding{51} & & & \ding{51} & & & \cite{arXiv:KARIMI:2022, IJ:CDS:AHMEDI:2022, IC:ICITACEE:Switrayana:2022}\\
\ding{51} & \ding{51} & & & & & & & \ding{51} & & \cite{IJ:IEEE:STERLING:2022, IJ:IEEE:HADHIATMA:2023,IJ:SHTI:GUO:2022}\\
\ding{51} & \ding{51} & & & & & \ding{51} & & \ding{51} & & \cite{IJ:IPM:Long:2024,ARXIV:Shen:2024,IJ:KIS:STERGIOPOULOS:2024}\\
\ding{51} & \ding{51} & & & \ding{51} & & \ding{51} & \ding{51} & \ding{51} & & \cite{IJ:ESWA:XI:2023, IJ:SC:XI:2024}\\
\ding{51} & \ding{51} & \ding{51} & & \ding{51} & & & \ding{51} & \ding{51} & & \cite{IJ:JS:Huang:2023, J:EIJ:Huang:2024}\\
\ding{51} & \ding{51} & & \ding{51} & & & & & \ding{51} & \ding{51} & \cite{IJ:ESWA:MEI:2022,IC:EITCE:NIU:2023}\\
\ding{51} & \ding{51} & \ding{51} & \ding{51} & & \ding{51} & & & & \ding{51} & \cite{IJ:ESWA:XIAO:2023, 
IJ:KBS:Xiao:2023}\\
& & & \ding{51} & & & & & \ding{51} & \ding{51} & \cite{IJ:Scientometrics:Kanwal:2024, IJ:mathematics:GUO:2024}\\
\ding{51} & \ding{51} & \ding{51} & & \ding{51} & & & \ding{51} & \ding{51} & & \cite{IC:KSEM:YU:2022}\\
\ding{51} & \ding{51} & \ding{51} & \ding{51} & & & \ding{51} & & \ding{51} & & \cite{IJ:AI:GAO:2023}\\
\ding{51} & \ding{51} & \ding{51} & \ding{51} & & & & & \ding{51} & & \cite{IJ:Scientometrics:Jiang:2023}\\
\ding{51} & \ding{51} & \ding{51} & & \ding{51} & & & \ding{51} & & & \cite{J:JIS:MUNGEN:2022}\\
\ding{51} & \ding{51} & & & \ding{51} & & & & \ding{51} & & \cite{IC:MLBDBI:GUAN:2022}\\
\ding{51} & \ding{51} & & \ding{51} & & & & & \ding{51} & & \cite{IJ:TIIS:GUO:2024}\\
\ding{51} & \ding{51} & & \ding{51} & & \ding{51} & \ding{51} & & \ding{51} & \ding{51} & \cite{IJ:AI:Thierry:2023}\\
\ding{51} & \ding{51} &  & \ding{51} & \ding{51} & \ding{51} & & \ding{51} & & & \cite{IJ:DATABASE:Kart:2022}\\
\ding{51} & & & & & & \ding{51} & & \ding{51} & & \cite{IC:ASSP:Shen:2024}\\
\ding{51} & & & & & & & & \ding{51} & & \cite{IJ:Heliyon:LI:2024}\\
\ding{51} & \ding{51} & & \ding{51} & & & \ding{51} & & \ding{51} & \ding{51} & \cite{IJ:AS:Wang:2024}\\
\ding{51} & \ding{51} & & \ding{51} & & \ding{51} & & & \ding{51} & \ding{51} & \cite{IJ:JOI:Yadav:2022}\\
\ding{51} & \ding{51} & & \ding{51} & & & \ding{51} & & \ding{51} & & \cite{IJ:Scientometrics:ALI:2022}\\
\ding{51} & \ding{51} & & \ding{51} & & & \ding{51} & & \ding{51} & & \cite{IJ:Scientometrics:Huang:2024}\\
\ding{51} & \ding{51} & & \ding{51} & \ding{51} & \ding{51} & & & \ding{51} & & \cite{PP:Mahesh:2023}\\
\ding{51} & & \ding{51} & \ding{51} & & \ding{51} & \ding{51} & & \ding{51} &  & \cite{CF:CSCW:LI:2022}\\
\ding{51} & \ding{51} & \ding{51} & & & & & & \ding{51} & & \cite{IJ:DSM:Sivasankari:2023}\\
\ding{51} & & & \ding{51} & & & \ding{51} & & \ding{51} & \ding{51} & \cite{IC:SEAI:Wang:2024}\\
\ding{51} & \ding{51} & & & \ding{51} & & \ding{51} & \ding{51} & & & \cite{ARXIV:Mohamed:2022}\\
\ding{51} & \ding{51} & & \ding{51} & & & \ding{51} & & \ding{51} & \ding{51} & \cite{IJ:IEEE:REN:2022}\\
& & & & \ding{51} & & \ding{51} & \ding{51} & & & \cite{IJ:ASC:Wang:2024}\\
& & & \ding{51} & & \ding{51} & \ding{51} & & \ding{51} & & \cite{IJ:ESWA:Wang:2023}\\
& & & & \ding{51} & & \ding{51} & \ding{51} & \ding{51} & & \cite{IS:CSET:WU:2022}\\
& \ding{51} & \ding{51} & \ding{51} & & \ding{51} & \ding{51} & & \ding{51} & & \cite{IC:ICDE:XIE:2022}\\
& & & & \ding{51} & & \ding{51} & \ding{51} & & & \cite{IJ:IS:PINEDO:2024}\\
& & & \ding{51} & & \ding{51} & \ding{51} & & \ding{51} & & \cite{IC:DASFAA:Wang:2022}\\
& & & \ding{51} & \ding{51} & \ding{51} & \ding{51} & \ding{51} & \ding{51} & & \cite{IC:IKT:JAFARI:2021}\\
& & & \ding{51} & \ding{51} & \ding{51} & \ding{51} & \ding{51} & \ding{51} & & \cite{IJ:AUT:Jafari:2023}\\
\bottomrule 
78.72\% & 72.34\% & 23.40\% & 51.06\% & 38.29\% & 25.53\% & 48.93\% & 34.04\% & 78.72\% & 23.40\%\\
\end{longtable}
\end{footnotesize}

In addition to the common data previously mentioned, other types of information, which are used to enrich or optimize the generation of recommendations, have also been identified in this review (see \tablename~\ref{tab:OtherInformation}).

\textit{\FoS} has been used to generate paper embeddings \cite{IJ:Scientometrics:Thierry:2023, IJ:AI:Thierry:2023}, capture researcher preferences in relation to papers and authors \cite{IJ:Scientometrics:ALI:2022}, and assist in clustering publications \cite{IJ:KIS:STERGIOPOULOS:2024}. 

\textit{Topics} are integrated into heterogeneous networks to obtain embeddings for papers and users \cite{IC:DASFAA:Wang:2022,IJ:ESWA:Wang:2023,IC:SEAI:Wang:2024}. \textit{Tags} serve as nodes for meta-path creation and application of random walk algorithms \cite{IC:IKT:JAFARI:2021, IJ:AUT:Jafari:2023}. They are also fused with other information contained in the papers to extract meaningful features \cite{IJ:ESWA:XI:2023,IJ:SC:XI:2024, ARXIV:Mohamed:2022, IS:CSET:WU:2022}.

\textit{Funding} and \textit{organization} data work similarly to \textit{tags}, supporting meta-path creation and embedding generation in \cite{IJ:AI:Xiao:2023, CF:CSCW:LI:2022}. \textit{Research groups} are used in group-aware systems to create embeddings for papers, users, and groups \cite{IJ:ASC:Wang:2024, IJ:SMC:WANG:2022}. \textit{Claims} are used as one part of an intra-document pair (\textit{e.g}., paired with the title/abstract) in a patent dataset to form positive samples for contrastive learning \cite{IJ:Scientometrics:TAN:2023}. \textit{Screenshots} capture visual layout preferences, helping users select papers based on style \cite{IC:ASSP:Shen:2024}. The \textit{year} indicates the novelty of the research \cite{IJ:AI:GAO:2023}, captures recency of papers, authors, and journals \cite{IJ:JOI:Yadav:2022}, and affects the decay of author interests, adjusting relevance in time-based heterogeneous graphs \cite{IJ:IEEE:REN:2022,IJ:Scientometrics:Huang:2024}. Finally, \textit{named entities} are used to create paper and user profiles \cite{IJ:IS:PINEDO:2024}.

%\tablename~\ref{tab:OtherInformation} summarizes other used information.

\begin{footnotesize}
\begin{longtable}{cp{5cm}c}
\caption{Other used information}\label{tab:OtherInformation}\\
\toprule
\textbf{Label} & \textbf{Works} & \textbf{pct}\\
\midrule
Year & \cite{IJ:AI:GAO:2023, IJ:JOI:Yadav:2022,IJ:IEEE:REN:2022,IJ:Scientometrics:Huang:2024, CF:CSCW:LI:2022,IJ:AS:Wang:2024,ARXIV:Shen:2024,IC:ICDE:XIE:2022} & 12.70\%\\
Tag & \cite{IC:IKT:JAFARI:2021, IJ:AUT:Jafari:2023, IJ:ESWA:XI:2023,IJ:SC:XI:2024,ARXIV:Mohamed:2022, IS:CSET:WU:2022} & 9.52\%\\
Field of Study & \cite{IJ:Scientometrics:Thierry:2023, IJ:AI:Thierry:2023, IJ:Scientometrics:ALI:2022,IJ:KIS:STERGIOPOULOS:2024} & 6.35\%\\
Topic & \cite{IC:DASFAA:Wang:2022,IJ:ESWA:Wang:2023,IC:SEAI:Wang:2024} & 4.76\%\\
Organization & \cite{IJ:AI:Xiao:2023,CF:CSCW:LI:2022} & 3.17\%\\
Group & \cite{IJ:ASC:Wang:2024, IJ:SMC:WANG:2022} & 3.17\%\\
Named entity & \cite{IJ:IS:PINEDO:2024} & 1.59\%\\
Fund & \cite{IJ:AI:Xiao:2023} & 1.59\%\\
Screenshot & \cite{IC:ASSP:Shen:2024} & 1.59\%\\
Claim & \cite{IJ:Scientometrics:TAN:2023} & 1.59\%\\
\bottomrule 
\end{longtable}
\end{footnotesize}

\subsubsection{How do the \RPRSS\ represent the papers?}\label{sec:PaperRepresentation}

When analyzing the  approaches implemented to represent the papers, it was observed that 11.11\% of the works encode papers using \textit{graph-based} strategies and 79.37\% use \textit{\VSM} representation techniques. Only one work represent the paper employing a \textit{text-based} approach using keyphrases \cite{IJ:IEEE:SARWAR:2021}.

Regarding \textit{graph-based} representation, only one work, \cite{IJ:AI:Xiao:2023}, focuses solely on textual content, representing papers as nodes along with other content-related entities such as authors, keywords, venues, funds, and organizations. In contrast, several works combine textual content with relational information. For example, in \cite{IC:IKT:JAFARI:2021, IJ:AUT:Jafari:2023}, \textit{\HINS} nodes include papers, authors, users, venues, and tags. \cite{IJ:Scientometrics:Huang:2024} represents papers, authors, words, and topics as nodes, while \cite{PP:Mahesh:2023} includes users, papers, terms, authors, and venues. Additionally, \cite{IJ:JOI:Yadav:2022} defines the popularity network by weighting citation links according to the difference in popularity scores between cited and citing papers, emphasizing connections involving more popular articles.

\textit{\VSM} approaches can also in turn be classified into subcategories considering the technique used. The irruption of \textit{\DL} has remarkably influenced the field of \RPRSS, as can be concluded from the fact that 50.79\% of the works represent papers using \textit{embeddings}.

Works that rely solely on textual content predominantly employ \textit{transformer-like} models. For instance, \cite{EC:ECIR:Takahashi:2022} leverage \textit{\LLMS} trained on scientific paper corpora, such as \textit{SciBERT}, alongside \textit{BERT}. In addition, \cite{IJ:Scientometrics:Thierry:2023} incorporate \textit{SciBERT} and \textit{\RNN}. Lastly, \cite{IC:WIST:Roßner:2023} adopt \textit{SPECTER}. However, some current works continue to apply classical embedding techniques such as \textit{Word2Vec} \cite{AC:ACIIDS:PAN:2024} and \textit{Doc2Vec} \cite{IJ:AS:Chen:2023, IJ:Scientometrics:GUNDOGAN:2022}, demonstrating their ongoing relevance alongside newer methods.

Within this content-based embedding landscape, different techniques are combined to generate \textit{embeddings}. \cite{IC:ASSP:Shen:2024} combine \textit{BERT} with \textit{ResNets}. \cite{arXiv:KARIMI:2022} integrate \textit{\RNNS} and \textit{\CNNS}. \cite{IJ:DSM:Sivasankari:2023} explore fusing \textit{Word2Vec embeddings} with \textit{\RNNS}.

Works that combine textual content with relational information primarily rely on \textit{transformers-based} models. Some use simpler approaches, employing only \textit{transformers} to generate paper \textit{embeddings} \cite{IJ:AI:GAO:2023, IJ:DATABASE:Kart:2022}. In contrast, others integrate \textit{transformers} with \textit{graph-based} methods. For example, several works \cite{IC:MLBDBI:GUAN:2022, IJ:AI:Thierry:2023, CF:CSCW:LI:2022, IC:ICDE:XIE:2022, ARXIV:Shen:2024, IC:EITCE:NIU:2023, IJ:Scientometrics:Jiang:2023} combine \textit{transformers} with \textit{\NN-based graph} approaches, while \cite{IJ:JES:LI:2024} pairs \textit{transformers} with \textit{meta-paths} to generate final paper representations. Additionally, \cite{IC:SEAI:Wang:2024} integrates \textit{transformers} with both \textit{\NN-based graph} and \textit{meta-path} techniques. On the other hand, works that rely more heavily on \textit{meta-paths} often combine them with various \textit{\NN} models. Specifically, \cite{IC:DASFAA:Wang:2022} integrate \textit{meta-paths} with \textit{\RNNS}. \cite{IS:CSET:WU:2022} combine \textit{meta-paths} with \textit{\GNNS}. \cite{IJ:ESWA:XIAO:2023} combine \textit{meta-paths} with both \textit{\GNNS} and \textit{\CNNS}. \cite{IJ:KBS:Xiao:2023} combine \textit{meta-paths} with \textit{\RNNS} and \textit{\GNNS}.

17.46\% of the works represent papers using \textit{\LFVS}. A purely content-based approach is presented in \cite{IJ:SMC:WANG:2022}, which uses \textit{TF-IDF}. Works that combine textual content with relational information typically use \textit{\MF} methods. Some apply \textit{\MF} techniques alone, such as \textit{\ALS}, \textit{\BPR}, and \textit{\PMF} \cite{J:EIJ:Huang:2024, IJ:IS:PINEDO:2024}. However, most combine \textit{\MF} with \textit{text-based} approaches to address the cold start problem. These include \textit{\TM} \cite{IJ:CDS:AHMEDI:2022, IJ:JS:Huang:2023, IJ:IEEE:REN:2022} and \textit{transformers} \cite{IC:ICITACEE:Switrayana:2022, IJ:TIIS:GUO:2024}. \textit{Auto-encoders} are also combined with \textit{\MF} \cite{IJ:ESWA:XI:2023, IJ:KIS:STERGIOPOULOS:2024}.

\textit{\FV} and \textit{\TD} represent 6.35\% and 4.76\% of the representations, respectively. For \textit{\FV}, \textit{TF-IDF} is the main technique used in works that combine textual content with relational information \cite{IJ:DATABASE:Kart:2022, IJ:SHTI:GUO:2022, J:JIS:MUNGEN:2022}. Regarding \textit{\TD}, some works use solely content-based information by applying \textit{\LDA} \cite{IJ:EIT:Smail:2023, IJ:ELECTRONICS:Blazevic:2023}, while others combine content-based and relational-based information \cite{IJ:IEEE:HADHIATMA:2023}.

Finally, some works produce different paper representations using \textit{embeddings} and \textit{\FV} techniques \cite{IJ:AMEC:KUS:2023}, which use purely content-based information, or \textit{\FV} and \textit{\LFV} techniques \cite{IJ:IS:PINEDO:2024}, which combine content-based and relational information, to create different recommendation sets that are then fused with \textit{hybridization} techniques.

In \tablename~\ref{tab:PaperSingleRepresentation}, a detailed summary of paper representation strategies is presented.

\begin{footnotesize}
\fontsize{6}{8}\selectfont
\begin{longtable}{llp{2cm}p{2cm}lp{1cm}l}
\caption{Classification according to paper single representation}\label{tab:PaperSingleRepresentation}\\

\toprule
\textbf{Label} & \textbf{Representation} & \textbf{Technique} & \textbf{Model} & \textbf{Graph type} & \textbf{Works}\\
\midrule
\endfirsthead

\multicolumn{6}{c}%
{{\bfseries \tablename\ \thetable{} -- continued from previous page}} \\
\toprule
\textbf{Label} & \textbf{Representation} & \textbf{Technique} & \textbf{Model} & \textbf{Graph type} & \textbf{Works}\\
\midrule
\endhead

\midrule
\multicolumn{6}{c}{{Continued on next page}} \\ 
\bottomrule
\endfoot

\endlastfoot

Graph-based & HIN & - & - & HIN & \cite{PP:Mahesh:2023,IJ:AI:Xiao:2023,IJ:AUT:Jafari:2023, IC:IKT:JAFARI:2021, IJ:Scientometrics:Huang:2024}\\
\cmidrule(lr){3-6}
\phantom{Graph-based} & \phantom{HIN} & Factor analysis & - & HoIN(PN) & \cite{IJ:JOI:Yadav:2022}\\
\cmidrule(lr){2-6}
\phantom{Graph-based} & HoIN & - & - & HoIN(CN) & \cite{IJ:Scientometrics:Kanwal:2024}\\ \midrule
VSM & Embedding & DL & Transformer, GNN & HoIN(CN) & \cite{IJ:Scientometrics:Jiang:2023, ARXIV:Shen:2024}\\
\cmidrule(lr){5-6}
\phantom{VSM} & \phantom{Embedding} & \phantom{DL} & \phantom{Transformer, GNN} & HIN & \cite{IJ:AI:Thierry:2023}\\
\phantom{VSM} & \phantom{Embedding} & \phantom{DL} & \phantom{Transformer, GNN} & BG & \cite{IC:MLBDBI:GUAN:2022}\\
\cmidrule(lr){4-6}
\phantom{VSM} & \phantom{Embedding} & \phantom{DL} & Transformer, GNN, Meta-path & HIN & \cite{IC:SEAI:Wang:2024}\\
\cmidrule(lr){4-6}
\phantom{VSM} & \phantom{Embedding} & \phantom{DL} & Transformer, GCN & HIN & \cite{CF:CSCW:LI:2022, IC:ICDE:XIE:2022}\\
\cmidrule(lr){4-6}
\phantom{VSM} & \phantom{Embedding} & \phantom{DL} & \phantom{Transformer, GCN} & HoIN(CN) & \cite{IC:EITCE:NIU:2023}\\
\cmidrule(lr){4-6}
\phantom{VSM} & \phantom{Embedding} & \phantom{DL} & Transformer, RNN, Meta-path, RW, TF-IDF & HIN & \cite{IJ:JES:LI:2024}\\
\cmidrule(lr){4-6}
\phantom{VSM} & \phantom{Embedding} & \phantom{DL} & Transformer, RNN, CNN  & HIN(KG) & \cite{IJ:AS:Wang:2024}\\
\cmidrule(lr){4-6}
\phantom{VSM} & \phantom{Embedding} & \phantom{DL} & Transformer, RNN & - & \cite{IJ:Scientometrics:Thierry:2023}\\
VSM & Embedding & DL & Transformer, MLP & HIN & \cite{IJ:Scientometrics:ALI:2022}\\
\cmidrule(lr){4-6}
\phantom{VSM} & \phantom{Embedding} & \phantom{DL} & Transformer, Residual network & - & \cite{IC:ASSP:Shen:2024}\\
\cmidrule(lr){4-6}
\phantom{VSM} & \phantom{Embedding} & \phantom{DL} & Transformer & - & \cite{IJ:DATABASE:Kart:2022,IC:WIST:Roßner:2023,IJ:AI:GAO:2023,IJ:IPM:Long:2024,EC:ECIR:Takahashi:2022}\\
\cmidrule(lr){4-6}
\phantom{VSM} & \phantom{Embedding} & \phantom{DL} & NN & - & \cite{IJ:Scientometrics:GUNDOGAN:2022, AC:ACIIDS:PAN:2024, IJ:AS:Chen:2023}\\
\cmidrule(lr){4-6}
\phantom{VSM} & \phantom{Embedding} & \phantom{DL} & NN, RNN & - & \cite{IJ:DSM:Sivasankari:2023}\\
\cmidrule(lr){4-6}
\phantom{VSM} & \phantom{Embedding} & \phantom{DL} & NN, Topic modeling & - & \cite{IJ:AMEC:KUS:2023}\\
\cmidrule(lr){4-6}
\phantom{VSM} & \phantom{Embedding} & \phantom{DL} & GCN & APRHG & \cite{IC:KSEM:YU:2022}\\
\phantom{VSM} & \phantom{Embedding} & \phantom{DL} & \phantom{GCN} & - & \cite{IJ:mathematics:GUO:2024}\\
\cmidrule(lr){4-6}
\phantom{VSM} & \phantom{Embedding} & \phantom{DL} & CNN & - & \cite{IJ:Scientometrics:TAN:2023}\\
\phantom{VSM} & \phantom{Embedding} & \phantom{DL} & GNN, CNN & HoIN(CoN) & \cite{IJ:ESWA:MEI:2022}\\
\phantom{VSM} & \phantom{Embedding} & \phantom{DL} & GNN, Meta-path & HIN & \cite{IS:CSET:WU:2022}\\
\phantom{VSM} & \phantom{Embedding} & \phantom{DL} & HGNN, RW & HIN & \cite{IJ:ASC:Wang:2024}\\
\cmidrule(lr){4-6}
\phantom{VSM} & \phantom{Embedding} & \phantom{DL} & RNN, Highway network, CNN & - & \cite{arXiv:KARIMI:2022}\\
\cmidrule(lr){4-6}
\phantom{VSM} & \phantom{Embedding} & \phantom{DL} & RNN, GNN, Meta-path & HIN & \cite{IJ:KBS:Xiao:2023}\\
\cmidrule(lr){4-6}
\phantom{VSM} & \phantom{Embedding} & \phantom{DL} & RNN, CNN, Meta-path & HIN(AHG) & \cite{IJ:ESWA:XIAO:2023}\\
\cmidrule(lr){4-6}
\phantom{VSM} & \phantom{Embedding} & \phantom{DL} & RNN, Meta-path & HIN & \cite{IC:DASFAA:Wang:2022}\\
\phantom{VSM} & \phantom{Embedding} & \phantom{DL} & RNN, GAT & HIN & \cite{IJ:ESWA:Wang:2023}\\
\cmidrule(lr){3-6}
\phantom{VSM} & \phantom{LFV} & DL,MF & HAN, MF & - & \cite{ARXIV:Mohamed:2022}\\
\cmidrule(lr){1-6}
VSM & LFV & DL,MF & Transformer, Cosine similarity, MF & Graph structure & \cite{IJ:TIIS:GUO:2024}\\
\cmidrule(lr){4-6}
\phantom{VSM} & \phantom{LFV} & \phantom{DL,MF} & MLP, AutoEncoder, MF & HIN(BG) & \cite{IJ:ESWA:XI:2023}\\
\cmidrule(lr){4-6}
\phantom{VSM} & \phantom{LFV} & \phantom{DL,MF} & Dual-AutoEncoder, MF & - & \cite{IJ:KIS:STERGIOPOULOS:2024}\\
\cmidrule(lr){4-6}
\phantom{VSM} & \phantom{LFV} & \phantom{DL,MF} & TF-IDF, GNN, MF & HoIN(CN) & \cite{IJ:SC:XI:2024}\\
\cmidrule(lr){4-6}
\phantom{VSM} & \phantom{LFV} & \phantom{DL,MF} & Transformer, MF, CNN & - & \cite{IC:ICITACEE:Switrayana:2022}\\
\cmidrule(lr){3-6}
\phantom{VSM} & \phantom{LFV} & MF & Topic modeling, MF & HoIN(CN) & \cite{IJ:JS:Huang:2023}\\
\cmidrule(lr){4-6}
\phantom{VSM} & \phantom{LFV} & \phantom{MF} & MF, Topic modeling & - & \cite{IJ:CDS:AHMEDI:2022}\\
\cmidrule(lr){4-6}
\phantom{VSM} & \phantom{LFV} & \phantom{MF} & MF, TF-IDF & - & \cite{IJ:SMC:WANG:2022}\\
\phantom{VSM} & \phantom{LFV} & \phantom{MF} & MF & - & \cite{J:EIJ:Huang:2024,IJ:IS:PINEDO:2024}\\
\cmidrule(lr){3-6}
\phantom{VSM} & \phantom{LFV} & MF, Statistical method & Topic modeling, MF & HIN & \cite{IJ:IEEE:REN:2022}\\
\cmidrule(lr){2-6}
\phantom{VSM} & FV & Weighted entity vector & - & - & \cite{IJ:IS:PINEDO:2024}\\
\cmidrule(lr){3-6}
\phantom{VSM} & \phantom{FV} & Statistical method & TF-IDF & - & \cite{IJ:DATABASE:Kart:2022, IJ:SHTI:GUO:2022, J:JIS:MUNGEN:2022}\\
\cmidrule(lr){4-6}
\phantom{VSM} & \phantom{FV} & \phantom{Statistical method} & TF-IDF, Topic modeling & - & \cite{IJ:AMEC:KUS:2023}\\
\cmidrule(lr){4-6}
\phantom{VSM} & \phantom{FV} & \phantom{Statistical method} & TF & - & \cite{IJ:SECS:Hamisu:2024}\\
\cmidrule(lr){3-6}
\phantom{VSM} & \phantom{FV} & Statistical method, Feature engineering & Topic modeling & - & \cite{IJ:IS:Chaudhuri:2022}\\
\cmidrule(lr){2-6}
\phantom{VSM} & TD & Statistical method & Topic modeling & - & \cite{IJ:IEEE:HADHIATMA:2023,IJ:ELECTRONICS:Blazevic:2023,IJ:EIT:Smail:2023}\\
\midrule
Text-based & keyphrases & - & Keyphrase extraction & - & \cite{IJ:IEEE:SARWAR:2021}\\
\bottomrule
\end{longtable}
\end{footnotesize}
% HAN = Hierarchical attention Network

It is worth mentioning that a very small percentage of works use some kind of multi-representation approach, that is, the same paper has different representations (see \tablename~\ref{tab:PaperMultiRepresentation}). \textit{\VSM} and \textit{graph-based} node representations are combined to produce combined recommendations based in \textit{cosine similarity} and \textit{citation similarity} \cite{IJ:IEEE:STERLING:2022, IJ:Heliyon:LI:2024}. 

\begin{footnotesize}
\fontsize{6}{8}\selectfont
\begin{longtable}{lp{3cm}lllll}
\caption{Classification according to paper multi-representation}\label{tab:PaperMultiRepresentation}\\
\toprule
\textbf{Label} & \textbf{Representation} & \textbf{Technique} & \textbf{Graph type} & \textbf{Work}\\
\midrule
Vector Space Model & Term frequency  & - & - & \cite{IJ:IEEE:STERLING:2022}\\
Graph-based & HoIN(CN) & - & HoIN(CN) & \cite{IJ:IEEE:STERLING:2022}\\
Vector Space Model & Feature vector  & - & - & \cite{IJ:Heliyon:LI:2024}\\
Graph-based & HoIN(CN) & - & HoIN(CN) & \cite{IJ:Heliyon:LI:2024}\\
\bottomrule 
\end{longtable}
\end{footnotesize}

\subsubsection{How do the \RPRSS\ represent the users?}\label{sec:User representation}

61.9\% of the analyzed works use some type of representation to model users that allows the generation of personalized recommendations.
10\% encode users using \textit{graph-based} strategies, 87.5\% use \textit{\VSM} representation techniques and only one work used other kind of representation by generating a vectorizer and a classifier to define or represent the user \cite{IJ:DATABASE:Kart:2022}.

Regarding \textit{graph-based} representation, unlike papers, users are represented only as nodes in a \textit{\HIN}, since \textit{\CNS} contain only paper nodes. User nodes in a \textit{\HIN} appear alongside other types of nodes, as already stated in Section~\ref{sec:PaperRepresentation}. 

The subcategories and percentages of usage identified in the \textit{\VSM} are very similar to those for papers, except that, in this case, there are no \textit{\TD} representations as none of the works analyzed use this approach to model the profile of the users. 50\% of the works represent users as \textit{embeddings}, 30\% use \textit{\LFV} to model users while 7.5\% of the works use \textit{\FV}.  

The works that use \textit{embedding} representation techniques to model users rely mainly in \textit{graph-based} techniques. \textit{\NN-based} approaches such as, \textit{\GCNS} \cite{IC:KSEM:YU:2022, IJ:mathematics:GUO:2024, IJ:AS:Chen:2023}, \textit{\GNNS} \cite{IC:MLBDBI:GUAN:2022,IJ:AI:GAO:2023}, \textit{\RNNS} \cite{IC:EITCE:NIU:2023, IJ:ESWA:Wang:2023}, \textit{\MLPS} \cite{IJ:AS:Wang:2024} or \textit{\RW} \cite{IJ:ASC:Wang:2024}. 
 \cite{CF:CSCW:LI:2022, IC:SEAI:Wang:2024, IC:ICDE:XIE:2022} use \textit{\GNNS} along with \textit{transformers} to get the user \textit{embeddings}.
Some works combine \textit{\RNNS} with \textit{meta-paths} in order to model the users \cite{IC:DASFAA:Wang:2022, IJ:ESWA:XIAO:2023, IJ:JES:LI:2024, IJ:KBS:Xiao:2023}. 

However, approaches that do not rely on \textit{graph-based} techniques can also be found among the works that rely on \textit{embedding} representations, namely,  \textit{Word2Vec} \cite{AC:ACIIDS:PAN:2024}, a combination of \textit{\RNN} with \textit{\CNN}  \cite{arXiv:KARIMI:2022} or \textit{transformers} used with \textit{\CNN}  \cite{IJ:Scientometrics:Jiang:2023}.

%In user representation, \LFVS\ approaches represent 30\% of the cases.
There are works that only rely on \textit{\MF} to represent users \cite{IC:ICITACEE:Switrayana:2022, ARXIV:Mohamed:2022, IJ:SMC:WANG:2022, J:EIJ:Huang:2024, IJ:IS:PINEDO:2024}. Other works combine  \textit{\MF} with \textit{\TM} \cite{IJ:JS:Huang:2023, IJ:IEEE:REN:2022, IJ:CDS:AHMEDI:2022}, with \textit{auto-encoders} \cite{IJ:ESWA:XI:2023, IJ:KIS:STERGIOPOULOS:2024}, transformers \cite{IJ:TIIS:GUO:2024} or \textit{\GNNS} \cite{IJ:SC:XI:2024}.

\textit{\FV} represent the 7.5\% of the approaches, where \textit{TF-IDF} \cite{J:JIS:MUNGEN:2022} and \textit{\TM} \cite{IJ:IS:Chaudhuri:2022} are the techniques used. 

\tablename~\ref{tab:UserRepresentation} presents a detailed summary of the user representation strategies identified in this analysis.

\begin{footnotesize}
\fontsize{6}{8}\selectfont
\begin{longtable}{llp{2cm}p{2cm}lp{1cm}}
\caption{Classification according to User representation}\label{tab:UserRepresentation}\\

\toprule
\textbf{Label} & \textbf{Representation} & \textbf{Technique} & \textbf{Model} & \textbf{Graph type} & \textbf{Works}\\
\midrule
\endfirsthead

\multicolumn{6}{c}%
{{\bfseries \tablename\ \thetable{} -- continued from previous page}} \\
\toprule
\textbf{Label} & \textbf{Representation} & \textbf{Technique} & \textbf{Model} & \textbf{Graph type} & \textbf{Works}\\
\midrule
\endhead

\midrule
\multicolumn{6}{c}{{Continued on next page}} \\ 
\bottomrule
\endfoot

\endlastfoot

Graph-based & HIN & - & - & HIN & \cite{PP:Mahesh:2023,IJ:AUT:Jafari:2023, IC:IKT:JAFARI:2021,IJ:Scientometrics:Huang:2024}\\
\midrule
VSM & Embedding & DL & Transformer, CNN & - & \cite{IJ:Scientometrics:Jiang:2023}\\
\cmidrule(lr){4-6}
\phantom{VSM} & \phantom{Embedding} & \phantom{DL} & Transformer, GNN, Meta-path & HIN & \cite{IC:SEAI:Wang:2024}\\
\cmidrule(lr){4-6}
\phantom{VSM} & \phantom{Embedding} & \phantom{DL} & Transformer, GCN & HIN & \cite{IC:ICDE:XIE:2022, CF:CSCW:LI:2022}\\
\cmidrule(lr){4-6}
\phantom{VSM} & \phantom{Embedding} & \phantom{DL} & RNN, Meta-path & HIN & \cite{IC:DASFAA:Wang:2022}\\
\phantom{VSM} & \phantom{Embedding} & \phantom{DL} & RNN, Meta-path, GNN & HIN & \cite{IJ:KBS:Xiao:2023}\\
\cmidrule(lr){4-6}
\phantom{VSM} & \phantom{Embedding} & \phantom{DL} & RNN, Meta-path, TF-IDF, RW & HIN & \cite{IJ:JES:LI:2024}\\
\cmidrule(lr){4-6}
\phantom{VSM} & \phantom{Embedding} & \phantom{DL} & RNN, Highway network, CNN & Graph structure & \cite{arXiv:KARIMI:2022}\\
\cmidrule(lr){4-6}
\phantom{VSM} & \phantom{Embedding} & \phantom{DL} & RNN, GAT & HIN & \cite{IJ:ESWA:Wang:2023}\\
\phantom{VSM} & \phantom{Embedding} & \phantom{DL} & RNN, GCN & HoIN(CN) & \cite{IC:EITCE:NIU:2023}\\
\cmidrule(lr){4-6}
\phantom{VSM} & \phantom{Embedding} & \phantom{DL} & GCN & APRHG & \cite{IC:KSEM:YU:2022}\\
\phantom{VSM} & \phantom{Embedding} & \phantom{DL} & \phantom{GCN} & - & \cite{IJ:mathematics:GUO:2024}\\
\phantom{VSM} & \phantom{Embedding} & \phantom{DL} & \phantom{GCN} & HIN(KG) & \cite{IJ:AS:Chen:2023}\\
\cmidrule(lr){4-6}
\phantom{VSM} & \phantom{Embedding} & \phantom{DL} & GNN & BG & \cite{IC:MLBDBI:GUAN:2022}\\
\phantom{VSM} & \phantom{Embedding} & \phantom{DL} & \phantom{GNN} & KG(HIN) & \cite{IJ:AI:GAO:2023}\\
\cmidrule(lr){4-6}
\phantom{VSM} & \phantom{Embedding} & \phantom{DL} & GNN, MLP  & HIN(KG) & \cite{IJ:AS:Wang:2024}\\
\phantom{VSM} & \phantom{Embedding} & \phantom{DL} & HGNN, RW & HIN & \cite{IJ:ASC:Wang:2024}\\
VSM & Embedding & DL & LSTM, CNN, Meta-path & HIN(AHG) & \cite{IJ:ESWA:XIAO:2023}\\
\cmidrule(lr){4-6}
\phantom{VSM} & \phantom{Embedding} & \phantom{DL} & NN & HIN & \cite{IS:CSET:WU:2022}\\
\phantom{VSM} & \phantom{Embedding} & \phantom{DL} & \phantom{NN} & - & \cite{AC:ACIIDS:PAN:2024}\\
\cmidrule(lr){2-6}
\phantom{VSM} & LFV & MF & MF & - & \cite{ARXIV:Mohamed:2022, J:EIJ:Huang:2024, IJ:IS:PINEDO:2024, IC:ICITACEE:Switrayana:2022, IJ:SMC:WANG:2022}\\
\cmidrule(lr){4-6}
\phantom{VSM} & \phantom{LFV} & \phantom{MF} & MF, Topic modeling & HoIN(CN) & \cite{IJ:JS:Huang:2023}\\
\cmidrule(lr){4-6}
\phantom{VSM} & \phantom{LFV} & \phantom{MF} & MF, Topic modeling, ARM & - & \cite{IJ:CDS:AHMEDI:2022}\\
\cmidrule(lr){4-6}
\phantom{VSM} & \phantom{LFV} & \phantom{DL,MF} & MF, Transformer, Cosine similarity & Graph type & \cite{IJ:TIIS:GUO:2024}\\
\cmidrule(lr){4-6}
\phantom{VSM} & \phantom{LFV} & \phantom{DL,MF} & MF, MLP, AutoEncoder & HIN(BG) & \cite{IJ:ESWA:XI:2023}\\
\cmidrule(lr){4-6}
\phantom{VSM} & \phantom{LFV} & \phantom{DL,MF} & MF, Dual-AutoEncoder & - & \cite{IJ:KIS:STERGIOPOULOS:2024}\\
\cmidrule(lr){4-6}
\phantom{VSM} & \phantom{LFV} & \phantom{DL,MF} & MF, GNN & HoIN(CN) & \cite{IJ:SC:XI:2024}\\
\cmidrule(lr){3-6}
\phantom{VSM} & \phantom{LFV} & MF, Statistical method & MF, Topic modeling & HIN & \cite{IJ:IEEE:REN:2022}\\
\cmidrule(lr){2-6}
\phantom{VSM} & FV & Statistical method, Feature engineering & Topic modeling & - & \cite{IJ:IS:Chaudhuri:2022}\\
\cmidrule(lr){1-6}
VSM & FV & Statistical method & TF-IDF & - & \cite{J:JIS:MUNGEN:2022}\\
\cmidrule(lr){3-6}
\phantom{VSM} & \phantom{FV} & Weighted entity vector & - & - & \cite{IJ:IS:PINEDO:2024}\\
\midrule
Other & Vectorized \& & - & - & - & \cite{IJ:DATABASE:Kart:2022}\\
& Clasiffier & & & \\
\bottomrule
\end{longtable}
\end{footnotesize}

\subsubsection{How do the \RPRSS\ generate the recommendations?}\label{sec:Recommendation procedure}

The recommendation generation step entails the last stage in \RPRSS, once the paper and/or user representations have been obtained. In the set of the analyzed works, four major categories have been identified: the \textit{inner product-based} methods are used in 36.51\% of the works analyzed, followed by \textit{similarity-based} methods with 28.57\%, \textit{\ML-based} methods with 19.05\%, and \textit{graph-based} methods with 7.94\%. However, with some works presenting alternative or hybrid approaches. 

All works that use the \textit{inner product-based} method generate personalized recommendations by calculating the similarity between user and paper vector representations through the dot product. 52.17\% of the works that use \textit{inner product} are based on embedding representations using graph-based techniques as a part of a complex model. In these models, various training strategies are employed, ranging from well-established methods such as \textit{\BPR\ loss} \cite{IC:MLBDBI:GUAN:2022, IJ:ASC:Wang:2024}, \textit{cross-entropy loss} \cite{IJ:Scientometrics:Jiang:2023, CF:CSCW:LI:2022, IJ:AI:GAO:2023, IC:DASFAA:Wang:2022}, and \textit{tuple-wise loss} \cite{IC:KSEM:YU:2022}, to more specialized approaches \cite{IJ:mathematics:GUO:2024, IJ:AS:Chen:2023, IJ:ESWA:Wang:2023}. Additionally, works utilizing transformer-based \cite{IC:ASSP:Shen:2024} and RNN-based embeddings \cite{arXiv:KARIMI:2022}, which  adopt \textit{\BPR\ loss} and \textit{\WMS\ loss} respectively as training strategies, have also been identified.

47.83\% of the works that utilize \textit{inner products} as part of the recommendation generation process involve complex models based on \MF\ to produce user and paper representations. Some of these incorporate \MF\ with auto-encoder architectures, either leveraging \textit{\ALS} to minimize the objective function \cite{IJ:ESWA:XI:2023} or using  the \textit{cross-entropy loss} \cite{IJ:KIS:STERGIOPOULOS:2024}. Others combine transformers with \MF\ models, applying \textit{\BPR\ loss} \cite{IJ:TIIS:GUO:2024} or \textit{\MSE\ loss} \cite{IC:ICITACEE:Switrayana:2022}. In certain cases, when recommendations are generated for a group of users, the \textit{\ER\ rule} is applied to aggregate member ratings \cite{IJ:SMC:WANG:2022} before computing the \textit{inner product}. Finally, works that incorporate \GNN\ and \HAN\ are integrated with \MF\ and \textit{inner products}, with \textit{cross-entropy loss} \cite{IJ:SC:XI:2024} applied in the former and \textit{\WARP\ loss} \cite{ARXIV:Mohamed:2022} in the latter.

19.05\% of the works utilize a \textit{\ML} approach, 66.6\% of them producing  personalized recommendations. Most of them employ a \textit{\MLP\ framework}, with several incorporating \textit{cross-entropy loss} \cite{IC:SEAI:Wang:2024, IJ:KBS:Xiao:2023}, while others opt for \textit{ad-hoc loss} functions \cite{IJ:AS:Wang:2024, IJ:ESWA:XIAO:2023}. In addition to \textit{\MLP}, a work combining \textit{\RNN} and \textit{\SVM} models \cite{IJ:IS:Chaudhuri:2022}, a \textit{\GCN} integrated with \textit{\BPR\ loss} for training \cite{IS:CSET:WU:2022}, and a \textit{\NN} with \textit{cross-entropy} loss \cite{IC:ICDE:XIE:2022} have also been found to yield personalized recommendations. Those works that do not provide personalized recommendations (33.3\%) typically employ various \textit{\NN} architectures. \cite{IJ:AI:Thierry:2023} uses  \textit{cross-entropy loss}, while \cite{IJ:Scientometrics:Thierry:2023} uses \textit{\MSE\ loss}, and \cite{ARXIV:Shen:2024} adopts a \textit{customized loss} function.

28.57\% of the works rely on \textit{similarity-based} methods, 77.78\% of them producing personalized recommendations. The most used is \textit{cosine similarity} \cite{IC:WIST:Roßner:2023, IJ:ESWA:MEI:2022, IJ:Scientometrics:GUNDOGAN:2022, IJ:SECS:Hamisu:2024, IJ:AMEC:KUS:2023, IJ:ELECTRONICS:Blazevic:2023, EC:ECIR:Takahashi:2022, AC:ACIIDS:PAN:2024}. \cite{IJ:Heliyon:LI:2024, IJ:IEEE:STERLING:2022} use both \textit{cosine similarity} and \textit{citation similarity}. \cite{IJ:IEEE:SARWAR:2021} applies \textit{cosine similarity} along with \textit{Jaccard similarity}. \cite{IJ:JES:LI:2024} also uses \textit{cosine similarity} after applying a clustering process to reduce the computation time. \cite{IJ:Scientometrics:TAN:2023} uses \textit{contrastive loss} with \textit{cosine similarity} to increase similarity within document paragraphs and decrease similarity across different documents. In \cite{IJ:IPM:Long:2024} the \textit{hierarchical loss} balances global themes and fine-grained aspects by jointly optimizing general and aspect-specific embeddings. Other similarity measures have also been identified in the analysis.  \cite{IJ:AI:Xiao:2023} combines \textit{PathSim} with \textit{\RW}.
\cite{IJ:EIT:Smail:2023} uses \textit{\JSD}, while \cite{IJ:IEEE:HADHIATMA:2023} integrates \textit{\JSD} with \textit{PageRank}.

Works employing \textit{graph-based} approaches to generate recommendations utilize \textit{\RW} algorithms with restart \cite{IC:IKT:JAFARI:2021, IJ:AUT:Jafari:2023, IJ:Scientometrics:Huang:2024} or without restart\cite{PP:Mahesh:2023}. All of them provide personalized recommendations. \cite{IJ:DSM:Sivasankari:2023} is the only \textit{graph-based} approach which does not provide personalized recommendations. This work uses the \textit{COOT optimizing} algorithm to efficiently identify highly influential papers.

Finally, some other approaches, which represent 7.94\% of the total analyzed works, have been identified. Only one of these works presents a personalized recommendation system. This work, \cite{IJ:IS:PINEDO:2024},  uses \textit{inner product} and a \textit{relevance score} to generate the recommendations.  
The rest of the works found in this group use \textit{co-citation} and \textit{bibliographic coupling} \cite{IJ:Scientometrics:Kanwal:2024}. The \textit{hybrid} solution presented in \cite{IJ:SHTI:GUO:2022} combines \textit{bibliographic coupling}, \textit{cosine similarity}, \textit{\PMRA}, and \textit{BM25}. Other approaches use a \textit{personalized score} \cite{IJ:JOI:Yadav:2022} or a \textit{Naïve Bayes classifier} \cite{IJ:DATABASE:Kart:2022}.

In \tablename~\ref{tab:RecommendationProcedure} a detailed summary of different recommendation procedure strategies is presented.

\begin{footnotesize}
\fontsize{6}{8}\selectfont
\begin{longtable}{lp{2cm}p{3cm}lp{2cm}l}
\caption{Classification according to the recommendation procedure strategy}\label{tab:RecommendationProcedure}\\
\toprule
\textbf{Label} & \textbf{Technique} & \textbf{Training features} & \textbf{Graph type} & \textbf{Works}\\
\midrule
\endfirsthead
\multicolumn{5}{c}
{{\bfseries \tablename\ \thetable{} -- continued from previous page}} \\
\toprule
\textbf{Label} & \textbf{Technique} & \textbf{Training features} & \textbf{Graph type} & \textbf{Works}\\
\midrule
\endhead
\midrule
\multicolumn{5}{c}{{Continued on next page}} \\ 
\bottomrule
\endfoot

\endlastfoot
Inner product & - & Cross-entropy loss & - & \cite{IJ:Scientometrics:Jiang:2023, IC:DASFAA:Wang:2022, IJ:AI:GAO:2023, CF:CSCW:LI:2022, IJ:SC:XI:2024}\\
\phantom{Inner product} & \phantom{-} & Tuple-wise loss & - & \cite{IC:KSEM:YU:2022}\\
\cmidrule(lr){3-5}
\phantom{Inner product} & \phantom{-} & WMS loss & - & \cite{arXiv:KARIMI:2022}\\
\cmidrule(lr){3-5}
\phantom{Inner product} & \phantom{-} & - & - & \cite{IJ:mathematics:GUO:2024, IJ:ESWA:XI:2023, IJ:JS:Huang:2023, IJ:ESWA:Wang:2023,IJ:IEEE:REN:2022, J:EIJ:Huang:2024,IJ:CDS:AHMEDI:2022, IJ:SMC:WANG:2022, IJ:KIS:STERGIOPOULOS:2024}\\
\cmidrule(lr){3-5}
\phantom{Inner product} & \phantom{-} & BRP loss & - & \cite{IJ:TIIS:GUO:2024}\\
\phantom{Inner product} & \phantom{-} & BPR loss & - & \cite{IC:ASSP:Shen:2024, IC:MLBDBI:GUAN:2022, IJ:ASC:Wang:2024}\\
\phantom{Inner product} & \phantom{-} & Cutomized loss & - & \cite{IJ:AS:Chen:2023}\\
\phantom{Inner product} & \phantom{-} & WARP loss & - & \cite{ARXIV:Mohamed:2022}\\
\phantom{Inner product} & \phantom{-} & MSE loss & - & \cite{IC:ICITACEE:Switrayana:2022}\\
\midrule
Similarity & Cosine similarity, Jaccard similarity & - & - & \cite{IJ:IEEE:SARWAR:2021}\\
\cmidrule(lr){2-5}
\phantom{Similarity} & Cosine similarity & Hierarchical loss & - & \cite{IJ:IPM:Long:2024}\\
\phantom{Similarity} & \phantom{Cosine similarity} & Contrastive loss & - & \cite{IJ:Scientometrics:TAN:2023}\\
\cmidrule(lr){3-5}
\phantom{Similarity} & \phantom{Cosine similarity} & - & - & \cite{IJ:Scientometrics:GUNDOGAN:2022,IJ:ESWA:MEI:2022,IJ:SECS:Hamisu:2024,IJ:ELECTRONICS:Blazevic:2023,IC:WIST:Roßner:2023,IJ:AMEC:KUS:2023,EC:ECIR:Takahashi:2022,AC:ACIIDS:PAN:2024}\\
\cmidrule(lr){2-5}
\phantom{Similarity} & Cosine similarity, Citation similarity & - & - & \cite{IJ:IEEE:STERLING:2022}\\
\cmidrule(lr){2-5}
\phantom{Similarity} & \JSD\ , PageRank & - & - & \cite{IJ:IEEE:HADHIATMA:2023}\\
\cmidrule(lr){2-5}
\phantom{Similarity} & Cosine similarity, Clustering & - & - & \cite{IJ:JES:LI:2024}\\
\cmidrule(lr){2-5}
\phantom{Similarity} & Cosine limilarity, Citation similarity & - & HoIN(CN) & \cite{IJ:Heliyon:LI:2024}\\
\cmidrule(lr){2-5}
\phantom{Similarity} & Meta-path, RW, PathSim & - & HIN & \cite{IJ:AI:Xiao:2023}\\
\cmidrule(lr){2-5}
\phantom{Similarity} & \JSD\ & - & - & \cite{IJ:EIT:Smail:2023}\\
\phantom{Similarity} & Term Frequency-based & - & - & \cite{J:JIS:MUNGEN:2022}\\
\midrule
ML-based & NN & Cross-entropy loss & - & \cite{IJ:KBS:Xiao:2023, IC:SEAI:Wang:2024, IJ:AI:Thierry:2023, IC:ICDE:XIE:2022}\\
\phantom{ML-based} & \phantom{NN} & Customized loss & - & \cite{IJ:ESWA:XIAO:2023, IJ:AS:Wang:2024}\\
\phantom{ML-based} & \phantom{NN} & MSE loss & - & \cite{IJ:Scientometrics:Thierry:2023}\\
\phantom{ML-based} & \phantom{NN} & - & - & \cite{ARXIV:Shen:2024}\\
\cmidrule(lr){2-5}
\phantom{ML-based} & Memory network & Cross-entropy loss & - & \cite{IJ:Scientometrics:ALI:2022}\\
\phantom{ML-based} & Auto-Encoder & Customized loss & - & \cite{IC:EITCE:NIU:2023}\\
\phantom{ML-based} & RNN, SVM & \phantom{Customized loss} & - & \cite{IJ:IS:Chaudhuri:2022}\\
\cmidrule(lr){2-5}
\phantom{ML-based} & GCN, Outer product & BPR loss & - & \cite{IS:CSET:WU:2022}\\
\midrule
Graph-based & Meta-path, RW & BPR loss & HIN & \cite{PP:Mahesh:2023}\\
\phantom{Graph-based} & COOT algorithm & - & CN & \cite{IJ:DSM:Sivasankari:2023}\\
\phantom{Graph-based} & Meta-path, RWR & - & HIN & \cite{IJ:AUT:Jafari:2023, IC:IKT:JAFARI:2021}\\
\phantom{Graph-based} & RWR & - & HIN & \cite{IJ:Scientometrics:Huang:2024}\\
Other & Co-citation, Bibliographic coupling & - & HoIN(CN) & \cite{IJ:Scientometrics:Kanwal:2024}\\
\phantom{Other} & Personalized(NaÏve Bayes classifier/Softmax) & - & - & \cite{IJ:DATABASE:Kart:2022}\\
\cmidrule(lr){2-5}
\phantom{Other} & Personalized score, PPR & - & - & \cite{IJ:JOI:Yadav:2022}\\
\cmidrule(lr){2-5}
\phantom{Other} & Bibliographic coupling, Similarity  & - & - & \cite{IJ:SHTI:GUO:2022}\\
\cmidrule(lr){2-5}
\phantom{Other} & Hybrid(Relevance score + Inner product) & - & - & \cite{IJ:IS:PINEDO:2024}\\
\bottomrule 
\end{longtable}
\end{footnotesize}

 \subsection{What datasets are used to train and evaluate the \RPRSS?}\label{sec:Datasets}
In this section, a detailed explanation of the datasets used  by the  analyzed works is provided. 

The datasets used to develop and evaluate systems, algorithms, and models on \RPRSS\ are crucial. The effectiveness of a solution often depends heavily on the data used. The profiles of users, researchers, and papers are influenced not only by the type of data but also by its quantity and, most importantly, its quality. The proposed solution (including its approach, techniques, etc.) and its purpose (such as context and scope) primarily determine the most suitable type of dataset to use.

A total of 72 different databases have been identified in the analyzed works. These datasets have been categorized based on the original dataset used. Four main dataset families (\textit{ACL}, \textit{DBLP}, \textit{AMiner} and \textit{CiteULike}) have been identified according to their origin, which are further divided into different versions depending on their intended purpose. However, some version or source used could not be determined due to a lack of information or inaccurate data. In those cases, the version of the dataset has been labeled as unknown (?). 

The following subsections describe each of these families and their different versions, if applicable. In most cases, when one of these datasets is used, it involves generating a customized version —either by adding additional information from other data sources or by applying filtering or cleaning procedures to obtain a partial version. Therefore, the datasets generated from the public versions are also detailed in additional tables. 

Besides the datasets corresponding to these four families, two other groups have been identified. On the one hand, \textit{publicly available datasets} which are rarely used  have been identified, typically being utilized only once —either in their original version or with modifications. These datasets represent 13.89\% of the total. On the other hand, the datasets that have been created \textit{ad-hoc} to develop or evaluate the proposed solutions can be considered. Those datasets represent 29.17\% of the total.

\subsubsection{ACL dataset}\label{sec:ACLdataset}

In the \textit{\ACL} ecosystem, the \textit{\ARC}\footnote{\url{https://www.sketchengine.eu/acl-anthology-reference-corpus-arc/}} and the \textit{\AAN}\footnote{\url{https://clair.eecs.umich.edu/aan/index.php}} datasets are available. The \textit{\ARC} is a structured collection of papers with metadata for content-based analysis, while the \textit{\AAN} is a citation network that captures relationships between papers for network and citation analysis. \tablename~\ref{tab:ACLDatasets} summarizes the details of both datasets.
\begin{footnotesize}
\fontsize{6}{8}\selectfont
\begin{longtable}{lc}
\caption{ACL datasets main characteristics}\label{tab:ACLDatasets}\\
\toprule
\multicolumn{2}{c}{\textbf{ACL Anthology Network (AAN)}}\\
\midrule
Year & \textless2015 \\
Disciplines & Computational linguistics; NLP\\
Papers & 23,766\\
Authors &	18,862\\
Venues &	373\\
Paper citations &	124,857 \\
Author collaborations	& 142,450\\
Citation network diameter	& 22\\
Collaboration network diameter	& 15\\
\toprule
\multicolumn{2}{c}{\textbf{ACL Anthology Reference Corpus (ARC)}}\\
\midrule
Year & 1979–2015\\
Disciplines & Computational linguistics; NLP\\
Papers & 18,288\\
Tokens & 74,875,938\\
Words & 62,196,334\\
Sentences & 2,380,457\\
\bottomrule
\end{longtable}
\end{footnotesize}

Among the works analyzed, 7.9\% utilize datasets from the \textit{\ACL} dataset ecosystem, with the \textit{\AAN} version being the most commonly used. It is worth mentioning that all these works apply some form of preprocessing, often involving pruning and refinement, to the original \textit{\ACL} dataset before using it in their experiments—meaning the datasets are never used in their original, unmodified form.

\tablename~\ref{tab:ACLDatasetsWorks} provides a summary of the various dataset versions derived and utilized by the analyzed works. This includes information on the specific original dataset versions, their principal characteristics, and the availability of the corresponding processed datasets. The availability is specified in the \textit{A?} column, where \textbf{\ding{51}} indicates that the dataset is publicly available, \textit{ur} denotes availability upon request, \textbf{\ding{55}} indicates that the dataset is not available, and \textit{?} signifies that the availability cannot be determined.

% Hemen aipatu behar dugu nola aurkeztuko dugun dataset hau nola erabili den lan bakoitzean. Zein dataset errabili dute? Iragazi al dute? Lortutako berstsioa atzigarri? ezaugarriak. Taularen egitura eta akronimoak edo erabiliz gero, azaldu zer adierazten duten.

\begin{footnotesize}
\fontsize{6}{8}\selectfont
\begin{longtable}{llllll}
\caption{ACL datasets-based customizations main characteristics}\label{tab:ACLDatasetsWorks}\\
\toprule
\textbf{Work} & \textbf{Version} & \textbf{A?} & \multicolumn{3}{p{8cm}}{\textbf{Content description provided by authors}}\\
\midrule
\endfirsthead
\multicolumn{6}{c}
{{\bfseries \tablename\ \thetable{} -- continued from previous page}} \\
\toprule
\textbf{Work} & \textbf{Version} & \textbf{A?} & \multicolumn{3}{p{8cm}}{\textbf{Content description provided by authors}}\\
\midrule
\endhead
\midrule
\multicolumn{6}{c}{{Continued on next page}} \\ 
\bottomrule
\endfoot
\endlastfoot        
\cite{IJ:ESWA:XIAO:2023} & ? & ur & \textbf{Papers}: 18,743 & \textbf{Authors}:  14,455 & \textbf{Venues}: 247 \\
& & & \textbf{Citations}:  96,209 & \textbf{Tags}:  27,570 \\[5pt]
\cite{IJ:Scientometrics:ALI:2022} & AAN & \ding{55} & \textbf{Papers}: 21,450 &  \textbf{Authors}: 17,335 &  \textbf{Venues}: 311\\
& & & \textbf{Citations}: 113,355\\ [5pt]   
\cite{IJ:ESWA:MEI:2022} & AAN & \ding{55} & \textbf{Papers}: 16,664 &  \textbf{Authors}: 13,934 & \textbf{Citations}: 81,379\\  [5pt]
\cite{IJ:TIIS:GUO:2024} & AAN \footnote{Extracted from publicly available  subdataset \url{https://github.com/AHULiuYang/WHIN-CSL} (\textbf{Release date}: 2013-02)} & \ding{55} & \textbf{Papers}: 13,375 &  \textbf{Authors}: 3,936 & \textbf{Citations}: 56,002 \\ 
& & & \textbf{Sparsity}: 99.89\% &  \multicolumn{2}{l}{\textbf{Cold-start papers}:6.25\%}  \\[5pt]
\cite{IJ:Scientometrics:Huang:2024} & AAN & \ding{55} & \textbf{Papers}: 13,426 &  \textbf{Authors}: 10,927 &  \textbf{Venues}: 297\\
\bottomrule 
\end{longtable}
\end{footnotesize}

\subsubsection{DBLP}\label{sec:DBPL}
The \textit{DBLP} collection is the source of two main datasets: the \textit{DBLP Bibliographic} dataset\footnote{\url{https://dblp.org}} and the \textit{DBLP Citation Network} dataset\footnote{\url{https://www.aminer.cn/citation}} (\tablename~\ref{tab:DBLPDataset}). The \textit{DBLP Bibliographic} dataset provides detailed metadata for research papers, including \textit{title}, \textit{doi}, \textit{authors}, \textit{publication year}, and \textit{conference or journal} information. The \textit{DBLP Citation Network} dataset, on the other hand, enriches this data by incorporating citation relationships from \textit{AMiner}, linking papers based on references to form a structured citation network. Both datasets are regularly updated to ensure they stay current.

\begin{footnotesize}
\fontsize{6}{8}\selectfont
\begin{longtable}{lc}
\caption{DBLP datasets main characteristics}\label{tab:DBLPDataset}\\
\toprule
\multicolumn{2}{c}{\textbf{DBLP (Digital Bibliography \& Library Project)}}\\
\midrule
 Release Date & 2023-06-28\\
Disciplines & Computer Science\\
 Papers &  \textgreater 7,600,000\\
 Authors &	\textgreater 3,600,000\\
 Venues &	\textgreater 6,800\\
 Journals &	\textgreater 1,800 \\
\toprule
\multicolumn{2}{c}{\textbf{DBLP-CN (DBLP Citation Network)}}\\
\midrule
Last Release Name & V14\\
Last Release Date & 2023-01-31\\
Disciplines & Computer science\\
Features & titles, abstracts, keywords, topics, and citation relationships\\
\#Papers & 5,259,858\\
\#Citation Relationship	&  	36,630,661\\
\bottomrule
\end{longtable}
\end{footnotesize}
Among the works analyzed, 34.92\% utilize some \textit{DBLP} version, accounting for 29.16\% of the total datasets used. Some of these works employ the original, unmodified versions of the \textit{DBLP} datasets, specifically versions \textit{V12} and \textit{V13}, which are also the most commonly used versions. These unmodified datasets represent 26.92\% of all \textit{DBLP} datasets used, while the remaining 73.08\% correspond to modified versions, where various forms of preprocessing—such as pruning and refinement—have been applied before using the datasets in their experiments.

\tablename~\ref{tab:DBLPDatasetsWorks} provides a summary of the various dataset versions derived and utilized by the analyzed works.

\begin{footnotesize}
\fontsize{6}{8}\selectfont
\begin{longtable}{llllll}
\caption{DBLP datasets-based customizations main characteristics}\label{tab:DBLPDatasetsWorks}\\
\toprule
\textbf{Work} & \textbf{Version} & \textbf{A?} & \multicolumn{3}{p{8cm}}{\textbf{Content description provided by authors}}\\
\midrule
\endfirsthead
\multicolumn{5}{c}
{{\bfseries \tablename\ \thetable{} -- continued from previous page}} \\
\toprule
\textbf{Work} & \textbf{Version} & \textbf{A?} & \multicolumn{3}{p{8cm}}{\textbf{Content description provided by authors}}\\
\midrule
\endhead
\midrule
\multicolumn{5}{c}{{Continued on next page}} \\ 
\bottomrule
\endfoot
\endlastfoot
\cite{IJ:AI:GAO:2023} &  DBLP & \ding{55} & \textbf{Papers}: 1,201,716 & \textbf{Authors}: 701,923\\   
& & & \textbf{Keywords}: 6,817,901 & \textbf{Citations}: 5,340,639\\
& & & \textbf{Papers-Keywords}: 8,037,502\\
& & & \textbf{Papers-Authors}: 3,721,531\\
\cite{IJ:AS:Chen:2023} &  DBLP & \ding{55} & \textbf{Papers}: 14,376 & \textbf{Authors}: 14,475\\
& & & \textbf{Phrases}: 8,920\\
& & & \textbf{Papers-Authors}: 41,794 & \textbf{Papers-Tags}: 114,624\\ 
& & & \textbf{Sub-disciplines}:\\
& & & databases, DM, AI, IR \\
\cite{IJ:JES:LI:2024} &  DBLP & \ding{55} & \textbf{Papers}: 21,044 & \textbf{Authors}: 28,646\\
& & & \textbf{Conferences}: 18\\
\cite{PP:Mahesh:2023} &  DBLP & \ding{55} & \textbf{Papers}: 2,126,267 & \textbf{Users}: 3,765\\
& & & \textbf{Authors}: 1,221,259 & \textbf{Venues}: 8,686\\ 
& & & \textbf{Terms}: 256,214 & \textbf{Paper-User}: 56,235\\ 
& & & \textbf{Paper-Paper}: 402,8245 & \textbf{Paper-Term}: 17,578,799\\
& & & \textbf{Paper-Venue}: 2,126,267 & \textbf{Paper-Author}: 5,772,357\\
\cite{IJ:KBS:Xiao:2023} &  DBLP & ur & \textbf{Papers: 60,034} & \textbf{Authors}: 196,957\\
& & & \textbf{Keywords}: 163,468 & \textbf{Journals}: 2,669\\
& & & \textbf{Author Relations}:\\
& & & \textbf{Co-author}: 20,585 & \textbf{Co-keyword}: 1,584,187\\ 
& & & \textbf{Co-journal}: 200,619\\ 
& & & \textbf{Paper Relations}:\\
& & & \textbf{Co-authored}: 36,622 & \textbf{Co-keyword}: 6,268,075\\ 
& & & \textbf{Co-journal}: 225,598 \\
\cite{IJ:ESWA:Wang:2023} &  DBLP-CN-? & \ding{55} & \textbf{Papers}: 10,291 & \textbf{User(Author)}: 17,616\\
& & & \textbf{Topics}: 3,990 & \textbf{Venues}: 747\\
& & & \textbf{User-Writing links}: 30,464 & \textbf{Paper-Paper}: 98,982\\
& & & \textbf{Paper-Topic}: 39,641 & \textbf{Paper-Venue}: 10,291\\
& & & \textbf{User-Reading links}: 135,881 &  \textbf{Years}: 1996–2005\\
& & & \textbf{Sub-disciplines}:\\
& & & computer graphics\\
\cite{IC:MLBDBI:GUAN:2022} &  DBLP-CN-? & \ding{55} &  \textbf{Papers}: 38,714 & \textbf{Users(Authors)}: 26,941\\
& & & \textbf{Citations}: 1,343,297 & \textbf{d}: 0,00129\\
\cite{IC:DASFAA:Wang:2022} &  DBLP-CN-? & \ding{55} & \textbf{Papers}: 9,566 & \textbf{Users(Authors)}: 16,863\\
& & & \textbf{tp}: 3,844 &  \textbf{v}: 752\\
& & & \textbf{u-p-p}: 29,266 &  \textbf{u-p-c}: 54,803\\
& & & \textbf{p-p}: 13,978 & \textbf{p-t}: 32,320\\
& & & \textbf{p-v}: 9,566 & \textbf{c.s-u}: 4,957\\
& & & \textbf{Years}: 2000-2005 \\
\cite{IC:DASFAA:Wang:2022} &  DBLP-CN-? & \ding{55} & \textbf{Papers}: 28,930 & \textbf{Users(Authors)}: 7,126\\
& & & \textbf{Topics}: 1,983 & \textbf{Venues}: 1,983\\
& & & \textbf{u-p-p}: 99,816 & \textbf{u-p-c}: 220,815\\
& & & \textbf{p-p}: 47,872 & \textbf{p-t}: 89,921\\
& & & \textbf{p-v}: 28,929 & \textbf{c.s-u}: 16,566\\
& & & \textbf{Years}: 2005-2015\\
\cite{IJ:IEEE:HADHIATMA:2023} &  DBLP-CN-V4 & \ding{55} & \textbf{Large-scale dataset}: & \textbf{Subset}:\\
& & & \textbf{Papers}: 653,506 &  \textbf{Papers}: 46,870\\
\cite{IJ:Scientometrics:TAN:2023} & DBLP-CN-V10 & \ding{51}\footnote{\url{https://github.com/opendata-ai/tr}} & \textbf{Papers: 21,000} & \textbf{Years}: 1967-2017\\
\cite{IJ:Scientometrics:Jiang:2023} &  DBLP-CN-V11 & \ding{55} & \textbf{Papers}: 23,012 & \textbf{s-r}: 1,037\\
& & & \textbf{j-r}: 2,200 & \textbf{r}: 146,635\\
\cite{IC:ASSP:Shen:2024} &  DBLP-CN-V11 & \ding{55} & \textbf{Papers}: 17,414 & \textbf{Users(Authors)}: 3,866\\
& & & \textbf{u-p-p}: 20,676 & \textbf{u-p-c}: 181,254\\
& & & \textbf{Sparcity}: 99.75\%\\
\cite{IJ:ESWA:XIAO:2023} &  DBLP-CN-V11 & ur & \textbf{Papers}: 3,212,633 & \textbf{Authors}: 219,732\\ 
\cite{IJ:AS:Wang:2024} &  DBLP-CN-V11 & ur & \textbf{Papers}: 716,549 & \textbf{Authors}: 976,544\\
& & & \textbf{Years}: 5 & \textbf{pi}: 4,023\\
& & & \textbf{a-pi}: 571,659\\        
\cite{IJ:Scientometrics:ALI:2022} &  DBLP-CN-V12 & \ding{55} & \textbf{Papers}: 3,501,133 & \textbf{Citations}: 25,022,314\\
& & & \textbf{Citations}: 21,143,557 & \textbf{Venues}: 14,135\\
& & & \textbf{Terms}: 3,513,508\\  
\cite{IJ:TIIS:GUO:2024} &  DBLP-CN-V12 & \ding{55} & \textbf{Papers}:14,545 & \textbf{Authors}:3,768\\
& & & \textbf{Citations}:10,921 & \textbf{Sparcity}: 99.98\%\\
& & & \textbf{Proportion of cold-start}\\
& & & \textbf{data in the test set}: 20\%\\     
\cite{IJ:Scientometrics:Huang:2024} &  DBLP-CN-V12 & \ding{55} & \textbf{Papers}: 60,407 & \textbf{Authors}: 77,172\\
& & & \textbf{Sub-disciplines}: & \textbf{c}: 15\\
& & & IR, ML, CV, Net, C.Security & \\    
\cite{IC:ASSP:Shen:2024} &  DBLP-CN-V13 & \ding{55} & \textbf{Papers}: 51,686 & \textbf{Users(Authors)}: 12,068\\
& & & \textbf{u-p-p}: 76,007 &  \textbf{u-p-c}: 809,986\\
\bottomrule 
\end{longtable}
\end{footnotesize}

A significant number of works have used published versions of this dataset collection (see \tablename~\ref{tab:DBLPDataset}) to conduct experiments without performing any data extraction or modifications. These include \textit{DBLP-CN-V12} \cite{IJ:Scientometrics:Kanwal:2024,IJ:Scientometrics:Thierry:2023,IJ:AI:Thierry:2023} and \textit{DBLP-CN-V13} \cite{IJ:DSM:Sivasankari:2023,IJ:Scientometrics:Thierry:2023,IJ:KIS:STERGIOPOULOS:2024,IJ:AI:Thierry:2023}

\subsubsection{AMiner}\label{sec:AMiner}

The \textit{AMiner} datasets repository\footnote{\url{https://cn.aminer.org/data}} is a collection of publicly available datasets focused on academic research, particularly in computer science and related fields (\tablename~\ref{tab:AMinerDataset}). It provides data on academic \textit{papers}, \textit{authors}, \textit{citations}, and \textit{conferences}, which support tasks such as citation analysis, \RSS\, and graph-based learning. These datasets help researchers advance work in \ML, \IR, and \NLP.

\begin{footnotesize}
\fontsize{6}{8}\selectfont
\begin{longtable}{lc}
\caption{AMiner dataset main characteristics}\label{tab:AMinerDataset}\\
\toprule
\multicolumn{2}{c}{\textbf{Academic Social Network (ASN)\footnote{\url{https://www.aminer.cn/aminernetwork}}}}\\
\midrule
Papers & 2,092,356\\
Authors & 1,712,433\\
Citations & 8,024,869\\
Co-authorships & 4,258,615\\
\bottomrule
\end{longtable}
\end{footnotesize}

11.11\% of the analyzed works employ an \textit{AMiner} dataset, representing 9.72\% of the total datasets utilized. Similar to the \textit{\ACL} datasets, every work applies some type of preprocessing to the original \textit{AMiner} dataset before utilizing it. In some cases, additional data from other sources is integrated to construct the final datasets.

\tablename~\ref{tab:AMinerDatasetWorks} provides a summary of the various dataset versions derived and utilized by the analyzed works. 

\begin{footnotesize}
\fontsize{6}{8}\selectfont
\begin{longtable}{llllll}
\caption{AMiner dataset-based customizations main characteristics}\label{tab:AMinerDatasetWorks}\\
\toprule
\textbf{Work} & \textbf{Version} & \textbf{A?} & \multicolumn{3}{p{8cm}}{\textbf{Content description provided by authors}}\\
\midrule
\endfirsthead
\multicolumn{5}{c}
{{\bfseries \tablename\ \thetable{} -- continued from previous page}} \\
\toprule
\textbf{Work} & \textbf{Version} & \textbf{A?} & \multicolumn{3}{p{8cm}}{\textbf{Content description provided by authors}}\\
\midrule
\endhead
\midrule
\multicolumn{5}{c}{{Continued on next page}} \\ 
\bottomrule
\endfoot
\endlastfoot               
    \cite{IJ:ESWA:XIAO:2023} & ? & ur & \textbf{Papers}: 3,499,032 & \textbf{Authors}: 325,341\\
    & & & \textbf{Venues}: 15,225 & \textbf{Citations}: 25,725,440\\ 
    & & & \textbf{Terms}: 4,872,411\\
    \cite{IC:EITCE:NIU:2023} & ? & \ding{55} & \textbf{Nodes}: 39,021 & \textbf{Edges}: 201,436\\
    & & & \textbf{Authors}: 1000,000 & \textbf{Years}: 2004-2014\\    
    \cite{IJ:AI:GAO:2023} & ? & \ding{55} & \textbf{Papers}: 961,254 & \textbf{Authors}: 623,706\\
    & & & \textbf{Keywords}: 4,121,783 & \textbf{Paper-Paper}: 2,811,720 \\
    & & & \textbf{Paper-Keyword}: 6,125,526 & \textbf{Paper-Author}: 2,193,514\\
    \cite{IJ:AS:Chen:2023} & ? & \ding{55} & \textbf{Papers}: 500,000 & \textbf{Keywords}: 9,668\\
    \cite{IJ:JOI:Yadav:2022} & ASN + Scopus & \ding{51}\footnote{\url{https://github.com/Pratyush1296/SPACE-R}} & \textbf{Papers}: 12,611 & \textbf{Authors}: 22,603\\
    & & & \textbf{Nodes}: 41,894 & \textbf{Edges}: 108,850\\
    & & & \textbf{Sparcity}: 0.999\% \\
    \cite{IJ:ASC:Wang:2024} & ASN & ur & \textbf{Papers}: 47,863 & \textbf{Researchers}: 2,030\\
    & & & \textbf{Groups}: 400 & \textbf{Researcher-Paper}: 584,807\\
    & & & \textbf{Group-Researcher}: 3,867\\    
    \cite{IJ:mathematics:GUO:2024} & ASN + AFT\footnote{\url{https://academictree.org/}} & \ding{55} & \textbf{Papers}: 89,013 & \textbf{Students}: 10,929\\
    & & & \textbf{Ratings}: 281,319 & \textbf{Social relations}: 90,686\\
    & & & \textbf{Rating density}: 0.0289\% \\
    & & & \textbf{Disciplines}: Computer Science\\
    & & & Neuroscience, Mathematics\\
    & & & Chemistry, Physics, Education\\
    \bottomrule 
\end{longtable}
\end{footnotesize}

\subsubsection{CiteULike}\label{sec:CiteULike}
\textit{CiteULike} was a free platform for managing references and bookmarking academic papers. It allowed users to organize, save, and share scholarly articles, facilitating the discovery of new content and collaboration through tagging and categorization. The service was discontinued in 2019. However, several publicly available datasets derived from \textit{CiteULike} data—such as bibliographic metadata, user interactions, and tags—remain accessible for research purposes. \textit{CiteULike-based} datasets, particularly the \textit{CiteULike-a}\footnote{\url{https://github.com/js05212/citeulike-a}} and \textit{CiteULike-t}\footnote{\url{https://github.com/js05212/citeulike-t}} releases, are among the most widely used in \RPRSS\, with both enriched by external citation data from \textit{Google Scholar} (\tablename~\ref{tab:CiteulikeDataset}).\\

\newpage
\begin{footnotesize}
\fontsize{6}{8}\selectfont
\begin{longtable}{lc}
\caption{CiteULike datasets main characteristics}\label{tab:CiteulikeDataset}\\
\toprule
\multicolumn{2}{c}{\textbf{CiteULike-a}}\\
\midrule
Disciplines & Multidisciplinary\\
Papers &  16,980\\
Users &	5,551\\
User-Paper interactions &  204,987\\
Tags & 46,391\\
Citations & 44,709\\
Sparsity & 99.78\%\\
\toprule
\multicolumn{2}{c}{\textbf{CiteULike-t}}\\
\midrule
Disciplines & Multidisciplinary\\
Papers & 25,975\\
Users & 7,947\\
User-Paper interactions &  134,860\\
Tags & 52,946\\
Citations & 32,565\\
Sparsity & 99.93\%\\
\bottomrule
\end{longtable}
\end{footnotesize}

Among the works analyzed, 26.98\% utilize some \textit{CiteULike} version, yet these represent only 8.3\% of the total datasets used, showing a greater reusability of the \textit{CiteULike} dataset compared to others. In fact, almost all works that use \textit{CiteULike} rely on \textit{CiteULike-a} \cite{J:EIJ:Huang:2024, IS:CSET:WU:2022, IJ:SC:XI:2024, IJ:CDS:AHMEDI:2022, IJ:ESWA:XI:2023, IC:MLBDBI:GUAN:2022, IJ:JS:Huang:2023, arXiv:KARIMI:2022, ARXIV:Mohamed:2022, IJ:JOI:Yadav:2022, IC:ICITACEE:Switrayana:2022} and \textit{CiteULike-t} \cite{J:EIJ:Huang:2024, IS:CSET:WU:2022, IJ:SC:XI:2024, IJ:ESWA:XI:2023, IC:MLBDBI:GUAN:2022, IJ:JS:Huang:2023, arXiv:KARIMI:2022, IC:ICITACEE:Switrayana:2022} without applying any filtering or modification.

\tablename~\ref{tab:CiteulikeDatasetWorks} provides a summary of the various dataset versions derived and utilized by the analyzed works. 

\begin{footnotesize}
\fontsize{6}{8}\selectfont
\begin{longtable}{llllll}
\caption{CiteULike dataset-based customizations main characteristics}\label{tab:CiteulikeDatasetWorks}\\
\toprule
\textbf{Work} & \textbf{Version} & \textbf{A?} & \multicolumn{3}{p{8cm}}{\textbf{Content description provided by authors}}\\
\midrule
\endfirsthead
\multicolumn{5}{c}
{{\bfseries \tablename\ \thetable{} -- continued from previous page}} \\
\toprule
\textbf{Work} & \textbf{Version} & \textbf{A?} & \multicolumn{3}{p{8cm}}{\textbf{Content description provided by authors}}\\
\midrule
\endhead
\midrule
\multicolumn{5}{c}{{Continued on next page}} \\ 
\bottomrule
\endfoot
\endlastfoot               
    \cite{IJ:ASC:Wang:2024,IJ:SMC:WANG:2022} & ? & ur & \textbf{Papers}: 82,376 & \textbf{Users}: 1,659\\
    & & & \textbf{Groups}: 718 & \textbf{User-Paper}:480,399\\
    & & & \textbf{User-Group}: 3,073 & \textbf{Group-Paper}: 198,744\\
    & & & \textbf{User-Paper sparcity}: 99.85\% & \textbf{User-Group sparcity}: 99.74\%\\    
    \cite{IC:KSEM:YU:2022} & CiteULike-a & \ding{55} & \textbf{Papers}:11,845 & \textbf{Users}: 4,880\\
    & & & \textbf{Interactions}: 172,267 & \textbf{Citations}: 31,517\\
    & & & \textbf{Keywords}: 582\\ 
    \cite{IJ:IS:Chaudhuri:2022} & CiteULike-a & \ding{55} & \textbf{Papers}: 1,000 & \textbf{Users}: 10,327\\
    & & & \textbf{Tags}: 1,050 & \textbf{User-Paper}: 14,500\\ 
    \cite{IC:IKT:JAFARI:2021,IJ:AUT:Jafari:2023} & CiteULike-a & \ding{51}\footnote{\url{ut-kdd/CiteUlike-SPIN-dataset (github.com)}} & \textbf{Papers}: 16,980 & \textbf{Users}: 5,551\\
    & & & \textbf{Authors}: 5,037 & \textbf{Tags}: 7,386\\
    & & & \textbf{Venues}: 1,230\\ 
    \bottomrule 
\end{longtable}
\end{footnotesize}

\subsubsection{Other source datasets}\label{sec:OtherDatasets}

In addition to the previously mentioned datasets, there exists another group of \textit{publicly available} datasets that, while published, are less frequently utilized in the analyzed works. Of the works analyzed, 17.46\% utilize some of these datasets, representing 13.89\% of the total datasets used; however, only three of these works employ the datasets in their original, unmodified form \cite{IJ:SECS:Hamisu:2024, AC:ACIIDS:PAN:2024, IJ:EIT:Smail:2023}.

\tablename~\ref{tab:OtherPublicDatasets} provides a summary of the datasets, either original or derived,  utilized by the analyzed works.

\begin{footnotesize}
\fontsize{6}{8}\selectfont
\begin{longtable}{llllll}
\caption{Publicly available other datasets main characteristics}\label{tab:OtherPublicDatasets}\\
\toprule
\textbf{Work} & \textbf{Version} & \textbf{A?} & \multicolumn{3}{p{4cm}}{\textbf{Content description provided by authors}}\\
\midrule
\endfirsthead
\multicolumn{5}{c}
{{\bfseries \tablename\ \thetable{} -- continued from previous page}} \\
\toprule
\textbf{Work} & \textbf{Version} & \textbf{A?} & \multicolumn{3}{p{4cm}}{\textbf{Content description provided by authors}}\\
\midrule
\endhead
\midrule
\multicolumn{5}{c}{{Continued on next page}} \\ 
\bottomrule
\endfoot
\endlastfoot    
    \cite{IJ:SECS:Hamisu:2024, AC:ACIIDS:PAN:2024} & Sugiyama\footnote{\url{http://scholarbank.nus.edu.sg/handle/10635/146027}}\cite{DT:Sugiyama:2013, JCDL:Sugiyama:2013}  & \ding{51} & \textbf{Researchers}: 50 & \textbf{Candidate Papers}: 100,351\\
     & & & \textbf{Average publication} & \textbf{Average citation}\\
     & & & \textbf{per researcher}: 10 & \textbf{per candidate}\\
     & & & \textbf{Average citation} & \textbf{papers}: 17.9\\
     & & & \textbf{per publication}: 14.8\\
     & & & \textbf{Average references}\\
     & & & \textbf{per publication}: 15.0\\  
    \cite{IJ:EIT:Smail:2023} & Elsevier OA\\ 
    & CC-BY Corpus\footnote{\url{https://researchcollaborations.elsevier.com/en/datasets/elsevier-oa-cc-by-corpus}} & \ding{51} & \textbf{Papers}: 40,091 & \textbf{Disciplines}: Biology,\\
    & & & & Chemistry, Engineering,\\
    & & & & Health Sciences,\\
    & & & & Mathematics,\\
    & & & & Physics\\
    & & & \textbf{Years}: 2014-2020\\
    \cite{IJ:SHTI:GUO:2022} & TREC 2005\\
    & Genomics Track\footnote{\url{https://dmice.ohsu.edu/trec-gen/2005data.html}} & \ding{55} & \textbf{Papers}: 3,098\\
    \cite{IC:ICDE:XIE:2022} & US patent\footnote{\url{https://developer.uspto.gov/data}} & \ding{55} &  \textbf{Papers}: 182,260 & \textbf{Authors}: 73,974\\
    & & & \textbf{Years}: 2017\\
    \cite{IJ:IPM:Long:2024} & PWC\footnote{\url{https://github.com/malteos/aspect-document-embeddings}}\cite{IJ:JCDL:Ostendorff:2022} & ur & \textbf{Papers}: 88,940 & \textbf{Disciplines}: Machine\\
    & & & & Learning\\
    \cite{IJ:IPM:Long:2024} & SCICITE\footnote{\url{https://github.com/allenai/scicite}} & ur & \textbf{Papers}: 8,045 & \textbf{Disciplines}:\\
    & & & & Computer Science, Medicine\\
    \cite{IJ:ESWA:Wang:2023} & Spotify\footnote{\url{https://www.kaggle.com/datasets/andrewmvd/spotify-playlists}} & \ding{55} & \textbf{Papers}: 11,703 & \textbf{Users}: 1,720\\
    \cite{IJ:ESWA:Wang:2023} & Yelp\footnote{\url{https://www.kaggle.com/datasets/yelp-dataset/yelp-dataset}} & \ding{55} & \textbf{Papers}: 19,479 & \textbf{Users}: 10,918\\
    \cite{IJ:AS:Chen:2023} & Metallurgical & ? & \textbf{Papers}: 5,702 & \textbf{Authors}: 3,268\\
    & DB & & \textbf{Phrases}: 2,000 & \textbf{Paper-Author}: 7,375\\
    & & & \textbf{Paper-Term}: 37,316\\
    \cite{IJ:ESWA:MEI:2022} & Arxiv HEP-TH\footnote{\url{https://snap.stanford.edu/data/cit-HepTh.html}} & \ding{55} & \textbf{Papers}: 18,066 & \textbf{Authors}: 11,879\\
    & & & \textbf{Citations}: 153,414\\
    \bottomrule
\end{longtable}
\end{footnotesize}

In many works, authors create \textit{ad-hoc} datasets specifically designed to train and evaluate their recommendation models. These datasets are often derived from well-established digital libraries, such as \textit{Scopus}\footnote{\url{https://dev.elsevier.com/sc_apis.html}}, \textit{Semantic Scholar}\footnote{\url{https://www.semanticscholar.org/product/api}}, \textit{ArXiv}\footnote{\url{https://info.arxiv.org/help/api/index.html}}, \textit{PubMed}\footnote{\url{https://www.ncbi.nlm.nih.gov/home/develop/api/}}, \textit{\WoS}\footnote{\url{https://developer.clarivate.com/apis/wos}}, and the \textit{ACM Digital Library}\footnote{\url{https://dl.acm.org/}}, all of which provide comprehensive collections of scholarly content. Within the works analyzed, 36.51\% utilize some \textit{ad-hoc} dataset, representing 31.94\% of the total datasets used.

\tablename~\ref{tab:adhocDatasets} provides a summary of the ad-hoc datasets created and utilized by the analyzed works.

\begin{footnotesize}
\fontsize{6}{8}\selectfont
\begin{longtable}{llllll}
\caption{Ad-hoc datasets main characteristics}\label{tab:adhocDatasets}\\
\toprule
\textbf{Work} & \textbf{Version} & \textbf{A?} & \multicolumn{3}{p{8cm}}{\textbf{Content description provided by authors}}\\
\midrule
\endfirsthead
\multicolumn{5}{c}
{{\bfseries \tablename\ \thetable{} -- continued from previous page}} \\
\toprule
\textbf{Work} & \textbf{Version} & \textbf{A?} & \multicolumn{3}{p{8cm}}{\textbf{Content description provided by authors}}\\
\midrule
\endhead
\midrule
\multicolumn{5}{c}{{Continued on next page}} \\ 
\bottomrule
\endfoot
\endlastfoot            
    \cite{IJ:IEEE:SARWAR:2021} & ? & ur & \textbf{Disciplines}: Computer Science\\ 
    \cite{IJ:IEEE:STERLING:2022} & ? & \ding{55} & \textbf{Papers}: 2,271 & \textbf{Disciplines}: Carbon,\\
    & & & & Chemical Physics,\\
    & & & & Nature Chemistry,\\
    & & & & Physics of Plasmas,\\
    & & & & X-Ray Spectrometry \\ 
    \cite{IJ:Heliyon:LI:2024} & DBLP & \ding{55} & \textbf{Papers}: 700\\ 
    \cite{IJ:AI:Xiao:2023} & CNKI\footnote{\url{https://eval.cnki.net/index/}} & ur & \textbf{Disciplines}:\\
    & & & Intergenerational \\
    & & & Mobility of education & \textbf{Years}: 1987-2021\\
    & & & \textbf{Paper-Author}:\\
    & & & \textbf{Nodes}:3,386-4,621 & \textbf{Edges}: 5,888\\
    & & & \textbf{Paper-Keyword}:\\
    & & & \textbf{Nodes}: 3,386-6,049 & \textbf{Edges}: 12,429\\
    & & & \textbf{Paper-Venue}:\\
    & & & \textbf{Nodes}: 3,386-1,312 & \textbf{Edges}: 3,386\\
    & & & \textbf{Paper-Organization}:\\
    & & & \textbf{Nodes}: 3,386-2,654 & \textbf{Edges}: 4,610\\
    & & & \textbf{Paper-Fund}:\\
    & & & \textbf{Nodes}: 3,386-3,202 & \textbf{Edges}: 4,800\\ 
    \cite{IJ:AI:Xiao:2023} & CNKI & ur & \textbf{Disciplines}: Data Mining\\
    & & & and Intelligent Media &  \textbf{Years}: 1997-2021\\
    & & & \textbf{Paper-Author}:\\ 
    & & & \textbf{Nodes}: 2,975-6,802 & \textbf{Edges}: 8,272\\
    & & & \textbf{Paper-Keyword}:\\ 
    & & & \textbf{Nodes}: 2,975-7,019 & \textbf{Edges}: 12,935\\
    & & & \textbf{Paper-Venue}:\\ 
    & & & \textbf{Nodes}: 2,975-724 & \textbf{Edges}: 3,340\\
    & & & \textbf{Paper-Organization}:\\ 
    & & & \textbf{Nodes}: 2,975-2,654 & \textbf{Edges}:4,762\\
    & & & \textbf{Paper-Funds}:\\ 
    & & & \textbf{Nodes}: 2,975-3,927 & \textbf{Edges}: 4,829\\ 
    \cite{IJ:JOI:Yadav:2022} & WoS & \ding{51}\footnote{\url{https://github.com/Pratyush1296/SPACE-R}} & \textbf{Papers}: 27,442 & \textbf{Authors}: 40,408\\
    & & & \textbf{Nodes}: 76,133 & \textbf{Edges}: 223,452\\
    & & & \textbf{Sparsity}: 0.999\% & \textbf{Disciplines}:\\
    & & & & Information systems,\\ 
    & & & & Physics\\ 
    \cite{IJ:KBS:Xiao:2023} &  WoS & ur & \textbf{Papers}: 6,037 & \textbf{Authors}: 80,375\\
    & & & \textbf{Keywords}: 10,295 & \textbf{Journals}: 235\\
    & & & \textbf{Author Relations}:\\ 
    & & & \textbf{Co-author}: 20,585 & \textbf{Co-keyword}: 1,584,187\\ 
    & & & \textbf{Co-journal}: 200,619\\
    & & & \textbf{Paper Relations}:\\
    & & & \textbf{Co-author}: 36,622 & \textbf{Co-keyword}: 6,268,075\\
    & & & \textbf{Co-journal}: 225,598 & \textbf{Disciplines}: Astronomy,\\
    & & & & Physics, Education,\\
    & & & & Psychology\\ 
    \cite{IJ:IEEE:REN:2022} & MAS API & \ding{55} & \textbf{Papers}: 2,794 & \textbf{Citations}: 764\\ 
    & & & \textbf{Years}: 2000-2010 & \textbf{Cold start ratio}: 72.66\%\\
    \cite{ARXIV:Shen:2024} & PaperWithCode +\\
    & ArXiv +\\
    & PubMed & \ding{55} & \textbf{Papers}: 313,278 & \textbf{Edges}: 2,233,780\\ 
    \cite{IJ:DATABASE:Kart:2022} & PubMed & \ding{51}\footnote{\url{https://github.com/bioinfcollab/emati}} & \textbf{Disciplines}:\\
    & & & Nature science\\
    & & & \textbf{Years}: \textgreater2018 &  \textbf{Users}: 6(artificial)\\ 
    \cite{IJ:ELECTRONICS:Blazevic:2023} & Springer Nature\\
    & API & \ding{55} & \textbf{Papers}: 87,196 & \textbf{Words}: 297,227\\
    \cite{IJ:AMEC:KUS:2023} & ArXiv & \ding{55} & \textbf{Papers}: 41,000 & \textbf{Users}: 24,707\\
    & & & \textbf{Years}: 1993-2018 & \textbf{Disciplines}:\\
    & & & & Computer Science,\\ 
    & & & & Economics, Mathematics,\\
    & & & & Physics, Electrical engineering,\\
    & & & & Systems science, Statistics\\ 
    \cite{IJ:Scientometrics:GUNDOGAN:2022} & ArXiv & \ding{55} & \textbf{Papers}: 122,825 & \textbf{Disciplines}: Computer science\\ 
    \cite{IJ:Scientometrics:TAN:2023} & ArXiv & \ding{51}\footnote{\url{https:// github. com/ opend ata- ai/ tr.}} & \textbf{Papers: 21,000} & \textbf{Years}: 1991-2016\\
    \cite{IC:ICDE:XIE:2022} & Scopus & \ding{55} & \textbf{Papers}: 1,304,907 & \textbf{Authors}: 482,602\\
    & & & \textbf{Years}: 2008-2017 & \textbf{Keywords}: 127,630\\
    & & & \textbf{Nenues}: 7,653 & \textbf{Disciplines}: 27\\  
    \cite{IJ:IS:Chaudhuri:2022} & Scopus & \ding{55} & \textbf{Size}: 2000 & \textbf{Authors}: 430\\
    & & &  \textbf{Tags}: 10 & \textbf{Disciplines}:\\
    & & & Computer Science\\ 
    \cite{CF:CSCW:LI:2022} & ACM & \ding{55} & \textbf{Papers}: 31,889 & \textbf{Users}:44,953\\ 
    & & & \textbf{Entities}: 148,376 & \textbf{Relations}: 7\\
    & & & \textbf{(e1,r,e2)}: 491,679\\
    \cite{IC:ICDE:XIE:2022} & ACM & \ding{55} & \textbf{Papers}: 3,056,388 & \textbf{Authors}: 1,752,401\\
    & & & \textbf{Keywords}: 354,693 & \textbf{Venues}: 11,397\\
    & & & \textbf{Clases}: 11 &  \textbf{Affiliation}: 15,376\\
    & & & \textbf{Years}: 2000-2019\\
    \cite{EC:ECIR:Takahashi:2022} & Semantic Scholar & \ding{55} & \textbf{Papers}: 805,063 & \textbf{Disciplines}: Engineering,\\
    & & & \textbf{Sub-disciplines}: 56 & Computer science\\
    \cite{IC:WIST:Roßner:2023} & Semantic Scholar + & \ding{51}\footnote{\url{https://opendata.iisys.de/opendata/Datasets/publication_similarity.zip}} & \textbf{Papers}: 15,359 & \textbf{Publication Pairs}:\\
    & ACM & & \textbf{Disciplines}: Deep Learning & 117,941,761\\
    & & & NLP, Hypertext,\\
    & & & Social media\\
    \cite{IJ:IS:PINEDO:2024} & Semantic Scholar + & \ding{51}\footnote{\url{https://github.com/IratxePinedoehu/arzigo}} & \textbf{Papers}: 1795 & \textbf{Users}: 736\\
    & artificial interactions & & \textbf{Interactions}: 350,000 & \textbf{Disciplines:} 10\\ 
    & & & & 7 Computer Science\\
    & & & & 3 Medicine\\
    \cite{IC:SEAI:Wang:2024} & ACM & \ding{55} & \textbf{Papers}: 3,025 & \textbf{Authors}: 5,835\\
    & & & \textbf{Topics}: 56\\
    \cite{IJ:Scientometrics:TAN:2023} & USPTO & \ding{51}\footnote{\url{https:// github. com/ opend ata- ai/ tr.}} & \textbf{Papers: 21,000} & \textbf{Years}: 2002-2018\\
    \bottomrule 
\end{longtable}
\end{footnotesize}

\subsection{How are the \RPRSS\ evaluated?}\label{sec:Evaluation}

Evaluating the performance of a \RPRS\ is essential to ensure it provides accurate and relevant article suggestions to users. The evaluation of a \RPRS\ helps assess the quality of the recommendations and identifies areas where the system can be improved to better meet the needs of the researchers.

Evaluating \RPRSS\ is a crucial task to ensure that they effectively support researchers in discovering relevant papers. The evaluation process involves several key components, starting with \textit{relevancy} and \textit{target} selection, which can either be based on \textit{pre-labeled datasets} or \textit{human assessments}. These processes help determine how accurately the system recommends papers that align with the research interest or queries of the user, with ground truth often being established through user interactions, relevance labels, or human judgments.

To assess the performance of \RPRSS, a variety of \textit{metrics} are used. These metrics measure different aspects of recommendation quality, including \textit{precision}, \textit{recall}, \textit{diversity}, \textit{novelty}, and \textit{user satisfaction}, each providing insights into how well the system is meeting user needs. The \textit{evaluation procedure} itself involves testing the system under various \textit{configurations}, performing \textit{ablation} studies to isolate the impact of specific components, and comparing results to \textit{baseline} models for a clearer understanding of improvements. Additionally, the choice of \textit{datasets} plays a crucial role in the evaluation process, as different datasets provide diverse benchmarks to test the \textit{robustness} and \textit{generalizability} of the system.

This paper explores these evaluation aspects in detail, discussing the methods used for assessing \textit{relevancy}, the specific \textit{metrics} applied, and the \textit{procedures} followed to comprehensively evaluate the effectiveness of \RPRSS.

\subsubsection{What are the relevance criteria used to evaluate the recommendations?}\label{sec:RelevanceCriteria}

Relevance criteria refer to the factors used to assess the quality and accuracy of the recommendations generated by \RPRSS . Several relevance criteria were identified in the analyzed works, including: \textit{paper references}, which use citations as a ground truth by considering the references cited by a paper as relevant (\textit{e.g.}, the references a paper cites are deemed relevant to that paper); \textit{user interactions}, where papers that the user has interacted with (\textit{e.g.}, clicked on, read, downloaded, or added to a personal library) are considered relevant based on their historical interactions; \textit{labeled papers}, where pre-classified documents serve as benchmarks for system performance (\textit{e.g.}, papers labeled as relevant or irrelevant, or assigned to specific topics or categories); and \textit{human judgment}, where experts evaluate the relevance of recommended papers based on their expertise (\textit{e.g.}, researchers manually reviewing suggestions).

44.29\% of the works use \textit{references} as a relevance criterion. 38.71\% of them consider the author as the user, so the papers referenced by authors in their publications are used as a basis for relevance \cite{IC:ASSP:Shen:2024, IC:DASFAA:Wang:2022, IC:EITCE:NIU:2023, IC:ICDE:XIE:2022, IJ:AI:GAO:2023, IJ:AS:Chen:2023, IJ:ESWA:Wang:2023, IJ:ESWA:XIAO:2023, IJ:IEEE:REN:2022, IJ:mathematics:GUO:2024, IJ:Scientometrics:Jiang:2023, IJ:TIIS:GUO:2024}. 38.71\% of the works using \textit{references} as relevance criteria rely on references found in the \textit{input paper} \cite{ARXIV:Shen:2024, IJ:AI:Thierry:2023, IJ:ESWA:MEI:2022, IJ:Heliyon:LI:2024, IJ:IEEE:HADHIATMA:2023, IJ:IEEE:STERLING:2022, IJ:JOI:Yadav:2022, IJ:Scientometrics:ALI:2022, IJ:Scientometrics:GUNDOGAN:2022, IJ:Scientometrics:Huang:2024, IJ:Scientometrics:Thierry:2023,IJ:Scientometrics:TAN:2023}. 12.9\% of the works using \textit{references}, even if the authors are not established as users, use datasets that only consider citation information for evaluating the system, so authors are considered the users, and their \textit{references} are treated as \textit{interactions} \cite{PP:Mahesh:2023, IC:MLBDBI:GUAN:2022, IJ:IS:Chaudhuri:2022, IJ:ASC:Wang:2024}. 6.45\% of the works use \textit{references} from the same paper that is used to extract the \textit{input text} \cite{EC:ECIR:Takahashi:2022, IJ:DSM:Sivasankari:2023}.
\cite{IJ:AS:Wang:2024} not only considers positive examples the articles \textit{referenced} by each author, but also the papers they have \textit{published}.
%In some cases, even the author's own papers are considered to be part of the relevant papers \cite{IJ:AS:Wang:2024}.

Out of the total works that use actual user \textit{interactions} as a relevance criterion, 85\% utilize the \textit{CiteULike} dataset. The remaining works either use \textit{non-RPRS-specific} datasets, such as \textit{Spotify} or \textit{Yelp} \cite{IJ:ESWA:Wang:2023}, or rely on artificially generated interactions \cite{IJ:KIS:STERGIOPOULOS:2024, IJ:DATABASE:Kart:2022}.

12.86\% of the works use some kind of \textit{label} to identify relevant papers. These labels can be \textit{discipline} classification labels that help match academic fields between users and recommended papers \cite{CF:CSCW:LI:2022, IC:SEAI:Wang:2024} or between \POI\ and recommended papers \cite{IJ:AMEC:KUS:2023, IJ:EIT:Smail:2023}. \textit{Citation intent}, which reflects the task or focus of the paper, is also used \cite{IJ:IPM:Long:2024}. In addition, more explicit labels are sometimes applied to indicate \textit{relevance} to particular authors \cite{AC:ACIIDS:PAN:2024, AC:ACIIDS:PAN:2024} or \textit{topics} \cite{IJ:SHTI:GUO:2022}.

Only 8.57\% of the works use \textit{human judgment} to evaluate the recommendations obtained. This relevance criterion can help assess the recommendations by better aligning them with actual needs, leading to more accurate evaluations. The choice of criteria depends on the level of detail required and the specific context in which the evaluation is conducted. Some systems use simple \textit{binary ratings}, such as categorizing publications as either relevant or irrelevant \cite{J:JIS:MUNGEN:2022}, or correct or incorrect with reasoning \cite{IJ:IS:PINEDO:2024}. Others implement more nuanced \textit{scales}, assigning points based on \textit{similarity} or \textit{relevance strength} \cite{IC:WIST:Roßner:2023, IJ:ELECTRONICS:Blazevic:2023, IJ:IEEE:SARWAR:2021}. In addition to rating systems, some evaluations involve assessing the \textit{satisfaction} of the users, highlighting the \textit{subjective} nature of \textit{relevance} judgment \cite{IJ:Scientometrics:Kanwal:2024}.

Finally, alternative relevancy criteria in evaluating recommendations include assessing \textit{co-authorship} potential by evaluating \textit{research compatibility} \cite{IJ:KBS:Xiao:2023}, determining success based on the inclusion of papers by scientific research users \cite{IJ:JES:LI:2024}, and matching recommendations with user input parameters such as research \textit{topics} and \textit{academic affiliations} \cite{IJ:AI:Xiao:2023}.

In \tablename~\ref{tab:RelevancyCriteria} a detailed summary of different relevancy criteria is presented.

\begin{footnotesize}
\fontsize{6}{8}\selectfont
\begin{longtable}{cp{5cm}c}
\caption{Classification according to relevance criteria}\label{tab:RelevancyCriteria}\\ 
\toprule
\textbf{Relevancy} & \textbf{Works} & \textbf{pct}\\
\midrule
\endfirsthead
\multicolumn{3}{r}
{{\bfseries \tablename\ \thetable{} -- continued from previous page}}\\
\toprule
\textbf{Relevancy} & \textbf{Works} & \textbf{pct}\\
\midrule
\endhead
\midrule
\multicolumn{3}{r}{{Continued on next page}} \\ 
\bottomrule
\endfoot
\endlastfoot               
    Paper references & \cite{ARXIV:Shen:2024, EC:ECIR:Takahashi:2022, IC:ASSP:Shen:2024, IC:DASFAA:Wang:2022, IC:EITCE:NIU:2023, IC:ICDE:XIE:2022, IC:MLBDBI:GUAN:2022, IJ:AI:GAO:2023, IJ:AI:Thierry:2023, IJ:AS:Chen:2023, IJ:AS:Wang:2024, IJ:ASC:Wang:2024, IJ:DSM:Sivasankari:2023, IJ:ESWA:MEI:2022, IJ:ESWA:Wang:2023, IJ:ESWA:XIAO:2023, IJ:Heliyon:LI:2024, IJ:IEEE:HADHIATMA:2023, IJ:IEEE:REN:2022, IJ:IEEE:STERLING:2022, IJ:IS:Chaudhuri:2022, IJ:JOI:Yadav:2022, IJ:mathematics:GUO:2024, IJ:Scientometrics:ALI:2022, IJ:Scientometrics:GUNDOGAN:2022, IJ:Scientometrics:Huang:2024, IJ:Scientometrics:Jiang:2023, IJ:Scientometrics:Thierry:2023,IJ:TIIS:GUO:2024, PP:Mahesh:2023,IJ:Scientometrics:TAN:2023} & 44.29\%\\
    Interactions & \cite{arXiv:KARIMI:2022, ARXIV:Mohamed:2022, IC:ICITACEE:Switrayana:2022, IC:IKT:JAFARI:2021, IC:KSEM:YU:2022, IC:MLBDBI:GUAN:2022, IJ:ASC:Wang:2024, IJ:AUT:Jafari:2023, IJ:CDS:AHMEDI:2022, IJ:DATABASE:Kart:2022, IJ:ESWA:Wang:2023, IJ:ESWA:XI:2023, IJ:IS:Chaudhuri:2022, IJ:JOI:Yadav:2022, IJ:JS:Huang:2023, IJ:KIS:STERGIOPOULOS:2024, IJ:SC:XI:2024, IJ:SMC:WANG:2022, IS:CSET:WU:2022, J:EIJ:Huang:2024} & 28.57\%\\
    Labeled Paper & \cite{AC:ACIIDS:PAN:2024, CF:CSCW:LI:2022, IC:SEAI:Wang:2024, IJ:AMEC:KUS:2023, IJ:EIT:Smail:2023, IJ:IPM:Long:2024, IJ:SECS:Hamisu:2024, IJ:SHTI:GUO:2022, IJ:Scientometrics:TAN:2023} & 12.86\%\\
    Human judment & \cite{IC:WIST:Roßner:2023, IJ:ELECTRONICS:Blazevic:2023, IJ:IEEE:SARWAR:2021, IJ:IS:PINEDO:2024, IJ:Scientometrics:Kanwal:2024, J:JIS:MUNGEN:2022} & 8.57\%\\
    Others & \cite{IJ:JES:LI:2024, IJ:KBS:Xiao:2023, IJ:AI:Xiao:2023} & 4.29\%\\
\bottomrule 
\end{longtable}
\end{footnotesize}

\subsubsection{What metrics are used in the evaluation process?}\label{sec:MeasureMetrics}
The analyzed works are mainly evaluated based on metrics that measure how accurately they suggest relevant papers. These include \textit{precision}, \textit{recall}, and \textit{F-score}, which assess the relevance of recommendations, as well as \textit{\MAP}, \textit{\NDCG}, and \textit{\MRR}, which evaluate ranking and the effectiveness of the recommendation order. These metrics typically depend on the k parameter, which sets the cut-off point for the top recommendations to evaluate.

Among the \textit{predictive} metrics, the most commonly used are \textit{recall}, \textit{precision}, and \textit{F-score}, with 73.02\%, 49.21\%, and 22.22\% of the works utilizing them, respectively. These metrics are widely employed to assess the accuracy and relevance of recommendations, with \textit{recall} focusing on the completeness of relevant items retrieved, \textit{precision} measuring the correctness of the recommendations, and \textit{F-score} providing a balance between \textit{precision} and \textit{recall}.

\textit{Ranking} metrics, which assess the ability of the system to order items based on their relevance to the user or \POI\, are also extensively used. Notably, \textit{\NDCG} is employed by 43.54\% of the works, \textit{\MRR} by 30.16\%, and \textit{\MAP} by 22.22\%. These metrics help evaluate the effectiveness of the recommendation ranking and the system's ability to present the most relevant items at the top.

11.11\% of the works evaluate their proposals using a \textit{single} metric \cite{IJ:IEEE:STERLING:2022, IJ:IEEE:SARWAR:2021, IJ:ELECTRONICS:Blazevic:2023, IJ:SHTI:GUO:2022, EC:ECIR:Takahashi:2022, ARXIV:Mohamed:2022, arXiv:KARIMI:2022}. 55.56\% of the works use both \textit{predictive} and \textit{ranking} quality metrics \cite{IC:IKT:JAFARI:2021, IJ:AUT:Jafari:2023, IJ:ESWA:MEI:2022, IJ:AI:GAO:2023, IJ:IS:Chaudhuri:2022, IJ:mathematics:GUO:2024, IC:KSEM:YU:2022, IJ:ASC:Wang:2024, IJ:AS:Wang:2024, IJ:ESWA:XIAO:2023, IJ:SMC:WANG:2022, IJ:JOI:Yadav:2022, IJ:Scientometrics:Jiang:2023, ARXIV:Shen:2024, IJ:DSM:Sivasankari:2023, AC:ACIIDS:PAN:2024, IJ:IS:PINEDO:2024, IJ:Scientometrics:ALI:2022, J:EIJ:Huang:2024, IJ:KIS:STERGIOPOULOS:2024, IJ:ESWA:XI:2023, IJ:TIIS:GUO:2024, IJ:SC:XI:2024, IC:EITCE:NIU:2023, IC:MLBDBI:GUAN:2022, IC:ICITACEE:Switrayana:2022, IJ:Scientometrics:Huang:2024, IJ:AI:Thierry:2023, IJ:IEEE:HADHIATMA:2023, IJ:Scientometrics:Thierry:2023, IJ:CDS:AHMEDI:2022, IC:ASSP:Shen:2024, IC:DASFAA:Wang:2022, IJ:ESWA:Wang:2023, IJ:Scientometrics:TAN:2023}. 31.75\% of the works use only \textit{predictive} metrics \cite{IJ:DATABASE:Kart:2022, J:JIS:MUNGEN:2022, IJ:KBS:Xiao:2023, IJ:SECS:Hamisu:2024, IJ:JES:LI:2024, IJ:AI:Xiao:2023, IJ:AMEC:KUS:2023, IJ:Heliyon:LI:2024, IJ:JS:Huang:2023, IJ:Scientometrics:GUNDOGAN:2022, PP:Mahesh:2023, IJ:EIT:Smail:2023, IJ:IEEE:STERLING:2022, IJ:ELECTRONICS:Blazevic:2023, IJ:IEEE:REN:2022, ARXIV:Mohamed:2022, arXiv:KARIMI:2022, IC:SEAI:Wang:2024, CF:CSCW:LI:2022, IJ:SHTI:GUO:2022}. Lastly, 11.11\% of the works employ only \textit{ranking} metrics \cite{IJ:Scientometrics:Kanwal:2024, IC:ICDE:XIE:2022, IJ:IPM:Long:2024, IJ:IEEE:SARWAR:2021, IS:CSET:WU:2022, IJ:AS:Chen:2023, EC:ECIR:Takahashi:2022}.

In \tablename~\ref{tab:Metrics} a detailed summary of used metrics is presented.

% https://www.evidentlyai.com/ranking-metrics/evaluating-recommender-systems

\begin{footnotesize}
\fontsize{6}{8}\selectfont
\begin{longtable}{cccccccccclc}
\caption{Classification according to measure metrics}\label{tab:Metrics}\\
\toprule
\textbf{P} & \textbf{R} & \textbf{F-score} & \textbf{nDCG}& \textbf{MRR} & \textbf{MAP} & \textbf{Others} & \textbf{Works}\\
\midrule
\endfirsthead
\multicolumn{8}{r}
{{\bfseries \tablename\ \thetable{} -- continued from previous page}} \\
\toprule
\textbf{P} & \textbf{R} & \textbf{F-score} & \textbf{nDCG}& \textbf{MRR} & \textbf{MAP} & \textbf{Others} & \textbf{Works}\\
\midrule
\endhead
\midrule
\multicolumn{8}{r}{{Continued on next page}} \\ 
\bottomrule
\endfoot
\endlastfoot   
    \ding{51} & \ding{51} & \ding{51} & \ding{51} & \ding{51} &  & \ding{51} & \cite{IC:IKT:JAFARI:2021, IJ:AUT:Jafari:2023}\\
    \ding{51} & \ding{51} & \ding{51} & \ding{51} &  & \ding{51} & \ding{51} & \cite{IJ:Scientometrics:TAN:2023}\\
    \ding{51} & \ding{51} & \ding{51} &  &  &  & \ding{51} & \cite{IJ:KBS:Xiao:2023}\\
    \ding{51} & \ding{51} & \ding{51} &  &  &  &  & \cite{ IJ:AI:Xiao:2023,IJ:SECS:Hamisu:2024,IJ:JES:LI:2024,IJ:DATABASE:Kart:2022,IJ:AMEC:KUS:2023,J:JIS:MUNGEN:2022}\\
    \ding{51} & \ding{51} &  & \ding{51} & \ding{51} & \ding{51} &  & \cite{IJ:ESWA:MEI:2022}\\
    \ding{51} & \ding{51} &  & \ding{51} & \ding{51} &  &  & \cite{IJ:AI:GAO:2023}\\
    \ding{51} & \ding{51} &  & \ding{51} &  &  &  & \cite{IJ:IS:Chaudhuri:2022, IC:KSEM:YU:2022, IJ:ASC:Wang:2024,IJ:mathematics:GUO:2024}\\
    \ding{51} & \ding{51} &  &  & \ding{51} & \ding{51} & \ding{51} & \cite{IJ:JOI:Yadav:2022}\\
    \ding{51} & \ding{51} &  &  & \ding{51} & \ding{51} &  & \cite{IJ:ESWA:XIAO:2023, IJ:SMC:WANG:2022}\\
    \ding{51} & \ding{51} &  &  & \ding{51} &  &  & \cite{ARXIV:Shen:2024, IJ:Scientometrics:Jiang:2023, IJ:DSM:Sivasankari:2023}\\
    \ding{51} & \ding{51} &  &  &  &  &  & \cite{IJ:EIT:Smail:2023, IJ:Heliyon:LI:2024, IJ:Scientometrics:GUNDOGAN:2022, PP:Mahesh:2023, IJ:JS:Huang:2023}\\
    & \ding{51} &  & \ding{51} &  & \ding{51} &  & \cite{IJ:Scientometrics:ALI:2022}\\
    & \ding{51} &  & \ding{51} &  &  &  & \cite{IJ:KIS:STERGIOPOULOS:2024, IJ:SC:XI:2024, IJ:TIIS:GUO:2024, IJ:ESWA:XI:2023, J:EIJ:Huang:2024,IC:MLBDBI:GUAN:2022}\\
    & \ding{51} &  &  & \ding{51} & \ding{51} &  & \cite{IJ:IEEE:HADHIATMA:2023, IJ:Scientometrics:Huang:2024, IJ:Scientometrics:Thierry:2023, IJ:AI:Thierry:2023}\\
    & \ding{51} &  &  & \ding{51} &  &  & \cite{IJ:CDS:AHMEDI:2022}\\
    & \ding{51} &  & \ding{51} &  &  &  & \cite{IC:ICITACEE:Switrayana:2022, IC:EITCE:NIU:2023}\\
    & \ding{51} &  &  &  &  &  & \cite{arXiv:KARIMI:2022, ARXIV:Mohamed:2022}\\
    & \ding{51} &  &  &  &  & \ding{51} & \cite{IC:ASSP:Shen:2024, IJ:IEEE:REN:2022}\\
    &  & \ding{51} & \ding{51} &  &  & \ding{51} & \cite{IC:DASFAA:Wang:2022, IJ:ESWA:Wang:2023}\\
    &  &  & \ding{51} & \ding{51} & \ding{51} &  & \cite{IJ:Scientometrics:Kanwal:2024,IC:ICDE:XIE:2022}\\
    &  &  & \ding{51} &  &  & \ding{51} & \cite{IJ:AS:Chen:2023, IS:CSET:WU:2022}\\
    &  &  & \ding{51} &  &  &  & \cite{IJ:IEEE:SARWAR:2021}\\
    &  &  &  &  &  & \ding{51} & \cite{IJ:ELECTRONICS:Blazevic:2023,IJ:SHTI:GUO:2022}\\
    &  & \ding{51} &  &  &  & \ding{51} & \cite{CF:CSCW:LI:2022,IC:SEAI:Wang:2024}\\
    \ding{51} & \ding{51} &  & \ding{51} &  &  & \ding{51} & \cite{IJ:AS:Wang:2024}\\
    &  &  & \ding{51} &  & \ding{51} &  & \cite{IJ:IPM:Long:2024}\\
    &  &  &  &  & \ding{51} &  & \cite{EC:ECIR:Takahashi:2022}\\
    \ding{51} &  &  &  & \ding{51} &  &  & \cite{IJ:IS:PINEDO:2024}\\
    \ding{51} &  &  & \ding{51} & \ding{51} &  &  & \cite{AC:ACIIDS:PAN:2024}\\
    \ding{51} &  &  &  &  &  &  & \cite{IJ:IEEE:STERLING:2022}\\
\bottomrule 
49.21\% & 73.02\% & 22.22\% & 44.44\% & 30.16\% & 22.22\% & 25.40\%
\end{longtable}
\end{footnotesize}

In addition to the previously mentioned metrics, other metrics have also been observed in 25.40\% of the works, with \textit{\AUC} and \textit{\HR} being the most commonly used. \textit{\AUC} evaluates the ability of the system to distinguish between relevant and irrelevant papers, while \textit{\HR} is the ratio of the number of relevant documents in the Top-N recommendation list to the total number of relevant documents in the dataset. Another, less known metric, \textit{bpref}, has been used in one work \cite{IJ:Scientometrics:TAN:2023}; it measures retrieval effectiveness by counting how often non-relevant documents are ranked before relevant ones using binary relevance judgments.

Only a small number of works use behavioral metrics that go 'beyond accuracy' and evaluate important qualities of a \RPRS, such as \textit{diversity}, \textit{recency}, and \textit{scalability}, among others.

\tablename~\ref{tab:OtherMetrics} summarizes other metrics found.

\begin{footnotesize}
\fontsize{6}{8}\selectfont
\begin{longtable}{cp{5cm}c}
\caption{Other measure metrics}\label{tab:OtherMetrics}\\
\toprule
\textbf{Label} & \textbf{Works} & \textbf{pct}\\
\midrule
\AUC & \cite{IJ:IEEE:REN:2022, IC:DASFAA:Wang:2022, IJ:ESWA:Wang:2023, IJ:KBS:Xiao:2023,CF:CSCW:LI:2022,IJ:SHTI:GUO:2022} & 9.52\%\\
\HR & \cite{IC:ASSP:Shen:2024, IJ:AS:Chen:2023, IJ:AS:Wang:2024,IS:CSET:WU:2022} & 6.35\%\\
ACC & \cite{IJ:KBS:Xiao:2023,IC:SEAI:Wang:2024} & 3.17\%\\
Success & \cite{IC:IKT:JAFARI:2021, IJ:AUT:Jafari:2023} & 3.17\%\\
Scalability & \cite{IJ:JOI:Yadav:2022} & 1.59\%\\
Recency & \cite{IJ:JOI:Yadav:2022} & 1.59\%\\
High citation popularity & \cite{IJ:JOI:Yadav:2022} & 1.59\%\\
Diversity & \cite{IJ:JOI:Yadav:2022} & 1.59\%\\
\TNR & \cite{IJ:KBS:Xiao:2023} & 1.59\%\\
\MSE & \cite{IJ:KBS:Xiao:2023} & 1.59\%\\
bpref & \cite{IJ:Scientometrics:TAN:2023} & 1.59\%\\
\bottomrule 
\end{longtable}
\end{footnotesize}

\subsubsection{How are the evaluations conducted?}\label{sec:EvaluationProcedure}

When analyzing the evaluation procedures, the various types of evaluations, experiments, and tests identified in the analyzed works have been considered. Both \textit{quantitative} evaluations, based on datasets, and \textit{qualitative} evaluations, derived from expert opinions through user studies or case studies, have been observed. In many instances, additional experiments and tests are carried out. On the one hand, \textit{ablation} studies are conducted, where specific components of the models are removed to assess partial system performance. On the other hand, tests involving different \textit{parameter configurations} are performed. Finally, in a few cases, \textit{runtime performance} is also assessed.

Moreover, a brief examination of the use of both \textit{datasets} and \textit{baselines} in these systems has been carried out, as both are crucial for the evaluation procedure.

76.19\% of the analyzed papers focus on \textit{quantitative} assessments. 31.25\% of them are complemented by both, \textit{ablation} studies and \textit{configuration} tests \cite{IJ:mathematics:GUO:2024, IC:KSEM:YU:2022, IJ:ASC:Wang:2024,IJ:ESWA:XIAO:2023,IJ:SMC:WANG:2022, IJ:Scientometrics:Jiang:2023, ARXIV:Shen:2024, IJ:Scientometrics:ALI:2022, J:EIJ:Huang:2024, IJ:Scientometrics:Huang:2024, IJ:AI:Thierry:2023, IJ:IEEE:HADHIATMA:2023, IC:DASFAA:Wang:2022, IC:ICDE:XIE:2022, IS:CSET:WU:2022}. 29.17\% are complemented by \textit{ablation} studies \cite{IC:IKT:JAFARI:2021, 
IJ:AUT:Jafari:2023, IJ:AS:Wang:2024, IJ:JOI:Yadav:2022, IJ:DSM:Sivasankari:2023, IC:EITCE:NIU:2023, IC:MLBDBI:GUAN:2022, IJ:Scientometrics:Thierry:2023, IC:ASSP:Shen:2024,IJ:KBS:Xiao:2023,IJ:JS:Huang:2023,IC:SEAI:Wang:2024,CF:CSCW:LI:2022,IJ:AS:Chen:2023} and 10.42\% by \textit{configuration} tests \cite{AC:ACIIDS:PAN:2024, IJ:TIIS:GUO:2024, IJ:Heliyon:LI:2024, IJ:IEEE:REN:2022, IJ:KIS:STERGIOPOULOS:2024}. Additionally, 8.33\% of the works that conduct exclusively \textit{quantitative} evaluations also assess \textit{execution time} \cite{J:EIJ:Huang:2024, IC:SEAI:Wang:2024, IJ:KIS:STERGIOPOULOS:2024, IJ:DATABASE:Kart:2022}.

14.29\% of the works carry out both \textit{quantitative} and \textit{qualitative} evaluation, of which 44.44\% \cite{IJ:ESWA:MEI:2022, IJ:ESWA:Wang:2023, IJ:SC:XI:2024, IJ:Scientometrics:TAN:2023} complement it with some type of \textit{ablation} study and 44.44\% \cite{IJ:AI:GAO:2023, IJ:IS:Chaudhuri:2022, IJ:ESWA:XI:2023, IJ:IPM:Long:2024} with an \textit{ablation} study and \textit{configuration} test.

9.52\% of the works conduct only a \textit{qualitative} evaluation \cite{IC:WIST:Roßner:2023, IJ:IS:PINEDO:2024, J:JIS:MUNGEN:2022, IJ:ELECTRONICS:Blazevic:2023, IJ:Scientometrics:Kanwal:2024, IJ:IEEE:SARWAR:2021}, only one includes an additional analysis \cite{IJ:IS:PINEDO:2024} to evaluate \textit{execution time} and only 2 conduct \textit{online} evaluations \cite{IJ:IS:Chaudhuri:2022, J:JIS:MUNGEN:2022}.

\tablename~\ref{tab:EvaluationProcedure} summarizes the evaluation procedures followed.

\begin{footnotesize}
\fontsize{6}{8}\selectfont
\begin{longtable}{ccccccccp{2cm}c}
\caption{Classification according to evaluation procedure}\label{tab:EvaluationProcedure}\\
\toprule
\textbf{Off.} & \textbf{On.} & \textbf{Quan.} & \textbf{Qual.} & \textbf{Abl.} & \textbf{Conf.}	& \textbf{Usr. Std.} &  \textbf{Exec.} & \textbf{Works}\\
\midrule
\endfirsthead
\multicolumn{9}{c}
{{\bfseries \tablename\ \thetable{} -- continued from previous page}} \\
\toprule
\textbf{Off.} & \textbf{On.} & \textbf{Quan.} & \textbf{Qual.} & \textbf{Abl.} & \textbf{Conf.}	& \textbf{Usr. Std.} &  \textbf{Exec.} & \textbf{Works}\\
\midrule
\endhead
\midrule
\multicolumn{9}{c}{{Continued on next page}} \\ 
\bottomrule
\endfoot
\endlastfoot   
    \ding{51} &  & \ding{51} & \ding{51} & \ding{51} & \ding{51} &  &  & \cite{IJ:AI:GAO:2023, IJ:ESWA:XI:2023, IJ:IPM:Long:2024}\\
    \ding{51} &  & \ding{51} & \ding{51} & \ding{51} &  &  &  & \cite{IJ:ESWA:MEI:2022, IJ:ESWA:Wang:2023,IJ:Scientometrics:TAN:2023}\\
    \ding{51} &  & \ding{51} & \ding{51} &  &  &  &  & \cite{IC:ICITACEE:Switrayana:2022}\\
    \ding{51} &  & \ding{51} &  & \ding{51} & \ding{51} &  &  & \cite{ARXIV:Shen:2024, IC:KSEM:YU:2022, IC:DASFAA:Wang:2022, IJ:Scientometrics:Huang:2024, IJ:ESWA:XIAO:2023, IJ:SMC:WANG:2022, IJ:Scientometrics:Jiang:2023, IJ:Scientometrics:ALI:2022, IJ:IEEE:HADHIATMA:2023, IJ:ASC:Wang:2024, IJ:mathematics:GUO:2024, IJ:AI:Thierry:2023,IS:CSET:WU:2022,IC:ICDE:XIE:2022}\\
    \ding{51} &  & \ding{51} &  & \ding{51} & \ding{51} &  & \ding{51} & \cite{J:EIJ:Huang:2024}\\
    \ding{51} &  & \ding{51} &  & \ding{51} &  &  &  & \cite{IJ:AS:Chen:2023, IC:IKT:JAFARI:2021, IJ:AUT:Jafari:2023, IJ:JOI:Yadav:2022, IJ:Scientometrics:Thierry:2023, IC:ASSP:Shen:2024, IJ:KBS:Xiao:2023, CF:CSCW:LI:2022, IJ:AS:Wang:2024, IJ:DSM:Sivasankari:2023, IC:MLBDBI:GUAN:2022,IJ:JS:Huang:2023,IC:EITCE:NIU:2023}\\
    \ding{51} &  & \ding{51} &  &  & \ding{51} &  &  & \cite{IJ:Heliyon:LI:2024, IJ:IEEE:REN:2022, IJ:TIIS:GUO:2024, AC:ACIIDS:PAN:2024}\\
    \ding{51} &  & \ding{51} &  &  &  &  & \ding{51} & \cite{IJ:DATABASE:Kart:2022}\\
    \ding{51} &  & \ding{51} &  &  &  &  &  & \cite{IJ:AI:Xiao:2023, IJ:Scientometrics:GUNDOGAN:2022, IJ:CDS:AHMEDI:2022, IJ:EIT:Smail:2023, IJ:IEEE:STERLING:2022,PP:Mahesh:2023, arXiv:KARIMI:2022, ARXIV:Mohamed:2022,IJ:SECS:Hamisu:2024,IJ:JES:LI:2024,IJ:SHTI:GUO:2022,IJ:AMEC:KUS:2023, EC:ECIR:Takahashi:2022}\\
    \ding{51} &  &  & \ding{51} &  &  & \ding{51} & & \cite{IJ:Scientometrics:Kanwal:2024}\\
    \ding{51} &  & \ding{51} & \ding{51} & \ding{51} &  & \ding{51} &  & \cite{IJ:SC:XI:2024}\\
    \ding{51} & \ding{51} & \ding{51} & \ding{51} & \ding{51} & \ding{51} & \ding{51} &  & \cite{IJ:IS:Chaudhuri:2022}\\
    \ding{51} &  &  & \ding{51} &  &  & \ding{51} &  & \cite{IC:WIST:Roßner:2023}\\
    \ding{51} &  &  & \ding{51} &  &  &  &  & \cite{IJ:ELECTRONICS:Blazevic:2023}\\
    \ding{51} &  & \ding{51} &  &  & \ding{51} &  & \ding{51} & \cite{IJ:KIS:STERGIOPOULOS:2024}\\
    \ding{51} &  &  & \ding{51} &  &  & \ding{51} &  & \cite{IJ:IEEE:SARWAR:2021}\\
    & \ding{51} &  & \ding{51} &  &  &  &  & \cite{J:JIS:MUNGEN:2022}\\
    \ding{51} &  &  & \ding{51} &  &  & \ding{51} & \ding{51} & \cite{IJ:IS:PINEDO:2024}\\
    \ding{51} &  & \ding{51} & & \ding{51} & & &  \ding{51} & \cite{IC:SEAI:Wang:2024}\\
\bottomrule 
98.41\% & 3.17\% & 92.06\% & 23.81\% & 58.73\% & 38.10\% & 9.52\% &  7.94\%\\
\end{longtable}
\end{footnotesize}

Two different analyses were conducted to study the use of \textit{baselines}. First, the \textit{number} of baselines used in the works was analyzed; second, the \textit{types} of baselines employed were examined. Three different types were identified: \textit{\RSS} or algorithms not focused on scientific articles, \textit{\RPRSS} or algorithms oriented toward scientific articles, and \textit{representation systems} or algorithms.

Only 12.70\% of the works do not compare their systems or algorithms with others \cite{IJ:Heliyon:LI:2024, IJ:SHTI:GUO:2022, IJ:AI:Xiao:2023,
IJ:DATABASE:Kart:2022, IC:WIST:Roßner:2023, IJ:IEEE:SARWAR:2021, IJ:ELECTRONICS:Blazevic:2023, IJ:IS:PINEDO:2024}, 50\% of which rely solely on \textit{qualitative} evaluations.

Another interesting aspect is that among the works using three or more \textit{baselines}, 92.68\% perform some type of complementary analysis, either \textit{ablation} or \textit{configuration}. In contrast, in works using fewer than three \textit{baselines}, this percentage drops to 21.43\%.

Regarding the \textit{types} of \textit{baselines} employed, 28.57\% of the works utilized a combination of \textit{\RPRS} and \textit{\RSS\ baselines} \cite{IC:MLBDBI:GUAN:2022, IS:CSET:WU:2022, IJ:AS:Chen:2023, IJ:KBS:Xiao:2023, IJ:Scientometrics:Jiang:2023, IJ:ASC:Wang:2024,
IJ:mathematics:GUO:2024, J:EIJ:Huang:2024, IC:IKT:JAFARI:2021, IC:ICDE:XIE:2022, IJ:IS:Chaudhuri:2022, IJ:AUT:Jafari:2023, IJ:ESWA:XI:2023, IJ:ESWA:XIAO:2023, IJ:JS:Huang:2023, PP:Mahesh:2023, IJ:SC:XI:2024, IJ:TIIS:GUO:2024}, followed by 25.40\% that employed only \textit{\RPRS\ baselines} \cite{IJ:ESWA:MEI:2022, IJ:JOI:Yadav:2022, IJ:AI:GAO:2023, IJ:Scientometrics:Huang:2024, arXiv:KARIMI:2022, IC:ICITACEE:Switrayana:2022, IJ:CDS:AHMEDI:2022, IJ:IEEE:REN:2022, IJ:Scientometrics:ALI:2022, IJ:Scientometrics:GUNDOGAN:2022, IJ:AI:Thierry:2023, IJ:DSM:Sivasankari:2023, IJ:EIT:Smail:2023, IJ:Scientometrics:Thierry:2023, IJ:Scientometrics:Kanwal:2024, IJ:SECS:Hamisu:2024}, and 17.46\% that used \textit{\RSS\ baselines} \cite{ARXIV:Mohamed:2022, CF:CSCW:LI:2022, IC:DASFAA:Wang:2022, IC:KSEM:YU:2022, IJ:SMC:WANG:2022, IC:EITCE:NIU:2023, IJ:ESWA:Wang:2023, IC:ASSP:Shen:2024, IJ:AS:Wang:2024, IJ:JES:LI:2024, IJ:KIS:STERGIOPOULOS:2024}. Notably, 15.87\% of the works did not compare recommendation algorithms but rather focused on paper and user \textit{representation} algorithms \cite{AC:ACIIDS:PAN:2024, ARXIV:Shen:2024, EC:ECIR:Takahashi:2022, IC:SEAI:Wang:2024, IJ:AMEC:KUS:2023, IJ:IEEE:HADHIATMA:2023, IJ:IEEE:STERLING:2022, IJ:IPM:Long:2024, J:JIS:MUNGEN:2022, IJ:Scientometrics:TAN:2023}. Although the analyzed time period is relatively brief, the data suggest a potential shift in trend towards the latter.

Regarding \textit{datasets}, the majority of works (79.37\%) use between one and two \textit{datasets}. More specifically, 47.62\% use only one \textit{dataset} for the evaluation \cite{IJ:Heliyon:LI:2024, IJ:SHTI:GUO:2022, IJ:DATABASE:Kart:2022, IC:WIST:Roßner:2023, IJ:IS:PINEDO:2024,AC:ACIIDS:PAN:2024, IJ:KIS:STERGIOPOULOS:2024, IJ:Scientometrics:GUNDOGAN:2022, IJ:IEEE:STERLING:2022, EC:ECIR:Takahashi:2022, IJ:CDS:AHMEDI:2022, IJ:IEEE:REN:2022,
IJ:EIT:Smail:2023, IJ:SECS:Hamisu:2024, IJ:JES:LI:2024, IJ:Scientometrics:Kanwal:2024, ARXIV:Shen:2024, IC:KSEM:YU:2022, IC:IKT:JAFARI:2021,
CF:CSCW:LI:2022, IJ:DSM:Sivasankari:2023, PP:Mahesh:2023, IJ:AMEC:KUS:2023, ARXIV:Mohamed:2022, IJ:AUT:Jafari:2023, IC:EITCE:NIU:2023,
IJ:Scientometrics:Jiang:2023, IJ:SMC:WANG:2022, IJ:AS:Wang:2024, IJ:mathematics:GUO:2024}, while 31.75\% use two \textit{datasets} \cite{IJ:AI:Xiao:2023,
IC:ICITACEE:Switrayana:2022, arXiv:KARIMI:2022, IJ:TIIS:GUO:2024, IJ:AI:GAO:2023, IC:DASFAA:Wang:2022, IJ:Scientometrics:Huang:2024, IC:SEAI:Wang:2024, IS:CSET:WU:2022, IJ:AI:Thierry:2023, IJ:Scientometrics:ALI:2022, IJ:Scientometrics:Thierry:2023, IJ:JS:Huang:2023,
IC:ASSP:Shen:2024, J:EIJ:Huang:2024, IJ:ESWA:MEI:2022, IJ:IPM:Long:2024, IJ:ASC:Wang:2024, IJ:IS:Chaudhuri:2022}. An additional 11.11\% use three \textit{datasets} \cite{IJ:AS:Chen:2023, IC:MLBDBI:GUAN:2022, IJ:ESWA:XIAO:2023, IJ:JOI:Yadav:2022, IC:ICDE:XIE:2022, IJ:ESWA:Wang:2023,
IJ:Scientometrics:TAN:2023}. It is worth noting that works utilizing more than four or eight \textit{datasets} to evaluate their results typically rely on a  base \textit{dataset}, building different versions of these to assess specific aspects. In \cite{IJ:IEEE:HADHIATMA:2023}, up to four versions of each dataset are generated by randomly selecting 25\%, 50\%, 75\%, and 100\% of the data, while in \cite{IJ:SC:XI:2024, IJ:KBS:Xiao:2023}, two versions of each base dataset are created to obtain dense and sparse versions.

\begin{figure}[H]
    \centering 
    \subfigure[Number of baselines used]{\includegraphics[width=0.4\textwidth]{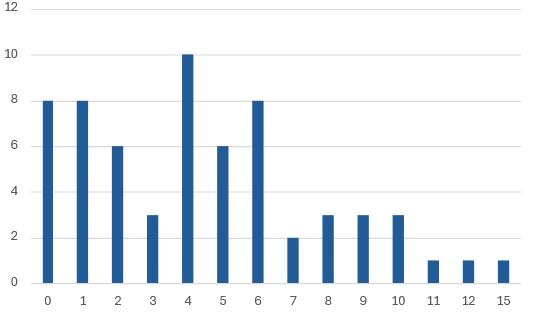}}
    \subfigure[Types of baselines used per year]{\includegraphics[width=0.4\textwidth]{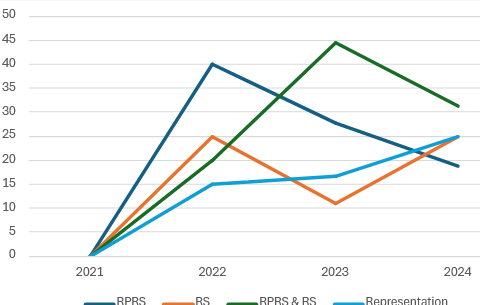}} 
    \subfigure[Number of datasets used]{\includegraphics[width=0.4\textwidth]{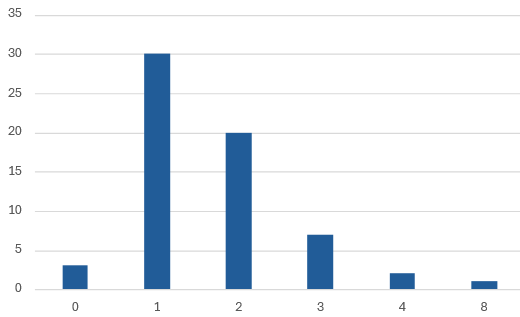}}    
    \caption{Baselines and datasets}
    \label{fig:EvaluationGraphics}
\end{figure}

\figurename~\ref{fig:EvaluationGraphics} presents the graphs which illustrate the use of commented datasets and baselines.

\subsection{What are the main challenges addressed by the \RPRSS?}\label{sec:MainObjectives}

Previous surveys in the field identified several open issues in their reviews \cite{IJ:DL:BEEL:2015, IJ:IEEE:BAI:2019, IJ:DL:KREUTZ:2022}. In the rapidly evolving field of \RPRSS\, significant advances have been made in recent years. However, despite these strides, several challenges persist that hinder the full realization of potential benefits.

\paragraph{User modeling.}
The challenge of disregarding user modeling in \RPRSS\ persists, as many systems rely on generic texts or specific papers provided by users, rather than incorporating personalized user profiles. This approach assumes that the information needs of the user are implicit and not tailored to individual preferences, which limits the overall effectiveness and relevance of recommendations. Personalization, which adapts the recommendations to the unique needs and behaviors of each user, plays a critical role in improving user satisfaction. Compared to \cite{IJ:DL:KREUTZ:2022}, the current work shows a 10\% increase in the adoption of user modeling, with 61.90\% of the reviewed works now addressing this issue and aiming to offer more personalized recommendations.

\paragraph{Evaluation metrics.}
Relying solely on \textit{precision} as the primary evaluation metric in \RPRSS\ overlooks key factors such as \textit{recall}, \textit{ranking}, \textit{novelty}, \textit{diversity}, and \textit{serendipity}. While \textit{precision} measures the accuracy of the top recommendations, \textit{recall} focuses on completeness, and ranking metrics like \textit{\NDCG}, \textit{\MAP}, and \textit{\MRR} assess how well the system orders relevant papers. However, other important aspects such as \textit{novelty} (the ability to suggest new and unexpected papers), \textit{diversity} (offering a broad range of topics), and \textit{serendipity} (providing surprisingly relevant recommendations) are often under-explored due to their complexity and rarity in use. These factors are crucial for ensuring that recommendations are not only relevant but also engaging and broad in scope, contributing to a richer user experience. Compared to \cite{IJ:DL:KREUTZ:2022}, there has been a 25.65\% increase in the use of \textit{recall} and a 18.12\% increase in \textit{\NDCG}, but the adoption of less common metrics has decreased by 21.21\%, with \textit{\AUC} and \textit{\HR} being the most widely used metrics at 9.52\% and 6.35\%, respectively.

\paragraph{Serendipity.}
In the field of \RPRSS\ serendipity refers to the process of suggesting unexpected yet valuable papers, thereby facilitating the discovery of novel ideas. However, challenges such as balancing novelty with relevance, assessing the impact of serendipitous recommendations, and personalizing the recommendation process have led to serendipity being underemphasized in recent literature, which has predominantly focused on improving accuracy and personalization. Despite these challenges, serendipity holds significant potential in fostering innovation and expanding the scope of research. Among the works reviewed, only one paper explicitly addresses the concept of serendipity \cite{IJ:IEEE:HADHIATMA:2023}. To address this challenge, in \cite{IJ:IEEE:HADHIATMA:2023} the authors propose a citation recommendation framework that integrates multi-topic communities into a modified version of PageRank.

\paragraph{Relevance and diversity.}
\cite{IJ:DL:KREUTZ:2022} discusses the balance between relevance and diversity in \RPRSS, highlighting that focusing only on relevance can lead to similar recommendations. Diversity in the generated recommendations can better satisfy the needs of the users. The use of \textit{\MRR} as a quality metric reflected the recognition of this issue, as \textit{\MRR} helps ensure that the suggestions are both relevant and diverse. A small number of recent works have specifically targeted the diversity issue in \RPRSS. \cite{AC:ACIIDS:PAN:2024} identifies a new feature capturing the diversity of research interests, moving beyond the recency of publications to model varied researcher interests. \cite{IJ:KIS:STERGIOPOULOS:2024} introduces a graph of \FoS, which uncovers latent knowledge paths to enhance recommendation diversity by discovering relationships in scientific areas. \cite{IJ:JOI:Yadav:2022} proposes measuring diversity by calculating the sum of shortest paths between pairs of items in the recommendation list, helping quantify diversity. \cite{IJ:mathematics:GUO:2024} focuses on enhancing diversity by utilizing indirect social ties within academic teams, narrowing down candidate papers based on shared preferences. \cite{IJ:IS:Chaudhuri:2022} promotes diversity through a \KDM, which evaluates paper relevance by the degree of keyword association, aiming to provide diverse topics and sub-topics. Lastly, \cite{IJ:AI:GAO:2023} employs a heterogeneous network and RippleNet to explore potential connections and expand user interests, enhancing the diversity of the recommendations. In the current survey, a small increment in the use of the \textit{\MRR} metric has been detected, making it the fourth most used metric, with 30.16\% of the works employing it.

\paragraph{Lack of detailed documentation.}
Many works in \RPRSS\ fail to provide sufficient documentation, such as details about experimental setups, algorithms, or data sources. This lack of transparency makes it difficult for other researchers to replicate or fully understand the methods and findings. A clear indication that this issue remains unresolved is that 21 works were excluded from further analysis in this survey due to missing essential information. It is worth noting that while \cite{IJ:DL:KREUTZ:2022} did not identify any works sharing their code, this survey has found 6 works that have made their developed code accessible \cite{IJ:IS:Chaudhuri:2022, IJ:DATABASE:Kart:2022, IJ:IEEE:SARWAR:2021, AC:ACIIDS:PAN:2024, IJ:KIS:STERGIOPOULOS:2024,IJ:Scientometrics:TAN:2023}.

\paragraph{Evaluation procedure.}
Evaluation remains a significant challenge in \RPRSS\ due to the subjective nature of relevance, the absence of a definitive ground truth, and the difficulties associated with applying traditional evaluation metrics. Conventional metrics such as \textit{precision} and \textit{recall} fail to account for critical factors such as \textit{novelty}, \textit{diversity}, and \textit{user satisfaction}. Furthermore, the cold start problem—where new users or papers lack sufficient interaction data—adds an additional layer of complexity to the evaluation process. As highlighted by \cite{IJ:DL:KREUTZ:2022}, user evaluation is essential for assessing how well recommendations align with individual research needs, yet obtaining direct user feedback in academic contexts remains a challenge. Moreover, given the rapid evolution of research fields and the often indirect nature of feedback, evaluating how well these systems meet the dynamic demands of researchers becomes particularly difficult. Due to these challenges, there remains a scarcity of works that incorporate user-centered evaluation approaches. Recent works on \RPRSS\ involve user studies with varying configurations. \cite{IJ:ELECTRONICS:Blazevic:2023} uses a three-point rating scale for paper similarity evaluation but provided no details about participants. \cite{IC:WIST:Roßner:2023} conducted a qualitative user study with four participants (two junior and two senior researchers) who evaluated publication similarity and complementarity across 20 starting points. \cite{IJ:Scientometrics:Kanwal:2024} involved ten researchers in assessing the ability of the system to recommend older papers and those with no citations, but no further participant details were provided. \cite{IJ:IS:PINEDO:2024} engaged 30 domain experts to evaluate recommendations across 10 research areas, with experts providing feedback on the relevance of the suggested articles. \cite{IJ:IS:Chaudhuri:2022} conducted an online study with 112 participants, using a 1-5 like scale to assess recommendation quality, but no expert information was provided . \cite{IJ:IEEE:SARWAR:2021} relied on experts from the Computer Science domain to manually rank articles for a chronological learning-based system, though no further expert details were given. \cite{J:JIS:MUNGEN:2022} used a survey for participants to evaluate publication relevance, but no participant or expert details were provided.

\paragraph{Research to practise.}
In previous surveys, it was highlighted that few approaches were available as prototypes, and there was often a disconnect between methods described in papers and real-world systems, a situation that remains unchanged to date. Among the works analyzed in this new survey, \cite{IJ:DATABASE:Kart:2022} emphasizes the scalability problem when applying state-of-the-art models. \cite{IJ:DATABASE:Kart:2022} states that while BERT models show outstanding performance in experiments using test data from model organism databases, the need for scalability led to the deployment of a simpler approach based on TF-IDF vectorization and a Naïve Bayes classifier. \cite{IJ:KIS:STERGIOPOULOS:2024} points out that state-of-the-art machine learning \RSS\ in the literature are typically limited to datasets with only a few thousand users and items, creating a significant gap between the datasets used in published systems and those available in real-world applications. \cite{IJ:JOI:Yadav:2022} demonstrates the applicability of community detection to various scenarios in order to achieve more scalable results, which is crucial in the current age of digital libraries . Currently, only 9.52\% of the works analyzed address this challenge by developing actual prototypes \cite{IJ:IS:PINEDO:2024, IJ:IS:Chaudhuri:2022, IC:WIST:Roßner:2023, IJ:IEEE:STERLING:2022, IJ:DATABASE:Kart:2022} or applying algorithms to real systems \cite{J:JIS:MUNGEN:2022}, while some others, plan to develop real web applications \cite{ARXIV:Shen:2024, IJ:Heliyon:LI:2024}

\paragraph{Cold start and sparsity.}
These issues are particularly evident in collaborative filtering systems, which rely on user interaction history to provide personalized suggestions. Recent advances have led to approaches that go beyond pure collaborative filtering, incorporating additional techniques to address these challenges. Current research follows the trends outlined in previous surveys. In recent research, 95.23\% of works utilize content-based information to mitigate the cold start and sparsity issues. A significant portion of these works specifically targets overcoming the cold start problem \cite{IJ:AS:Chen:2023,IJ:ESWA:MEI:2022,IJ:IEEE:REN:2022,IJ:SC:XI:2024,IJ:Scientometrics:Jiang:2023,IJ:SECS:Hamisu:2024,IJ:JS:Huang:2023,IJ:AI:GAO:2023,IJ:AUT:Jafari:2023,IJ:EIT:Smail:2023,IJ:IS:PINEDO:2024} or the sparsity problem \cite{IC:ICITACEE:Switrayana:2022,J:EIJ:Huang:2024}. Another set of works attempts to tackle both problems \cite{IJ:ESWA:XI:2023,IJ:TIIS:GUO:2024,IJ:DSM:Sivasankari:2023,IC:ICITACEE:Switrayana:2022,IJ:KIS:STERGIOPOULOS:2024}. In some cases, test performance is carried out in cold start and sparse scenarios \cite{IJ:ESWA:XIAO:2023,IJ:Scientometrics:ALI:2022,IJ:TIIS:GUO:2024,arXiv:KARIMI:2022,IJ:KIS:STERGIOPOULOS:2024}

\paragraph{Privacy.}
The main privacy challenges in \RPRSS\ include protecting sensitive user data, ensuring proper anonymization without losing personalization, obtaining informed user consent, and securing data against breaches. Additionally, there is a lack of transparency in how user data is used, complicating privacy management. Compliance with privacy regulations, such as the General Data Protection Regulation (GDPR), adds further complexity, and there are concerns about potential biases and discrimination arising from how data is handled. The current situation remains largely the same, as many methods still overlook privacy concerns in their designs. No privacy conflicts were found in the works analyzed for this survey, as the data used in the various works are either public (see \tablename~\ref{sec:ACLdataset}, \tablename~\ref{sec:AMiner},  \tablename~\ref{sec:DBPL}), labeled \cite{JCDL:Sugiyama:2013}, anonymized (\tablename~\ref{sec:CiteULike}), or artificial \cite{IJ:DATABASE:Kart:2022, IJ:KIS:STERGIOPOULOS:2024} .

\paragraph{Target audience.}
As stated in \cite{IJ:DL:KREUTZ:2022}, numerous \RPRSS\ failed to clearly define their target audience, thus overlooking the distinct needs of various user groups in their evaluations. While some works identified specific audiences, such as \textit{junior} or \textit{senior researchers}, most did not adequately consider how these differences influenced the recommendation process. Recent works, however, provide some examples that address this issue. \cite{IJ:mathematics:GUO:2024} introduces a student academic paper social recommendation model to address the interaction challenges between students and academic papers. \cite{IJ:Scientometrics:Jiang:2023} focuses on cold start \textit{junior researchers}, recognizing the difficulty in learning research interests for users with limited publications. \cite{IJ:AI:Xiao:2023} \textit{targets postgraduates}, \cite{IJ:DATABASE:Kart:2022} \textit{researches} and \textit{curators}, and \cite{IC:IKT:JAFARI:2021} emphasize the need to differentiate between \textit{novice} and \textit{experienced researchers} in \RPRSS. \cite{IJ:ASC:Wang:2024} study researchers within scientific social networks, and \cite{IJ:IS:Chaudhuri:2022} propose a ranking approach that distinguishes between \textit{new} and \textit{experienced users}. \cite{IC:ASSP:Shen:2024} focuses on recommending papers for researchers to cite. While these recent works reflect a growing awareness of the importance of target audience considerations, only about 12.69\% of the analyzed works discuss or mention the target audience in their evaluations, highlighting that very few works address this aspect.

\paragraph{Explainability.}
\cite{IJ:DL:KREUTZ:2022} underlines the importance of explainability in \RPRSS\, noting that while accuracy often takes priority, confidence and trust in the system are crucial for user satisfaction. Explainability enhances trust and engagement by making recommendations more transparent and understandable. However, in the works surveyed in this study, explainability remains largely unaddressed, and the increasing complexity of the proposed systems (see Section~\ref{sec:PaperRepresentation}, Section~\ref{sec:User representation}, Section~\ref{sec:Recommendation procedure}) further complicates the provision of clear and understandable justifications for recommendations.

\paragraph{Datasets.}
\cite{IJ:DL:KREUTZ:2022} highlights that \RPRSS\ often rely on a variety of datasets, many of which are proprietary and not publicly accessible. Although some approaches incorporate open datasets for evaluation, a considerable portion still fails to prioritize this aspect. They emphasize the importance of utilizing publicly available datasets or developing datasets that are shared and accessible to enhance the reproducibility of research in this domain. Even when public datasets are employed, modifications resulting from filtration or cleansing procedures—frequently lacking adequate documentation—compromise the ability to reproduce findings (see \tablename~\ref{sec:Datasets}). Specifically, 9.90\% of datasets are either modified or created by the authors and made publicly available, while 13.86\% are accessible upon request. Furthermore, 27.72\% of datasets, originally publicly available, are employed without any further modifications, and the remaining 47.52\% consist of datasets that are either originally publicly available or ad-hoc constructed datasets, but are not accessible. These final datasets represent nearly 50\% of all datasets employed, highlighting a significant barrier to reproducibility in the field.

\paragraph{Comparability.}
\cite{IJ:DL:KREUTZ:2022} discusses comparability issues in \RPRSS\, noting that many approaches identify as \RPRSS\ but are often evaluated against general \RSS\ or methodologies from different sub-domains. Emphasizing that even when approaches claim broader applicability, they should still be compared with systems specifically designed for paper recommendation to assess their relevance. Furthermore, cases where baselines are not established systems but artificial ones or entirely absent are highlighted. \cite{IJ:DL:KREUTZ:2022} stresses that evaluations of \RPRSS\ should always include at least one dedicated \RPRS\ to ensure a clear assessment of relevance within the field. Regarding the current situation, 28.57\% of works utilize a combination of \RPRSS\ and general \RSS\ baselines, followed by 25.40\% that employ only \RPRSS\ baselines, and 17.46\% that use \RSS\ baselines. Notably, 15.87\% of works do not compare recommendation algorithms but focus instead on paper and user representation algorithms. These trends still reflect ongoing challenges in ensuring proper comparability and comprehensive evaluation in the field.

\subsection{What new challenges have arisen?}\label{sec:challenges}

New challenges have emerged during the course of this survey, including dataset-related issues namely multidisciplinary scenarios, data augmentation and \textit{what-if} scenarios, as well as time-awareness considerations, group or scientific social network dynamics, and multi-task settings. By synthesizing these findings, this section aims to offer a comprehensive overview of the current landscape, highlighting key areas that warrant further exploration and development.

\paragraph{Multidisciplinarity.}
Although this issue was initially highlighted in the work of \cite{IJ:DL:BEEL:2015}, it has not remained a primary focus in subsequent research. However, with the increasing interdisciplinarity of modern research, the need to accommodate the diversity of disciplines in \RPRSS\ has become an emerging challenge. Research on \RPRSS\ often relies on datasets that are limited to specific fields of study, which can hinder the generalization and applicability of these systems across various academic and research domains. Datasets containing papers from multiple disciplines are crucial for developing \RPRSS\ that can provide valuable suggestions to a broader audience. The research carried out indicates that only 28.91\% of datasets are multidisciplinary, while 59.03\% are not, and 12.04\% do not provide information on the disciplines represented in the dataset. Furthermore, many works identify this issue as an aspect to be addressed in future research \cite{IJ:AI:Xiao:2023,IJ:Heliyon:LI:2024,IJ:Scientometrics:Jiang:2023, IJ:IEEE:SARWAR:2021,IJ:TIIS:GUO:2024, IC:SEAI:Wang:2024}. This challenge should be reintroduced into the discussion, as it remains critical for the development of more generalizable and robust systems.

\paragraph{Dynamic preferences.}
Dynamic preferences refer to the evolving interests and needs of users over time. In the context of \RPRSS\, academic interests of the researchers are subject to change, often influenced by factors such as new developments within their primary field, emerging topics, or the exploration of adjacent areas of study. As these preferences shift, \RPRSS\ must be capable of adapting to these changes in order to provide timely and relevant suggestions. Traditional \RPRSS\, however, tend to assume a degree of stability in user preferences, which can limit their ability to reflect the dynamic and interdisciplinary nature of the academic journey of the researcher. To address this, \RPRSS\ need mechanisms that can track and predict the evolution of user preferences, ensuring recommendations align with the changing academic trajectories of users. In the current research, some works focus on incorporating dynamic trends in user interests by leveraging temporal data and heterogeneous information networks, while others explore the integration of attention mechanisms to capture evolving patterns in user behavior. Despite these efforts, only 15.87\% of the works analyzed specifically tackle the issue of dynamic preferences \cite{IC:DASFAA:Wang:2022,IJ:ESWA:Wang:2023, IJ:IEEE:REN:2022, IJ:Scientometrics:Jiang:2023, J:EIJ:Huang:2024, AC:ACIIDS:PAN:2024, IJ:AS:Wang:2024, IC:ASSP:Shen:2024, IJ:KBS:Xiao:2023, IJ:IS:Chaudhuri:2022}.

\paragraph{Emergence of social media.}
The increasing availability of data from social media platforms offers significant opportunities to enhance \RPRSS . Social media data, such as the online activity of the researchers, collaborations, and discussions, can provide valuable insights into dynamic academic trends and shifting interests, complementing traditional sources such as citations and publication history. However, integrating such data presents challenges, including ensuring data quality, managing diverse and noisy information, and addressing privacy concerns. Recently, some works have begun exploring the use of social media and structural data in \RPRSS, focusing on group dynamics and collaboration networks to enhance recommendation accuracy and mitigate issues such as sparsity and cold-start problems. These works utilize techniques such as heterogeneous graph neural networks with hierarchical attention mechanisms \cite{IJ:ASC:Wang:2024}, \PMF\ with evidential reasoning \cite{IJ:SMC:WANG:2022}, and hypergraph convolutional networks to model complex academic social structures \cite{IJ:mathematics:GUO:2024}. The emergence of social media opens up new opportunities for research groups that are worth exploiting.

\paragraph{Multi-task recommendations.}
Multitasking in \RPRSS\ refers to the simultaneous execution of multiple related tasks within a single model to enhance the accuracy and relevance of paper recommendations. This approach goes beyond simply recommending papers based on user preferences by integrating additional tasks. In the analyzed works, only 4 papers employed such characteristics: \cite{arXiv:KARIMI:2022} proposes a multi-task \RPRS\ that simultaneously predicts paper recommendations and generates meta-data, such as keywords. Similarly, \cite{IC:EITCE:NIU:2023} combines models for academic paper link prediction and paper recommendation. Furthermore, \cite{ARXIV:Mohamed:2022} suggests the potential of extending their approach by integrating multi-task learning, where a single model is trained for both tag prediction and paper recommendation tasks. More recently, \cite{IC:SEAI:Wang:2024} introduces a graph neural network model to handle different tasks such as paper and co-author recommendation. While multitasking has been a concept scarcely mentioned in prior \RPRSS, it now appears to be gaining traction. This shift opens up new opportunities in the field, although it also introduces challenges, such as balancing task priorities, ensuring efficient resource use, and preventing interference between tasks.

Handling what-if scenarios in \RPRSS\ is a key consideration for the development of systems capable of predicting or providing recommendations under hypothetical or alternative academic contexts. These scenarios may involve changes in user interests, shifts in collaboration networks, the emergence of new academic fields, or the need for larger datasets to address scalability challenges. For \RPRSS\ to effectively respond to such dynamic conditions, the datasets employed must reflect realistic academic environments that are able to simulate these variations. However, many existing datasets lack the necessary complexity to capture these dynamics, which limits the adaptability of \RPRSS\ to evolving trends and changing user needs. When algorithms or models are designed to tackle specific challenges within \RPRSS, it is therefore essential that the datasets used accurately represent the corresponding scenarios. Despite its relevance for modeling controlled or hypothetical academic settings, synthetic data has not been explicitly considered in the reviewed works. Exploring its full potential may enable the development of more adaptable and scalable \RPRSS.

\section{Discussion}\label{sec:Discussion}

As global scientific publications continue to grow, researchers face challenges like information overload and difficulty finding relevant papers. \RPRSS\ help alleviate these issues by providing tailored research suggestions. This survey offers an in-depth analysis of the current state of \RPRSS\ and explores advances in the field from November 2021 to December 2024.

To address these challenges, this survey investigates key aspects related to the techniques used, datasets employed, evaluation methods, and the challenges faced by \RPRSS. By prioritizing the techniques and specifying where these are applied within the recommendation process, this work fills the gap left by prior surveys, which did not explicitly explore these stages. Specifically, \textit{RQ1} examines how \RPRSS\ generate recommendations, breaking this down into specific questions regarding inputs, information used, representations of papers and users, and the recommendation generation process. \textit{RQ2} explores the datasets used for training and evaluation, with an emphasis on their characteristics and availability. \textit{RQ3} delves into the evaluation methods, focusing on relevance criteria, metrics, and procedures. \textit{RQ4} addresses the key challenges \RPRSS\ aim to solve, such as scalability and the cold start problem, while \textit{RQ5} identifies emerging challenges related to the adoption of advanced technologies.

\begin{itemize}
   
    \item On \textit{How do the \RPRSS\ generate the recommendations? (RQ1)}. Since the previous surveys did not always specify the stages at which each technique was applied, comparing their findings with ours is somewhat challenging. In the following, we address the last three key aspects by highlighting the most relevant findings from our survey.

    \vspace{0.5\baselineskip}
    \subitem $\circ$ On \textit{What type of input do the \RPRSS\ use to generate the recommendations? (RQ1.1)}, we can conclude that, compared to the previous survey, the current analysis shows an increase in the adoption of personalized input. Specifically, 53.97\% of the systems now use user input, up from 36.92\%. The use of paper input has slightly decreased from 30.77\% to 30.15\%, while text input has dropped from 12.31\% to just 4.76\%. Additionally, the other category, which includes various combinations of inputs, has increased from 10.77\% to 15.87\%.

    \vspace{0.5\baselineskip}
    \subitem $\circ$ On \textit{What kinds of information do the \RPRSS\ use to generate the recommendations?(RQ1.2)}. Regarding content data, there has been a notable increase in the use of several content-based features. The inclusion of paper titles in recommendation models rose substantially from 40\% in the previous survey to approximately 76.19\% in the current survey. Similarly, abstracts saw increased usage, climbing from 32.31\% to 71.43\%. Conversely, the use of keywords decreased from 33.85\% to 22.22\%, and the utilization of text dropped sharply from 43.08\% to 7.94\%. These changes likely stem from differences in how data usage was reported: earlier works often lacked clarity on which paper content was used, leading to higher reports of text or keyword usage, while current works specify content sources more clearly, focusing on titles and abstracts. Additionally, the current survey reports the use of new content-related features that were not documented in the previous survey: \textit{user} information is utilized in about 30.16\% of works, \textit{author} information in 47.62\%, and \textit{venue} information in 22.22\%.
    
    In terms of relational data, the previous survey showed that such data—en\-com\-pass\-ing citations, interactions, and co-authorship—was used in 81.5\% of the analyzed works. The current survey finds a slight decrease in relational data usage, now present in 74.60\% of works. This decline is primarily attributed to a significant drop in the use of interaction data, which fell from 44.6\% to 25.39\%. Citations, however, have remained essential, with their usage rising from 53.84\% to 58.73\%, underscoring their ongoing importance in assessing the relevance of research papers. Additionally, co-authorship data, which was not reported in the prior survey, has now been utilized in 17.46\% of the analyzed works, signaling a growing interest in collaborative research and author networks. This shift could reflect an increased awareness of privacy concerns, as the collection and use of interaction data may pose greater risks to user privacy compared to more established forms of relational data, such as citations.

    \vspace{0.5\baselineskip}
    \subitem $\circ$ On \textit{How do the \RPRSS\ represent the papers? (RQ1.3)}. The majority of works (79.37\%) use \VSM\ techniques for paper representation, while 11.11\% employ graph-based methods, with some works combining both approaches. Graph-based techniques often represent papers as nodes in \textit{\HINS}, involving entities such as \textit{authors}, \textit{venues}, and \textit{topics}. \textit{\VSM} methods predominantly rely on deep learning, with 68\% using embeddings, including models like \textit{Word2Vec}, \textit{Doc2Vec}, and \textit{BERT}. Additionally, some works integrate \textit{text} with \textit{visual} or \textit{graph-based} techniques. \textit{\LFVS} and \textit{\MF} techniques, employed in 17.56\% of works, are often combined with other methods to address cold start issues. \textit{\FV} and \textit{\TD} representations are less common, relying on \textit{TF-IDF} and \textit{\LDA}, respectively.

    \vspace{0.5\baselineskip}
    \subitem $\circ$ On \textit{How do the \RPRSS\ represent the users? (RQ1.4)}. Approximately 63.49\% of the analyzed works focus on user modeling for personalized recommendations, with 10\% using \textit{graph-based} strategies in \textit{\HINS}, where users are represented as nodes alongside entities such as papers and authors. The majority (87.5\%) of the works \textit{\VSM} techniques, with embeddings being the most popular representation method (50\%), followed by \textit{\LFV} at 30\%, and \textit{\FV} at 7.5\%. \textit{Embedding-based} approaches, often combined with \textit{graph-based} techniques such as \textit{\GCNS} and \textit{\GNNS}, dominate, reflecting the growing use of \textit{\DL} for more semantically aware user representations. Despite this shift, \textit{\MF} and simpler techniques like \textit{TF-IDF} are still in use, indicating a continued reliance on hybrid and traditional approaches.

    \vspace{0.5\baselineskip}
    \subitem $\circ$ On \textit{How do the \RPRSS\ generate the recommendations? (RQ1.5)}. The methods for generating recommendations are divided into 4 major categories: \textit{Inner product} methods (36.51\%), \textit{similarity-based} methods (28.57\%), \textit{\ML-based} methods (19.05\%), and \textit{graph-based} methods (7.94\%), with some works using hybrid or alternative approaches. \textit{Inner product} methods calculate the similarity between user and paper vector representations through the dot product, with 52.17\% using \textit{embeddings} and \textit{graph-based} techniques as part of more complex models. Training strategies include \textit{\BPR\ loss}, \textit{cross-entropy loss}, and \textit{tuple-wise loss}. \textit{\ML-based} methods, mainly using \textit{\MLP} architectures, account for 66.6\% of personalized recommendations, employing \textit{cross-entropy loss} and \textit{ad-hoc loss} 
    functions. \textit{Similarity-based} methods account for 28.57\% of the works, with 77.78\% of them generating personalized recommendations, often using \textit{cosine similarity}. Other similarity measures, such as \textit{citation similarity} or \textit{Jaccard similarity}, are also used in some cases. \textit{Graph-based} methods, including \textit{\RW} algorithms with or without restart, are employed by 7.94\% of works, with most providing personalized recommendations. Alternative methods, such as \textit{co-citation}, \textit{bibliographic coupling}, and \textit{Naïve Bayes classifiers}, are also found, with only one work offering personalized recommendations.    

    \vspace{0.5\baselineskip}
    \item On \textit{What datasets are used to train and evaluate the \RPRSS? (RQ2)}. The analysis of datasets in the surveyed works reveals key trends and challenges. A total of 72 datasets were identified, most of which originated from four main families: \textit{ACL}, \textit{DBLP}, \textit{AMiner}, and \textit{CiteULike}, often with customized versions for specific research needs. Some works relied on \textit{publicly available} datasets, typically used only once, while others created \textit{ad-hoc} datasets for their works. A notable challenge was the lack of transparency regarding dataset versions and sources, making replication difficult. Additionally, many works utilized modified versions of public datasets, complicating efforts to evaluate the generalizability of results. A recurring issue in the literature is the scalability of models. While advanced models perform well in research settings, they often struggle with larger, real-world datasets. This scalability problem highlights the need for datasets that better reflect real-world conditions, stressing that, for research to be applicable in practical environments, models must be able to scale effectively while using data that mirrors actual user behavior and content distribution.

    \vspace{0.5\baselineskip}
    \item On \textit{How are the \RPRSS\ evaluated? (RQ3)}. Evaluating \RPRSS\ is essential to assess their ability to provide accurate and relevant recommendations. The evaluation process involves determining \textit{relevance criteria}, selecting appropriate \textit{metrics}, and conducting rigorous \textit{procedure}. Additionally, the choice of datasets plays a key role in ensuring the robustness of the system. RQ3 addresses these aspects through three specific questions: the \textit{relevance criteria} for the recommendations, the \textit{metrics} employed, and the \textit{evaluation procedure} used.

    \vspace{0.5\baselineskip}
    \subitem $\circ$ On \textit{What are the relevance criteria used to evaluate the recommendations? (RQ3.1)} \textit{Relevance criteria} in \RPRSS\ have evolved over time. When comparing the data from the current survey with those of a previous survey \cite{IJ:DL:KREUTZ:2022}, several notable differences emerge. Paper \textit{references} are the most frequently used criterion in the current survey, accounting for 44.29\% of the methods, compared to 24.07\% in the previous survey. This increase likely reflects a growing reliance on citation-based relevance, driven by the increasing importance of citation metrics and the availability of citation-based datasets, which are more accessible and scalable compared to user interaction data. \textit{User interactions}, including clicks, reads, and downloads, are utilized by 28.57\% of the systems in the current survey, showing a slight decline from 31.48\% in the earlier survey. This decrease can be attributed to the challenges of acquiring real user interaction data, particularly in large-scale settings, and the increasing concern around user privacy. \textit{Human judgment} was employed in 8.57\% of the methods in the current survey, compared to 27.78\% in the previous survey. This decline reflects the resource-intensive nature of manual evaluation, leading to a preference for automated or semi-automated evaluation methods that can handle larger datasets more efficiently. \textit{Labeled papers}, now at 12.86\%, were previously included in the \textit{other} category. Finally, the \textit{other} category has seen a significant reduction, from 24.07\% in the previous survey to just 4.29\% in the current one. This shift indicates a movement toward more explicit, standardized, and structured evaluation approaches, as researchers are increasingly prioritizing well-defined, reproducible criteria. Overall, the increase in the use of paper \textit{references} points to a growing focus on citation-based relevance, while the decline in user \textit{interactions} and \textit{human judgment} reflects the growing emphasis on scalability, data availability, and the challenges of manual evaluation.

    \vspace{0.5\baselineskip}
    \subitem $\circ$ On \textit{What metrics are used in the evaluation process? (RQ3.2)}. The metrics used to evaluate \RPRSS\ primarily include \textit{precision}, \textit{recall}, \textit{F-score}, \textit{\NDCG}, \textit{\MRR}, and \textit{\MAP}, each assessing different aspects of recommendation quality. \textit{Precision}, employed in 49.21\% of the works in the current survey, is slightly higher than the 45.61\% from the previous survey, confirming its fundamental role in evaluating the accuracy of recommendations. \textit{Recall}, which measures the completeness of relevant papers, shows a significant increase to 73.02\% in the current survey, compared to 47.37\% in the previous one, suggesting a growing emphasis on retrieving a broader set of relevant recommendations. \textit{F-score}, used in 22.22\% of the works in the current survey, remains stable, slightly decreasing from 22.81\% in the earlier survey, reflecting its ongoing role in balancing \textit{precision} and \textit{recall}, albeit less prominently. \textit{\NDCG}, which evaluates the ranking quality of recommendations, has increased substantially to 44.44\% from 26.32\%, indicating a stronger focus on the ordering of recommendations in the current evaluation landscape. \textit{\MRR}, used in 30.16\% of the works, has slightly increased from 26.32\%, showing continued importance in evaluating the effectiveness of ranking the most relevant recommendations at the top. \textit{\MAP} remains stable at 22.22\% in the current survey, showing consistency in its use for assessing ranking performance. Meanwhile, the category of \textit{other} metrics, which previously accounted for 45.61\% of the works, has dropped significantly to 25.40\%, suggesting that the use of more standard, widely accepted metrics has become more prominent in the current survey.

    \vspace{0.5\baselineskip}
    \subitem $\circ$ On \textit{How are the evaluations conducted? (RQ3.3)}. The evaluation procedures in the current survey encompass a broader range of methodologies, including both \textit{quantitative} and \textit{qualitative} approaches, as well as more specialized techniques such as \textit{ablation} studies and \textit{configuration} tests. \textit{Quantitative} evaluation procedures are featured in 90.06\% of the works, a dimension not specifically analyzed in the previous survey, where data on this procedure was not provided. Similarly, \textit{qualitative} evaluation procedures are used in 23.81\% of the current works, an aspect that was not covered in the prior survey. Ablation studies are present in 58.73\% of the works in the current survey, while the previous survey did not include analysis of this procedure. \textit{Configuration tests} are employed in 38.10\% of the current works, a dimension also not addressed in the prior survey. In contrast, \textit{user studies} are featured in 9.52\% of the works in the current survey, showing a decrease from the 21.05\% observed in the previous survey. \textit{Execution time} analysis, which appears in 7.94\% of the works in the current survey, reflects an increased emphasis on system efficiency, a dimension not covered in the earlier survey. Regarding \textit{baselines}, 12.70\% of the works in the current survey did not include \textit{baseline} comparisons, with half of these relying solely on \textit{qualitative} evaluations. This aspect was not specifically examined in the prior survey. In terms of \textit{datasets}, 79.37\% of the works in the current survey utilize one to two \textit{datasets}, though this dimension was not addressed in the previous survey. Therefore, while the current survey provides a more comprehensive overview of evaluation procedure dimensions, the absence of specific data in the previous survey precludes a more detailed comparative analysis of these evaluation procedure dimensions.

    \vspace{0.5\baselineskip}
    \item On \textit{What are the main challenges addressed by the \RPRSS? (RQ4)}. \RPRSS\ continue to face several key challenges that hinder their full potential. One prominent issue is the insufficient incorporation of \textit{user modeling}, as many systems still rely on text-based content or specific papers rather than personalized user profiles, although there has been a 10\% increase in the adoption of \textit{user modeling} in recent years. Additionally, the overreliance on \textit{precision} as the primary evaluation metric persists, despite growing recognition of the need for more comprehensive metrics such as \textit{recall}, \textit{\NDCG}, \textit{novelty}, and \textit{user satisfaction}, which help ensure relevance, diversity, and temporal accuracy \cite{IJ:DL:BEEL:2015, IJ:IEEE:BAI:2019, IJ:DL:KREUTZ:2022}. The gap between \textit{theoretical} advancements and \textit{practical} application remains significant, with only 9.52\% of works focusing on real-world prototypes or systems, indicating \textit{scalability} challenges. Similarly, the \textit{lack of detailed documentation} persists, with many works failing to provide adequate transparency for replicability. \textit{Cold start} and \textit{sparsity} issues are ongoing, with a dominant focus on content-based methods to address these challenges, reflecting the continued complexity of these problems in real-world settings. \textit{Privacy} concerns also remain underaddressed, as many works still overlook how to protect sensitive data and ensure proper anonymization. \textit{Serendipity}, though crucial for fostering innovation, has not received sufficient attention, with few works explicitly focusing on it. The \textit{evaluation} of \RPRSS\ remains limited to traditional accuracy metrics, necessitating a broader adoption of metrics that account for \textit{novelty}, \textit{diversity}, and \textit{user satisfaction}, as highlighted by recent works in the field. Furthermore, the lack of a clearly defined \textit{target audience} in many systems continues to hinder the design of user-centered recommendations, with only 12.69\% of works explicitly considering audience differentiation. Lastly, \textit{dataset accessibility} issues persist, with a significant portion of datasets being either proprietary, modified, or inaccessible, complicating reproducibility efforts. These ongoing challenges indicate that while progress has been made, there is still much to be addressed for \RPRSS\ to achieve their full potential.

    \vspace{0.5\baselineskip}
    \item On \textit{What new challenges have arisen? (RQ5)}. A primary issue is the lack of \textit{multidisciplinary datasets}, which limits the generalizability of recommendations across different academic domains, hindering the development of more robust systems. Additionally, \RPRSS\ must address \textit{dynamic user preferences}, as interests of the researchers evolve over time, necessitating systems that can adapt to these shifts. The integration of \textit{social media} data also presents both opportunities and challenges; while it can provide valuable insights into collaboration networks and academic trends, it introduces complexities related to data quality, privacy, and integration. Moreover, \textit{multi-task learning} in \RPRSS—where multiple related tasks are performed simultaneously—offers potential for enhancing recommendation accuracy but raises issues concerning task prioritization and resource allocation. Finally, \textit{what-if} scenarios are key for adapting \RPRSS\ to changing academic contexts. While synthetic data has been explored, its full potential for enhancing system scalability and adaptability remains underutilized.

\end{itemize}

\section{Conclusion}\label{sec:Conclusion}
In this survey, a deep and detailed analysis of \RPRSS\ from November 2021 to December 2024, building upon the previous surveys by \cite{IJ:DL:BEEL:2015, IJ:IEEE:BAI:2019,IJ:DL:KREUTZ:2022} is provided. A total of 63 works have been analyzed in depth, providing a comprehensive snapshot of the current trends and challenges in \RPRSS. Unlike prior surveys, this survey goes beyond simply examining the techniques and models, offering a through exploration of how these methods are applied at various stages of the recommendation process, which was not fully addressed in previous surveys. Additionally, multiple tables that organize and summarize the key elements of \RPRSS\ are presented, providing valuable insights into the current status of the field. By addressing key research questions—such as how \RPRSS\ generate recommendations, the datasets used, evaluation methods, the challenges faced, and the emerging challenges in the field—a comprehensive understanding of the current status of \RPRSS\ is gained.

Although \RPRSS\ have made significant progress, several challenges remain. \textit{User modeling}, although improving with a 10\% increase in its use compared to previous surveys, still faces limitations as many systems continue to rely on generic approaches rather than fully personalized profiles, which impacts recommendation relevance. Evaluation metrics have seen some progress, with increased use of \textit{recall} and \textit{ranking} measures. However, there is still limited adoption of more nuanced metrics like \textit{novelty}, \textit{serendipity}, and \textit{user judgment} metrics, which are crucial for a richer user experience.

Beyond these aspects, other critical issues remain underexplored. The lack of \textit{detailed documentation}, the \textit{privacy} challenges, and the \textit{target audience} misalignment are prominent areas requiring attention for further development. Moreover, despite the recognition of the importance of \textit{diversity} and \textit{fairness} in recommendations, very few works explicitly tackle these concerns. 

A key challenge is the \textit{research-to-practice} gap, primarily driven by \textit{scalability} concerns. While there have been significant advancements in the technical aspects of \RPRSS, translating these developments into real-world applications remains difficult. The challenge lies in applying state-of-the-art models at scale, which limits the practical deployment of these systems. Bridging this gap will be critical to ensuring the real-world impact and long-term success of \RPRSS.

With the incorporation of new advances and sources of information, new challenges also arise, alongside the need to revisit and push forward aspects that have not received enough attention. Key emerging challenges include the need for \textit{multidisciplinary} datasets, the ability to adapt to \textit{dynamic user preferences}, and the integration of \textit{social media} data. Additionally, the rise of \textit{multi-task} learning and the potential of synthetic data for \textit{what-if} scenarios introduce new opportunities for enhancing recommendation accuracy, adaptability, and scalability. Addressing these issues, while ensuring practical application, will be crucial for advancing the field of \RPRSS.

\section*{Acknowledgements}
Funding: This work is supported by RTI2018-096846-B-C21 (MCIU/ AEI/FEDER, UE – Spanish Ministry of Science, Innovation and University/Spanish State Research Agency/European Regional Development Fund), the project DeepMinor
(CNS2023-144375) funded by MCIN/AEI/10.13039/501100011033 and European
Union NextGenerationEU/PRTR and ADIAN grant IT-1437-22 (Basque Government).

\bibliography{sn-bibliography}% common bib file

\end{document}